\documentclass[adp,a4paper,fleqn%
]{w-art-modifiedbyrs}
\usepackage{w-thm}
\usepackage[]{graphicx}
\usepackage[]{epsfig}
\usepackage{amsmath}
\usepackage{amssymb}

\usepackage{cite}

\newcommand{\kb}{k_{\text{B}}}

\newcommand{\beq}{\begin{eqnarray} }
\newcommand{\eeq}{\end{eqnarray}}

\newcommand{\punkt}{\, .}
\newcommand{\komma}{\, ,}
\newcommand{\summe}[1]{\sum\limits_{#1}}
\newcommand{\sumijs}{ \sum\limits_{\langle i \neq j \rangle\sigma}}
\newcommand{\cplus}[1]{{\bf c}^+_{#1}}
\newcommand{\cminus}[1]{{\bf c}_{#1}}
\newcommand{\nsigma}[1]{{\bf n}_{#1 \sigma}}
\newcommand{\noperator}[1]{{\bf n}_{#1 }}

\newcommand{\hamilton}{{\bf H}}

\newcommand{\opS}{\mbox{\boldmath $ S $}}

\newcommand{\ket}[1]{| #1 \rangle}

\newcommand{\mean}[1]{\langle #1 \rangle}

\begin{document}

\renewcommand{\copyrightyear}{2008}
\pagespan{3}{}
\Reviseddate{}
\Dateposted{}



\title[Extended Hubbard model on a cubic clusters]{The Hubbard model extended by 
nearest-neighbor Coulomb and exchange interaction on a cubic cluster - 
rigorous and exact results}

\author[Schumann]{Rolf Schumann \inst{1,}%
  \footnote{Corresponding author\quad E-mail:~\textsf{schumann@theory.phy.tu-dresden.de}, 
            Phone: +49\,351\,463\,33644, 
            Fax: +49\,351\,463\,37079}}
\address[\inst{1}]{Institute for Theoretical Physics, TU Dresden, Dresden D-01062, Germany}


\author[Zwicker]{David Zwicker \inst{1} }


\begin{abstract}
The Hubbard model on a cube was revisited and extended by both
nearest-neighbor (nn) Coulomb correlation $W$ and {nearest-neighbor}
Heisenberg exchange $J$.
The complete eigensystem was computed exactly for all electron occupancies
and all model parameters ranging from minus infinity to plus infinity.
For two electrons on the cluster the eigensystem 
is given in analytical form. For six  electrons and infinite on-site correlation $U$
we determinded the groundstate and the groundstate energy of the 
pure Hubbard model analytically.
For fixed electron numbers we found a multitude of ground state level crossings.
in dependence on the various model parameters.
Furthermore the  groundstates  of the pure Hubbard model in dependence 
on a magnetic field $h$ coupled to the spins are shown for the complete $U-h$ plane. 
The critical magnetic field, where the zero spin groundstate breaks down is
given for four and six electrons.
Suprisingly we found parameter
regions, where the ground state spin does not depend monotonously on $J$ in the extended model.
For the cubic cluster gas, i.e. an ensemble of clusters coupled to an electron bath, 
we calculated  the density $n(\mu,T,h)$ and the thermodynamical density of states 
from the grand potential.
The ground states and the various spin-spin correlation functions are studied for both attractive and repulsive values of the three interaction constants.
We determined the  various  anomalous degeneration lines, where $n(\mu,T=0,h=0)$ steps more than one, since in this parameter regions exotic phenomena as phase separation are to expect in extended models.
For the cases where these lines end in triple points, i.e.  groundstates of three different occupation numbers are degenerated, we give the related parameter values.
Regarding the influence of the nn-exchange and the nn-Coulomb correlation onto the
anomalous degeneration we find both lifting and inducing of degeneracies 
in dependence on the parameter values.
\end{abstract}

\maketitle                   





%
%
\section{Introduction}
The conspicuous renaissance  in the  study of small Hubbard clusters is mainly  due  to two reasons.
The first is the availability and success of cluster methods, which were developed within 
the context of strong electron correlation during the last decade
\cite{Senechal00,Maier05,Wang05,Potthoff03,Potthoff07,Balzer08} and applied to
problems of high-$T_c$ superconductivity. A considerable
amount of insight into the hot topics of pairing mechanisms, spin-charge separation,
charge ordering and pseudo gap behavior has been reached 
by detailed studies of small Hubbard clusters
\cite{Schumann02,Kocharian06a,Tsai06,Kocharian06b,Palandage07,Kocharian08,Tsai08,Kocharian09}.
The other source is the technical possibility to produce and
reproduce high quality nanostructures which makes it possible to study 
the behavior of  quantum dot clusters 
\cite{Georges99,Craig04,Chen04,Vidan04,Vidan05,Kikoin06,Zhao08}
or even clusters of a few atoms contingently coupled to organic molecules
\cite{Cuniberti05,Heersche06,Donarini06}.
Of course, a detailed knowledge of the cluster physics is inevitable in both fields.
If one aims at clusters with strongly correlated electrons, the Hubbard model,
occasionally extended by nn-Coulomb- and/or exchange interaction,
is often a reasonable starting point.
Complete analytical solutions (in the sense, that all eigenvalues and all eigenvectors are
determined) exist for the Hubbard model with more or less additional terms 
for the dumpbell, the triangle, the square, the tetrahedron, the square pyramid and the
octahedron
\cite{Falicov69,Falicov84,Falicov88,Callaway87,Schumann02,Schumann08,Schumann09}.
Incomplete (but very valueable) solutions, e.g. restricted to the ground state  or to half or low filling,
big external fields etc. exist for $n$-site rings \cite{Lieb68,Korepin00}.
For other models analytical expressions are available 
for special electron (or hole) occupations and/or parameter sets, e.g. for the square with 
next nearest neighbor hopping $t'$  equal to nearest neighbor hopping $t$.

The present paper is very related to our former studies
of  the (extended) Hubbard model on
three and four-site clusters and the related cluster gases \cite{Schumann02,Schumann08,Schumann09}.
It is rather astonishing to what a degree it is possible to understand 
qualitatively the rather involved phase diagram of the $T_c$ superconductors on the bases of the
square-cluster gas model. 
The main reason for that is the degeneration of cluster states with different particle numbers and spin, 
which has a great influence of the coupling of spin- and 
charge degrees of freedom, and indeed, the degeneracy of the ground states for
the four-site cluster occupied either with two or four electrons, is responsible for
the charge separation and inhomogeneities in the pseudogap phase 
of the cuprates, whereas
their bosonic character accounts for the superconducting phase
\cite{Kocharian06a,Kocharian06b}. Also the d-wave symmetry of
the gap seems to be a direct consequence of the symmetry of 
the four-site cluster ground states \cite{Tsai06,Tsai08}.
As was shown in \cite{Schumann08,Schumann09}, the addition of 
Coulomb or Heisenberg interaction to the pure Hubbard Hamiltonian influences 
this degeneration considerably.

In the present paper we re-examine the Hubbard model on a cube to study the influence
of additional nearest-neighbor (nn) Coulomb and nn-exchange interaction, which are added to the
standard Hubbard model. The main focus of the present paper lies on the grand canonical
spectrum and its degeneracies between states differing by more than one electron. 
We want to see, where such 
regions in the complete parameter space of the pure Hubbard model are and how they are influenced
by the additional interactions.
After introducing the model in the next section,
the third section will be based on a numerical calculation of the complete canonical 
potential for every occupation number.
The fourth section is devoted to ensembles of clusters coupled 
to an electron reservoir, the cubic cluster gas.
Here, the focus lies on ground state phase diagrams, which show the degenerations differing by
more than one electron.
The behavior in the vicinity of the degeneration points or lines may give
rise to a likewise rich physics in the ``cluster gas'' or extended systems 
built from these
clusters as it was observed in the relation of the four-site Hubbard 
model on a square to the square lattice.
In a concluding section we will discuss the results.
\section{The Hamiltonian}
In the following, we consider the model
\begin{align}
	\hamilton &= \hamilton_H+\hamilton_C+\hamilton_J
	\label{eqn:hamiltonian_total}
\intertext{with $\hamilton_H$ being the pure Hubbard model:}
	\hamilton_H &= t \sumijs \cplus{i \sigma}\cminus{j \sigma} + 
		\frac{1}{2}\summe{i \sigma} \bigl(
			U\nsigma{i}\noperator{i-\sigma}- 2\mu\nsigma{i} - \sigma h \nsigma{i} 
		\bigr ) 
	\komma 
	\label{eqn:hamiltonian_simple}
\intertext{$\hamilton_C$  is the nn-Coulomb repulsion}
	\hamilton_C &= \frac{W}{2}\,\sum_{\langle i \neq j \rangle} \noperator{i} \noperator{j} \punkt
	\label{eqn:hamiltonian_nn_coulomb}
\intertext{Furthermore a nn-Heisenberg exchange term $\hamilton_J$ }
	\hamilton_J &= \frac{J}{2}\,\sum_{\langle i \neq j \rangle} \opS_i \opS_j
	\label{eqn:hamiltonian_exchange}
\end{align}
was added.
Here $\cplus{i \sigma}$ and $\cminus{i \sigma}$ are the creation and 
destruction operators of electrons at site $i$ with spin $\sigma$,
$\noperator{i \sigma}=\cplus{i \sigma}\cminus{i \sigma}$ and
$\noperator{i}=\sum_{\sigma} \noperator{i \sigma}$. $\opS_i$ indicates
the local spin operator at site $i$.
For a detailed physical reasoning of the additional terms in the Hamiltonian see 
e.g. \cite{FuldeBuch}. Our Hamiltonian is an extension of the Hamiltonian used
e.g. in \cite{Davoudi06}, 
since we treat the nearest-neighbor 
correlation $W$ and the nearest-neighbor exchange constant $J$ as independent parameters.
This is done mainly to cover pure Heisenberg type models by setting $W/t=0,\, J/t\ne0$.
The chemical potential $\mu$  and the magnetic 
field $h$ in $z$-direction are introduced
to take into account the effects of doping and applying external magnetic fields.
Please note, that the signs in front of the hopping and exchange term are positive,
thus one has to be carefully while comparing with papers where other conventions
are used.
There are $4^8$ states within that model. Utilizing all symmetries, we reduced the Hamilton matrix
to a block diagonal form. It has to be noted, that the particle-hole-symmetry is broken in case of
$W/t \neq 0$. For details of this symmetry reduction we refer to \cite{Schumann02}.
In contrast to our previous works, the biggest block has dimension 88, what
prevented a closed form solution. Nevertheless, the smallness of the analytically given block
matrices allows a very quick and extremely precise evaluation of the eigensystem.
The results of the pure Hubbard model were compared to the results of Callaway et al.
\cite{Callaway87}
and to a complete independent pure numerical calculation of the spectrum \cite{Richter08}.

One interesting property of the cluster is its equivalence to the simple cubic lattice, using
periodic boundary conditions in every other bond with kinetic energy set to $t/2$ \cite{FreFal90}.
All following results are applicable to that system as well.

\begin{table}
  \begin{tabular}{@{}lllllllllll@{}}
  Irreducible representation: &
  $\Gamma_1$ & $\Gamma_2$ & $\Gamma_3$ & $\Gamma_4$ & $\Gamma_5$ & 
  $\Gamma_6$ & $\Gamma_7$ & $\Gamma_8$ & $\Gamma_9$ & $\Gamma_{10}$ \\
  Dimension: &
  1 & 1 & 2 & 3 & 3 & 1 & 1 & 2 & 3 & 3 \\
  \end{tabular}
  \caption{Dimensions of the irreducible representations of the cubic point group
$\text{O}_{\text{h}}$.}
  \label{tbl:irrep_dimensions}
\end{table}

\section{Fixed electron occupations}
\subsection{Rigorous results}
For two electrons on the cluster the Hamilton matrix  can be
brought to block diagonal form by applying all the U-independent symmetries,
with the maximum block dimension of four. 
Thus we have for that case 
both the eigenvalues and eigenvectors as closed analytical expressions.
We mention here, that an alternate exact solution to the two-electron 
problem in the domain $s_z=0$ in d-dimensional hypercubes
was given before in Ref. \cite{Caffarel98}. Although we expect that our complete solution
for the nondegenerate part of the spectrum  coincides  with their solution in the threedimensional case in  the considered domain, a comparison was prevented by the very different conventions characterizing the eigenstates  and a  different set of symmetries employed. In the appendix
we give the symmetries of the groundstates and the corresponding energies 
in dependence on the model parameters in a closed form.\\
For the pure Hubbard model (i.e. J=0 and W=0) in the infinite $U$ limit we have investigated
the Nagaoka state \cite{Nagaoka65,Tasaki89}, which is quite trivial since it has spin $\frac{7}{2}$ and $\Gamma_1$ symmetry.  It is an eigenstate of the Hamiltonian for finite $U$ also, despite the
fact, that it does not depend on the on-site correlation.
The related energy is $-3t$ which becomes the groundstate for $U\ge 39.641741191(1)$. 
For $N=6$ in the infinite $U$ limit the groundstate is highly non-trivial. It  has again spin zero and
$\Gamma_1$ symmetry. 
In contrast to the Nagaoka state it depends monotonously on the on-site correlation.
It becomes the groundstate for $U\ge61.312646262(1) t$, where the related 
energy is $-4.90564t$ and goes to $-2 \sqrt{4 +\sqrt{3}}\,t$ if the on-site correlation
goes to infinity. The letter value agrees to 6 digits with the numerical calculated number given by Takahashi \cite{Takahashi79}.

\subsection{Pure Hubbard model}
The pure Hubbard model for the isolated cubic cluster has been discussed convincingly by
Callaway et al. \cite{Callaway87}. We were able to reproduce all figures in the reference and
could approve all properties mentioned in the text. The ground state spin, given in Table 1 of the
reference has been extended to negative correlation {parameters $U$}. The only difference occurs for
$n=5$, where the ground state has minimal spin $S=\frac{1}{2}$ for $U<0t$. For all other occupation
numbers, the spin for negative $U$ is the same as for vanishing $U$.

It is due to the rapidly increasing computer power over more than twenty years, that we are able now to draw a more detailed picture of the magnetic properties
of the system with respect to an applied  magnetic field $h$. For better comparison, the complete parameter
space is shown in the same manner as in \cite{Schumann08} using the following mappings:
\begin{align}
  U'&:=\frac{U}{6t+|U|}
  &
  h'&:=\frac{h}{t+|h|}
  &
  J'&:=\frac{J}{t+|J|}
  &
  W'&:=\frac{W}{t+|W|}
  \label{eqn:scaling_function}
\end{align}
Note, that the correlation parameter has been scaled differently compared to the other
parameters, mainly to emphasize the transition region between weak and strong correlation, which is
located around $U=6t$, the band width of the simple cubic lattice.

\newlength{\figurewidth}
\newlength{\figureheight}
\begin{figure}
  \setlength{\figurewidth}{45mm}
  \includegraphics[width=\figurewidth]{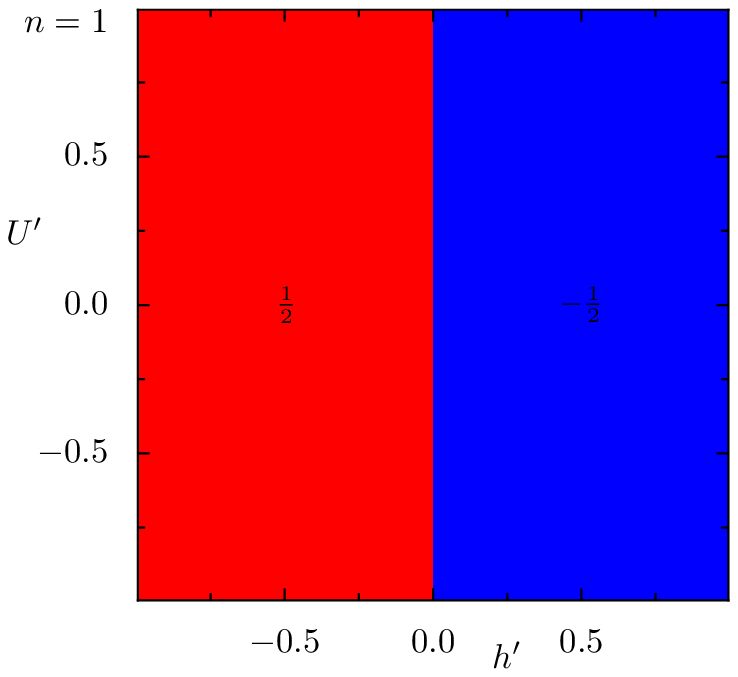}˜a)
  \hfil
  \includegraphics[width=\figurewidth]{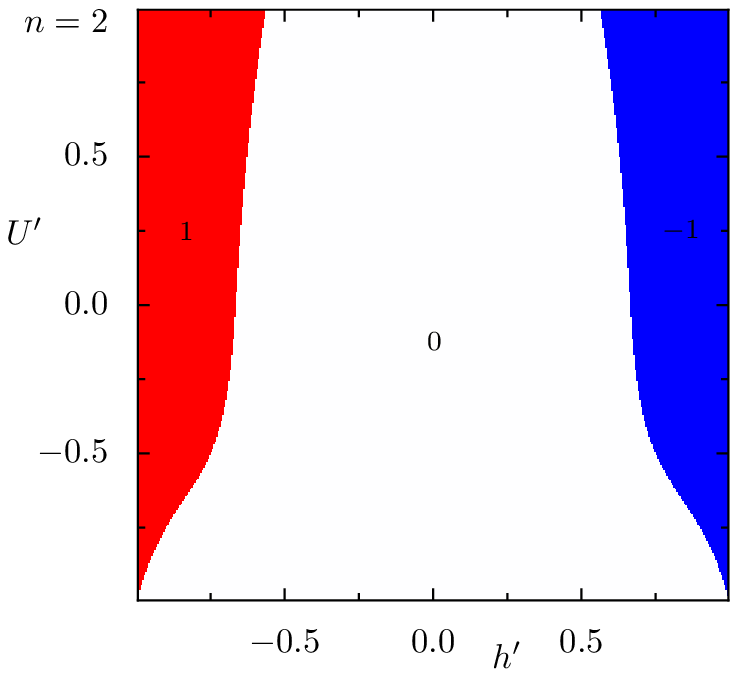}˜b)
  \hfil
  \includegraphics[width=\figurewidth]{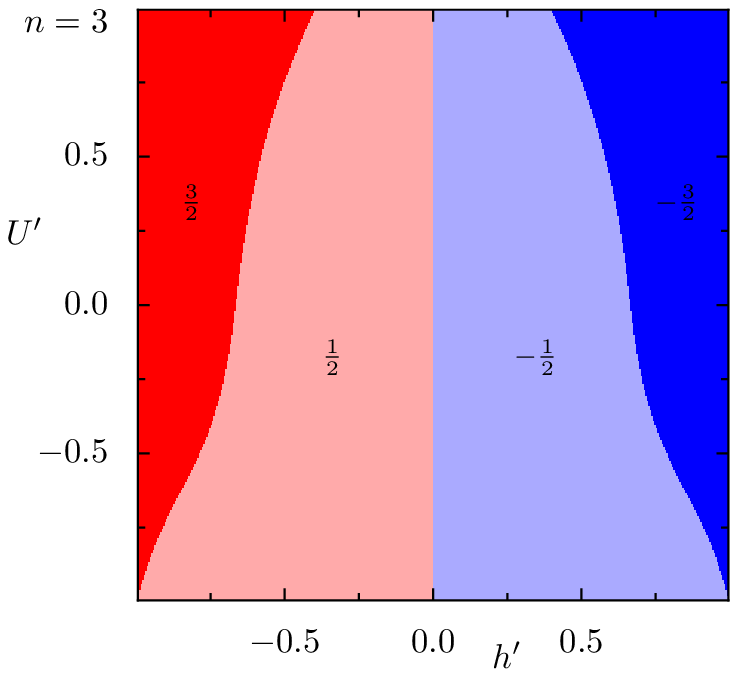}˜c)
  \\[3mm]
  \includegraphics[width=\figurewidth]{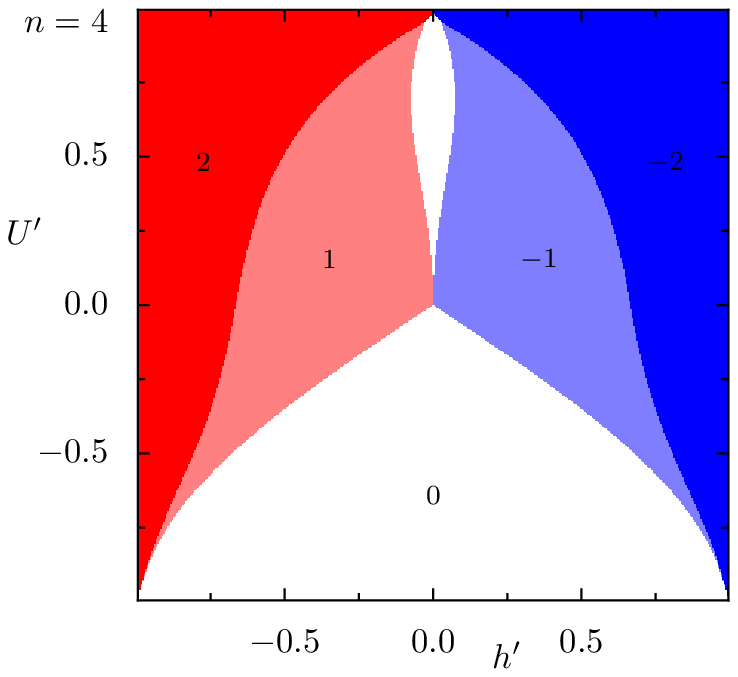}˜d)
  \hfil
  \includegraphics[width=\figurewidth]{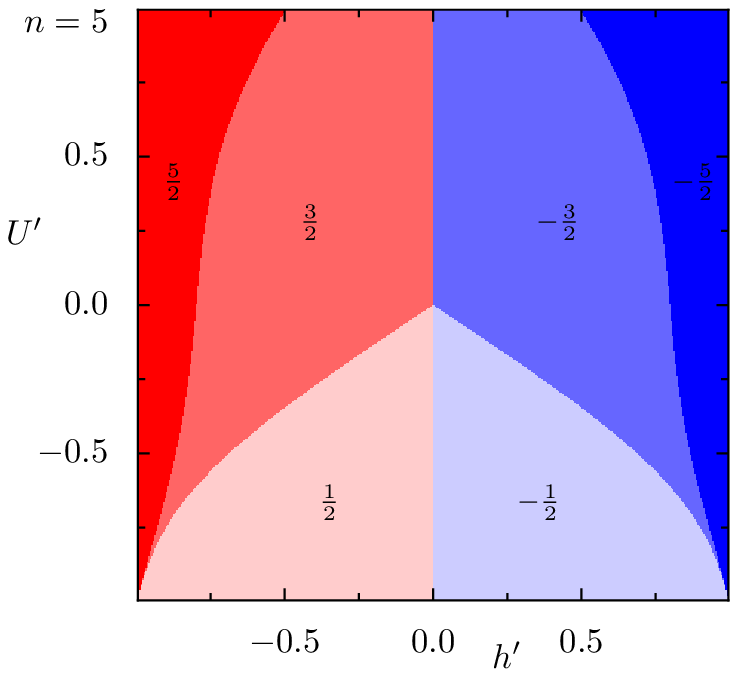}˜e)
  \hfil
  \includegraphics[width=\figurewidth]{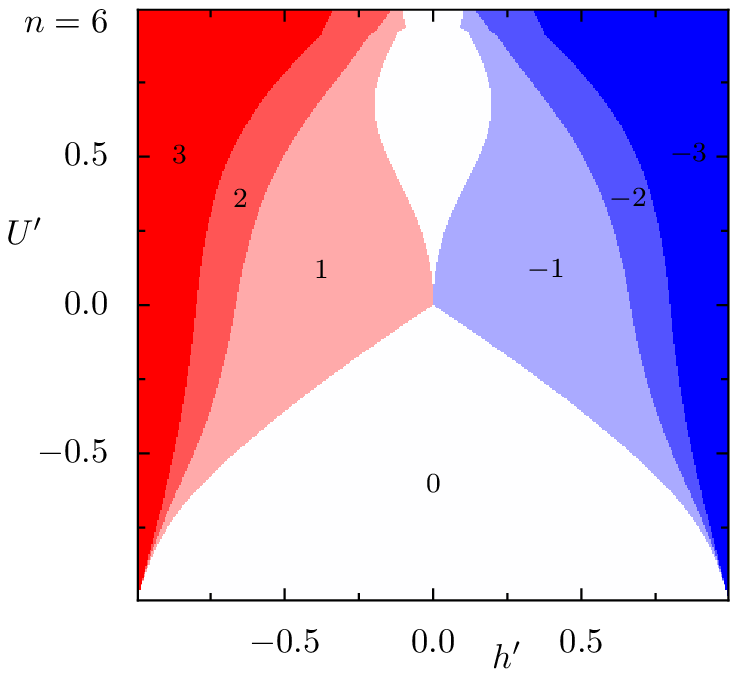}˜f)
  \\[3mm]
  \includegraphics[width=\figurewidth]{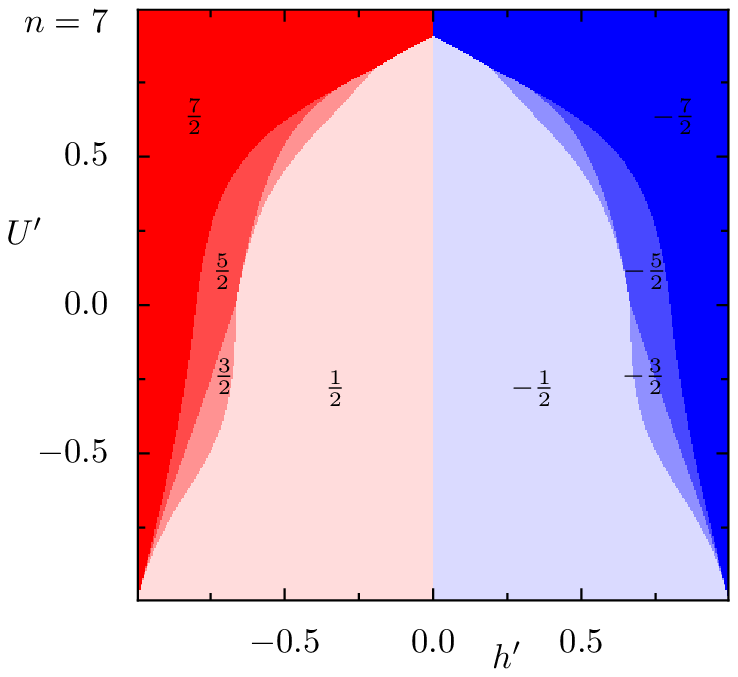}˜g)
  \hfil
  \includegraphics[width=\figurewidth]{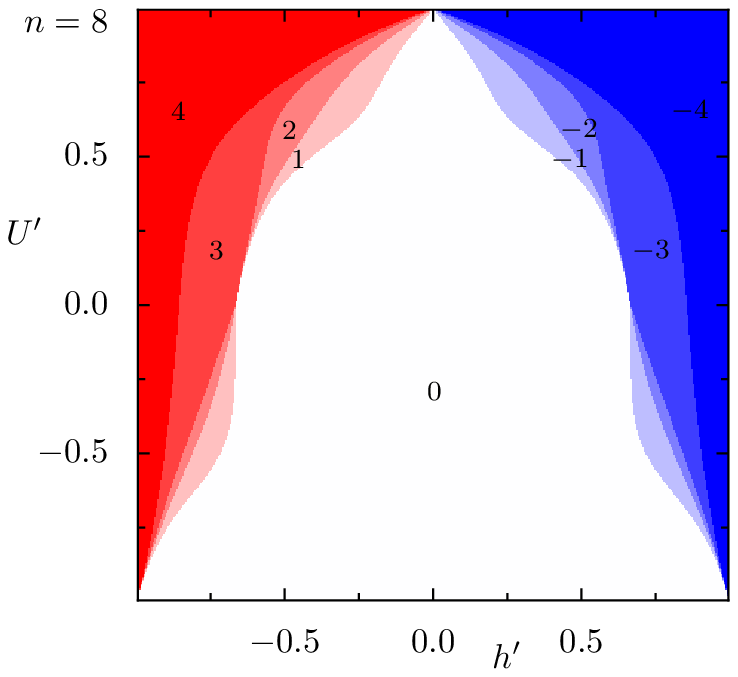}˜h)
  \hfil
  \includegraphics[width=\figurewidth]{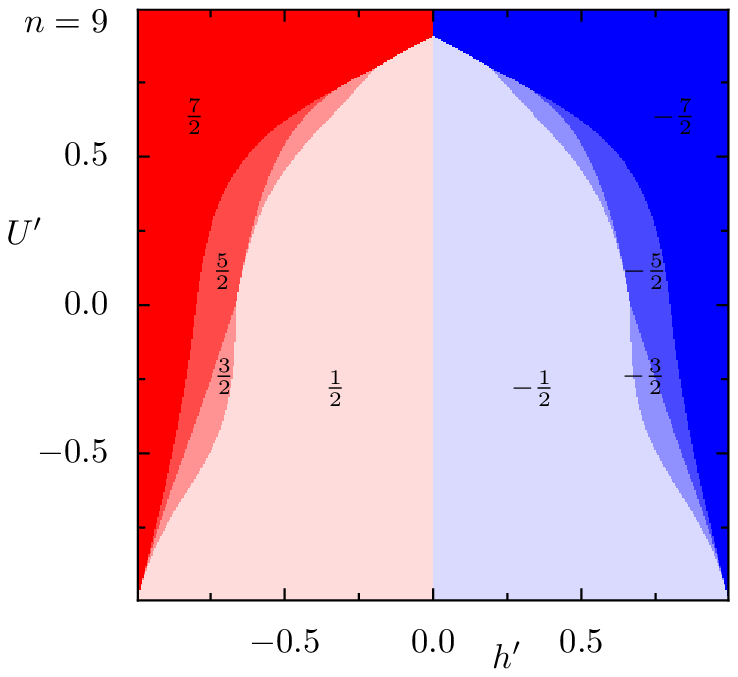}˜i)
    
  \caption
  {Total spin projection $S_z$ of the canonical ground states of the isolated cubic cluster in
dependence on the on-site correlation $U$ and the magnetic field $h$ for all occupation numbers in
increasing order, from a) $n=1$ to i) $n=9$. The colored areas denote a constant value of $S_z$ with
the inscribed value. To show the complete parameter range, the axes have been transformed
non-linearly using eq. \eqref{eqn:scaling_function}.}
  \label{fig:canonical_simple_Sz_Uh}
\end{figure}

Although states of high energy might be interesting for e.g. spectroscopy, the present paper focuses
on the ground states and their dependence on the various parameters. The quantum numbers of the
lowest states for each set of parameters will be shown in a two-dimensional plot, allowing for
studying the dependence on two parameters simultaneously. In analogy to thermodynamics of extended
systems, these diagrams are called \emph{ground state phase diagrams} (GSPDs), although they display
no phase transitions, in the sense of a diverging correlation length. The GSPDs utilize colored
areas to depict regions of constant ground state quantum numbers. The degeneracy of the
respective ground states may be calculated by multiplying the spin degeneration $2S+1$ by the
spatial one, which is equal to the dimension of the irreducible representations
printed in Table \ref{tbl:irrep_dimensions}. If a magnetic field is applied, the spin degeneracy is lifted  of course.

The first GSPDs in Fig. \ref{fig:canonical_simple_Sz_Uh} analyze the isolated cubic cluster with
respect to the on-site correlation $U$ and an external magnetic field $h$ for all occupation
numbers.
They obviously are symmetric with respect to the magnetic field, as expected. Additionally, the case
g) $n=7$ is identical to i) $n=9$ as is required by the particle-hole-symmetry. 
Since coding errors usually destroy this symmetry, we calculated the pictures
independently, thus increasing the confidence in our results.
Nevertheless, in what follows only 
the bottom half of the occupation numbers will be shown, if particle-hole-symmetry holds, what is the case for $W/t=0$.

\begin{figure}
  \setlength{\figureheight}{30mm}
  \sidecaption
  \includegraphics[height=\figureheight]{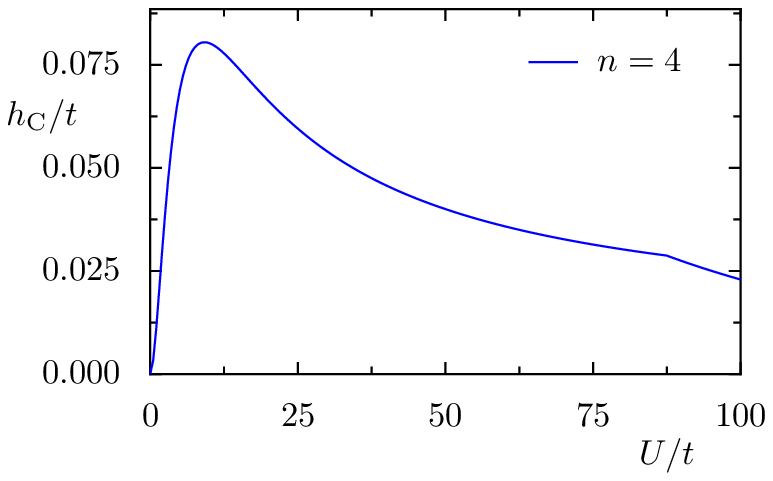}
  \hfil
  \includegraphics[height=\figureheight]{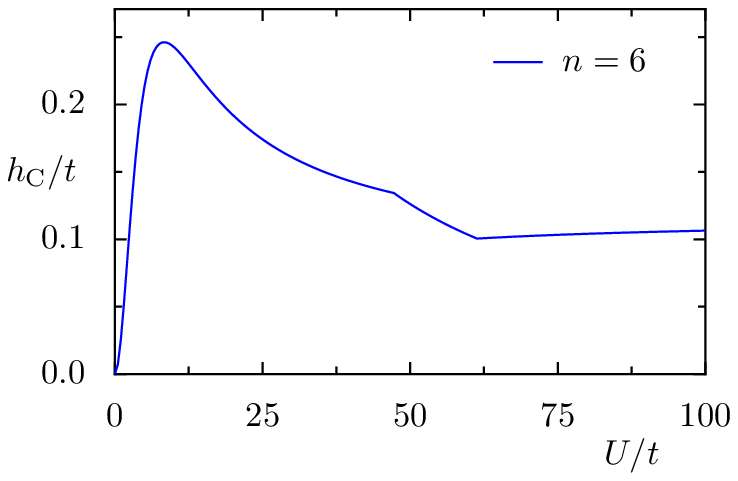}
  \caption
  {Critical magnetic field $h_\text{C}$, where the ground states with $S_z=0$ and $-1$ are
degenerated, in dependence on the on-site correlation $U$. The kink
on the right panel is caused by the spatial transition at $U=61.313t$.}
  \label{fig:canonical_simple_hCrit}
\end{figure}
Even occupation numbers exhibit a non-magnetic ground state for absent magnetic field, which is in
accordance to the picture of an antiferromagnetic state. Though, it is an interesting feature, that
in some cases, e.g. $n=4$ and $n=6$, a very small magnetic field is sufficient at
small on-site correlation to produce a magnetic ground state, whereas in the other cases $n=2$
and $n=8$ no such behavior may be seen. 
For the interesting cases $n=4$ and $n=6$, Fig. \ref{fig:canonical_simple_hCrit} shows the critical
magnetic field $h_{\text{C}}$ destroying the non-magnetic ground state in dependence on the on-site
correlation, e.g. the line between the white and light red region of figures d) and f).
The states with $S_z=0$ are most stable in an intermediate region of $U$. Similar
behavior has been found for the tetrahedron at $n=4$, but not for the hexagon, showing
that this is not generic for small clusters. 
The kink in the $n=6$ image is caused by the symmetry transition from a  two fold degenerated $\Gamma_3$ groundstate with spin zero below 
$U=61.313t$ to the singlet $\Gamma_1$ above \cite{Callaway87}.
In case of an odd occupation number, the picture is much more uniform, in that there cannot be any
states with vanishing $S_z$ and the magnetization rises with either increasing magnetic field
strength $|h|$ or the on-site correlation $U$. At fixed $U$, the system runs through all possible
spin projection values $S_z$ with increasing strength of the magnetic field until saturation.
\subsection{Extended Hubbard model}
\setlength{\figurewidth}{72mm}
\begin{figure}[p]
  \includegraphics[width=\figurewidth]{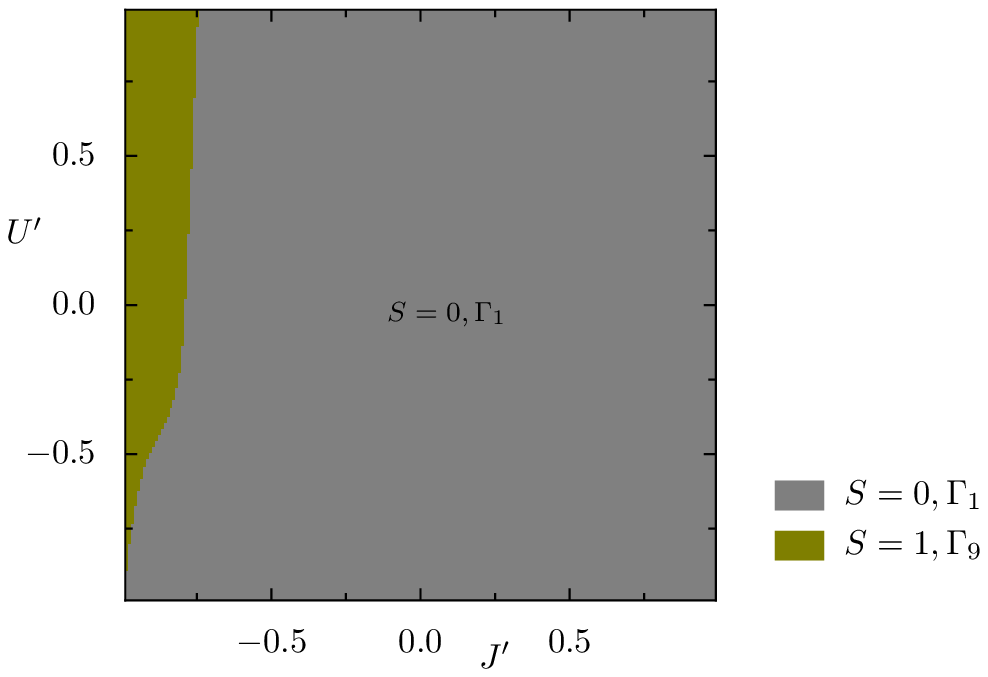}
  \hfil
  \includegraphics[width=\figurewidth]{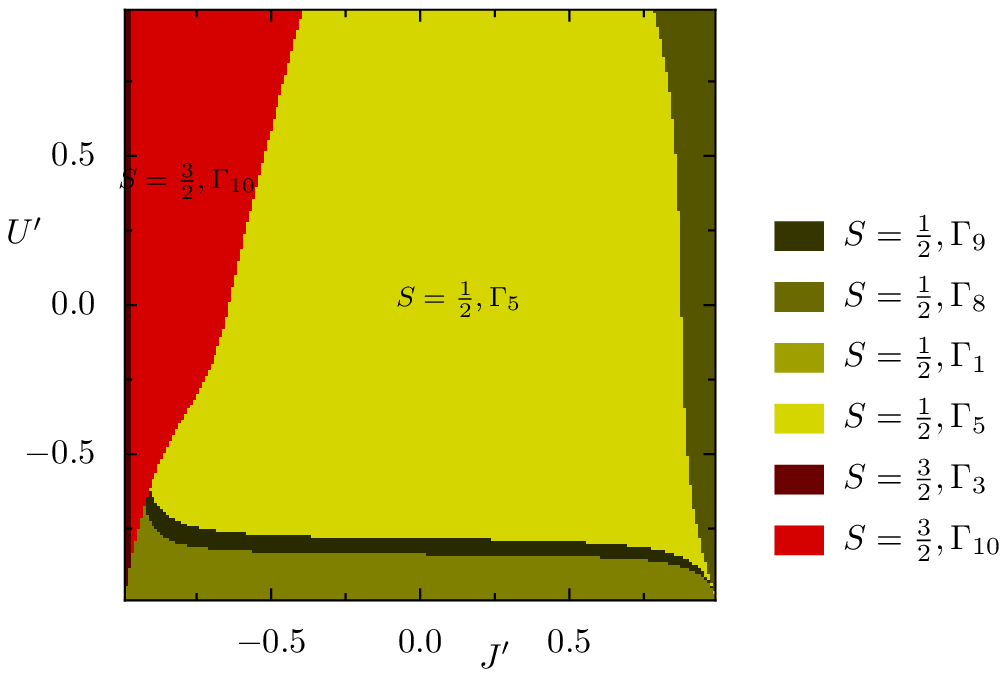}
  \\[3mm]
  \includegraphics[width=\figurewidth]{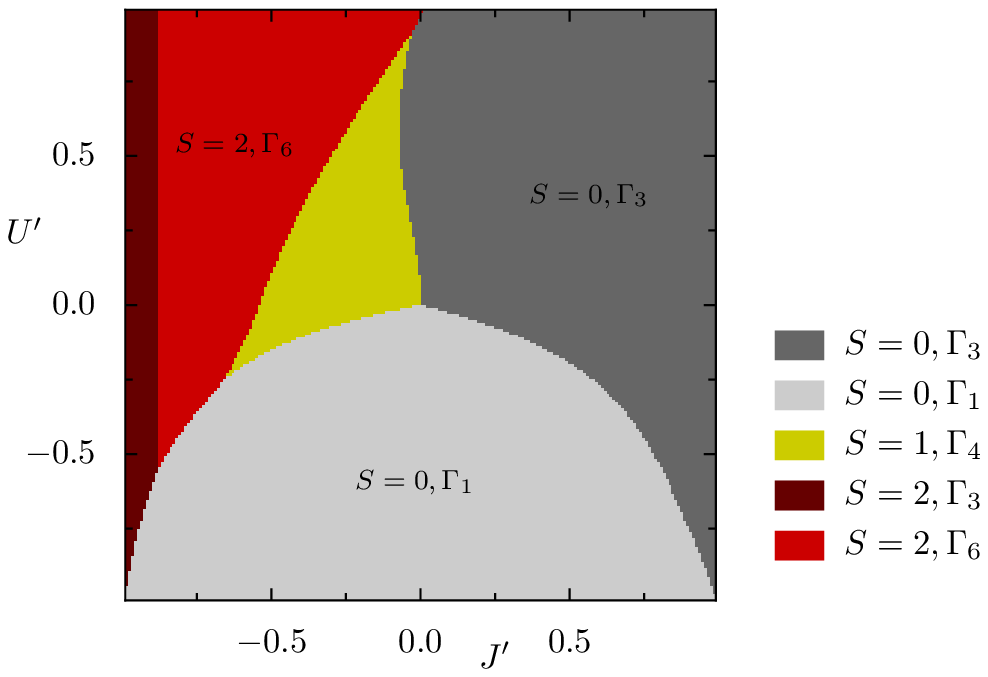}
  \hfil
  \includegraphics[width=\figurewidth]{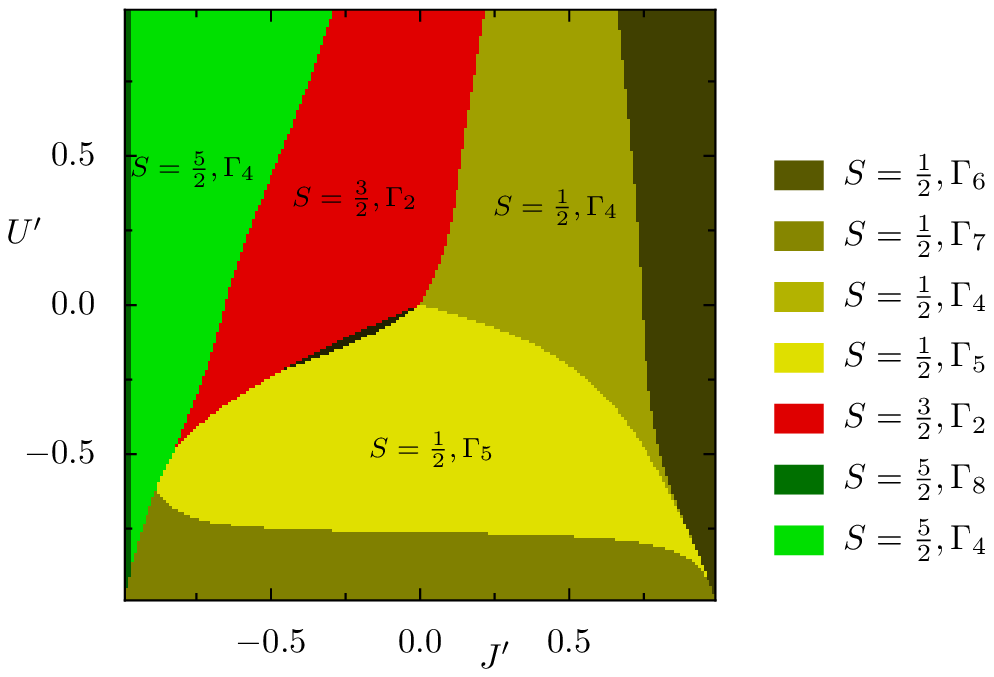}
  \\[3mm]
  \includegraphics[width=\figurewidth]{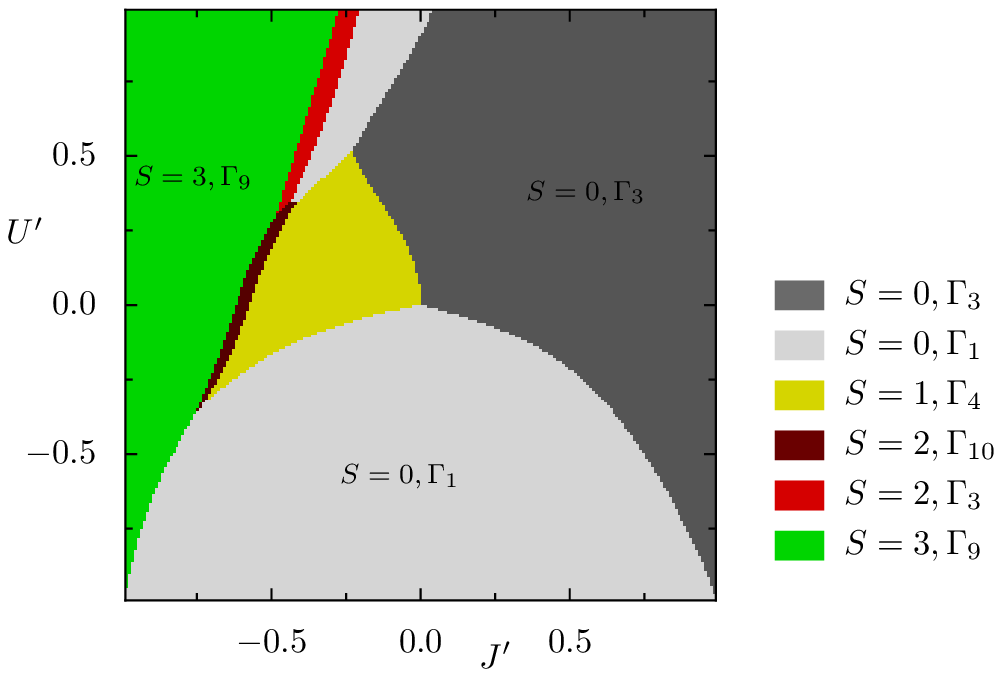}
  \hfil
  \includegraphics[width=\figurewidth]{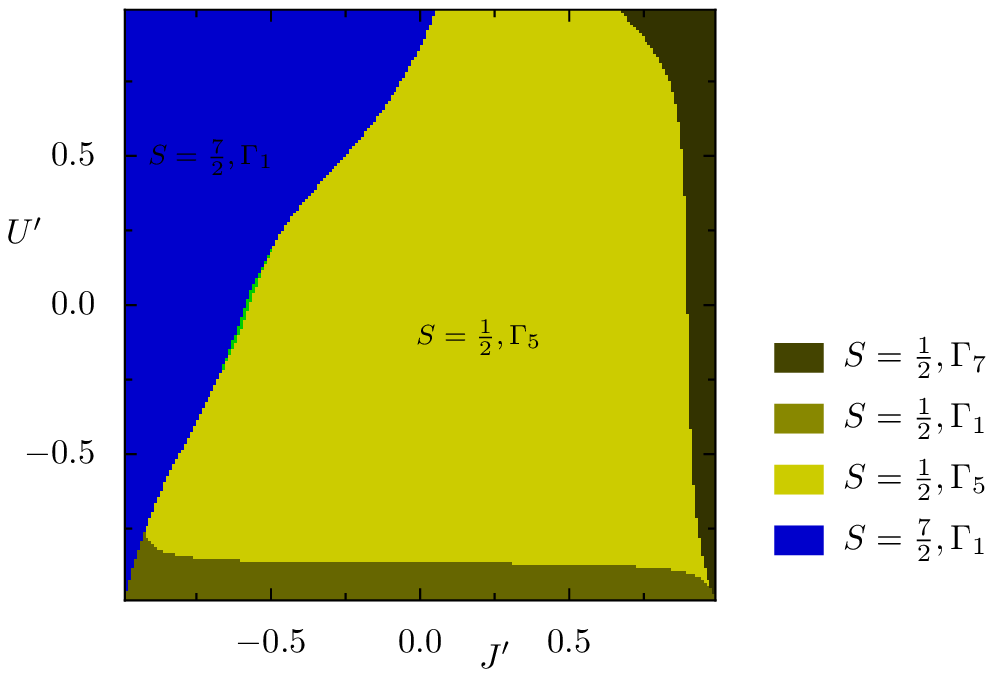}
  \\[3mm]
  \sidecaption
  \includegraphics[width=\figurewidth]{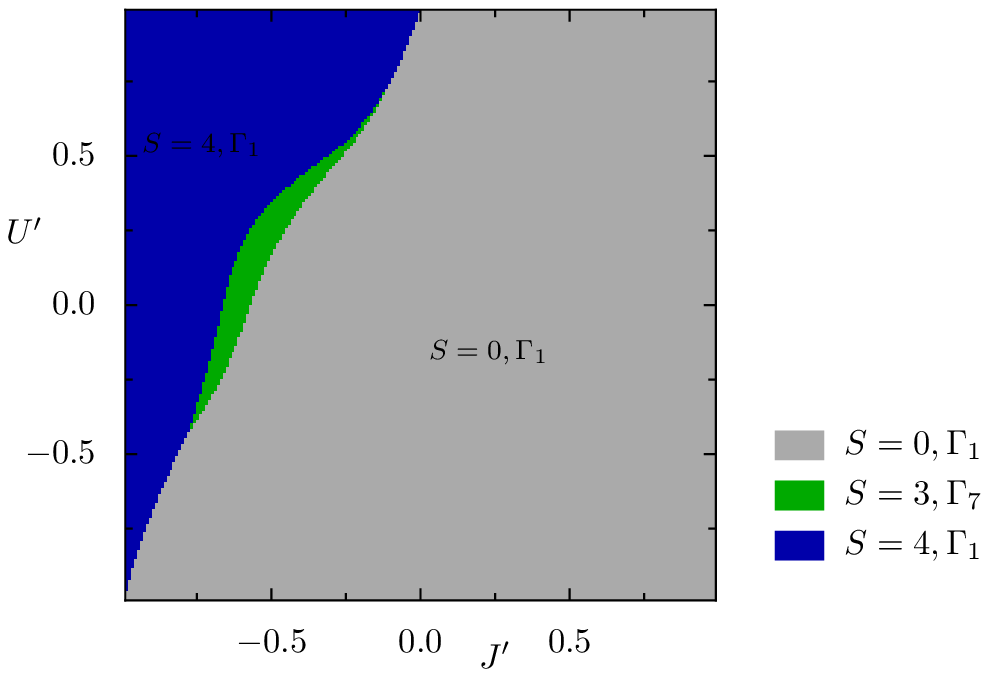}
  \hspace{7mm}
  \caption
  {Canonical GSPDs in dependence on $U$ and spin exchange parameter $J$
for the occupation numbers $n=2\ldots8$. Colored areas stand for ground
states of constant quantum numbers, which are listed on the right of each figure. Because no
magnetic field is applied, all states are degenerated with respect to $S_z$. Degenerations may be
read off by multiplying the spin degeneracy by the dimension of the irreducible representation
$\Gamma_i$.}
  \label{fig:canonical_extended_groundstates_UJ}
\end{figure}

\begin{figure}[p]
  \includegraphics[width=\figurewidth]{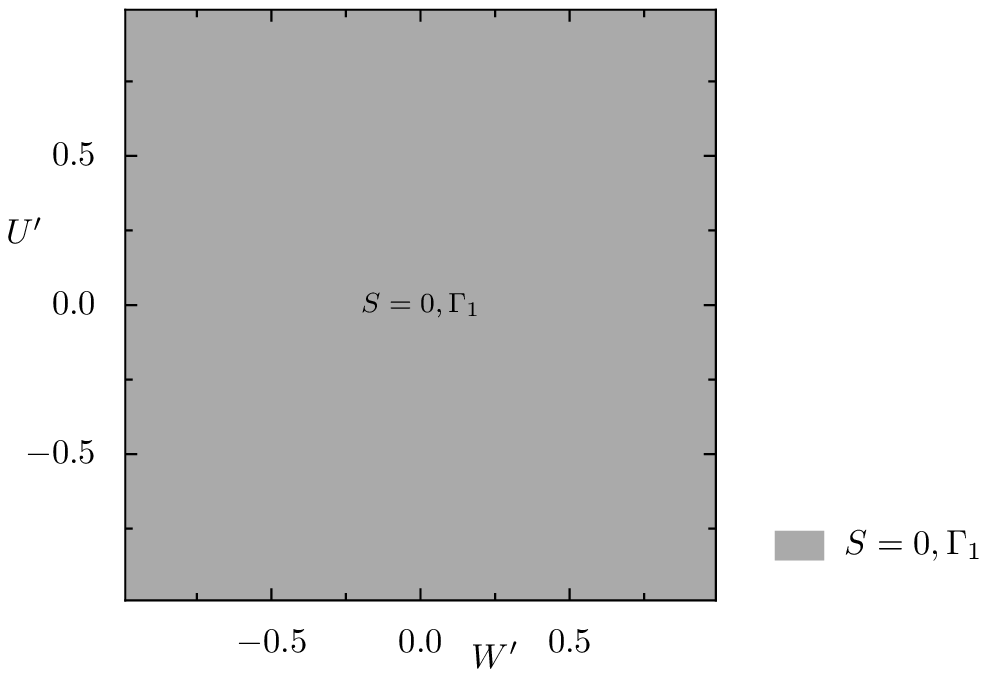}
  \hfil
  \includegraphics[width=\figurewidth]{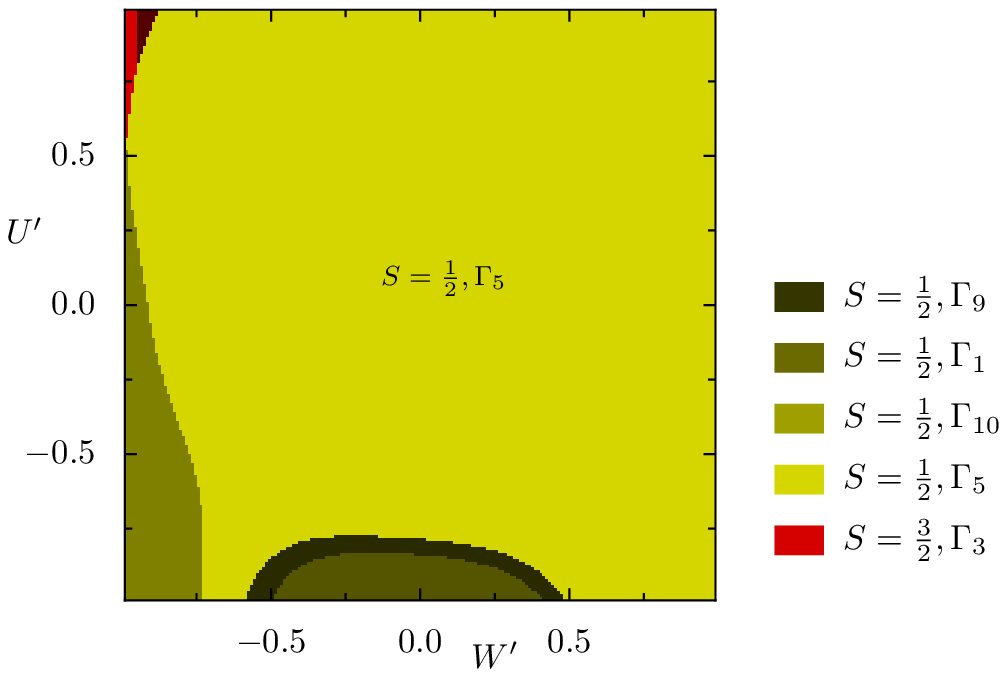}
  \\[3mm]
  \includegraphics[width=\figurewidth]{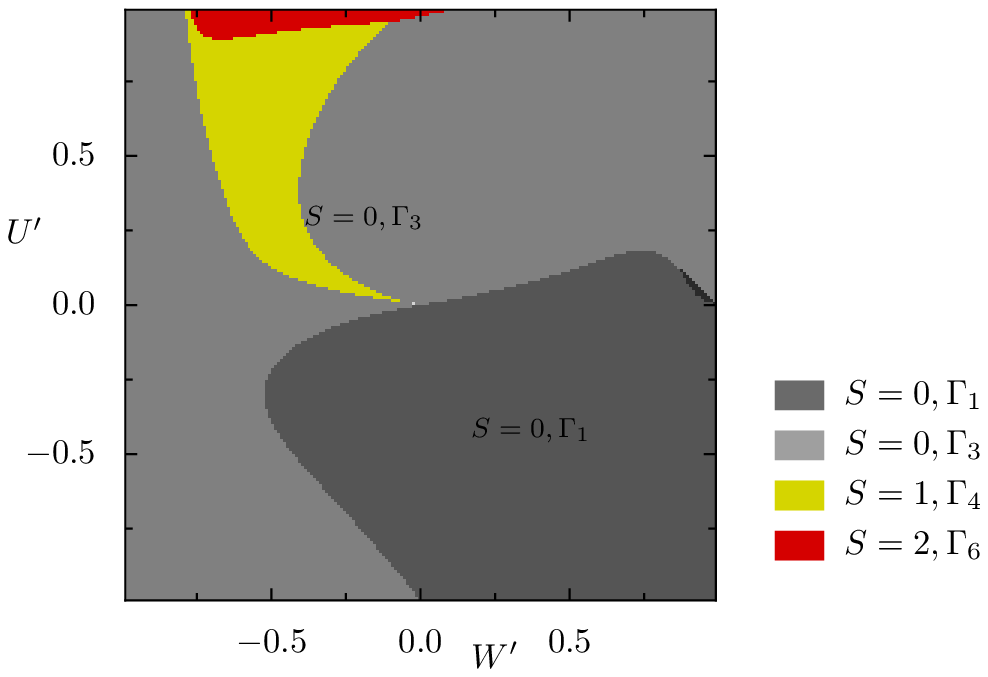}
  \hfil
  \includegraphics[width=\figurewidth]{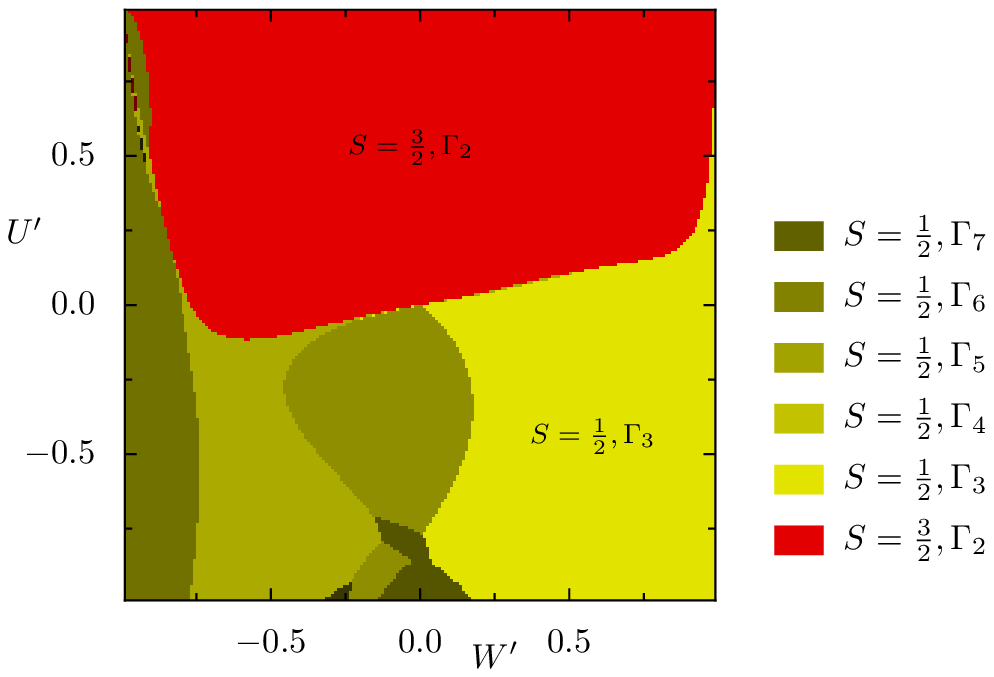}
  \\[3mm]
  \includegraphics[width=\figurewidth]{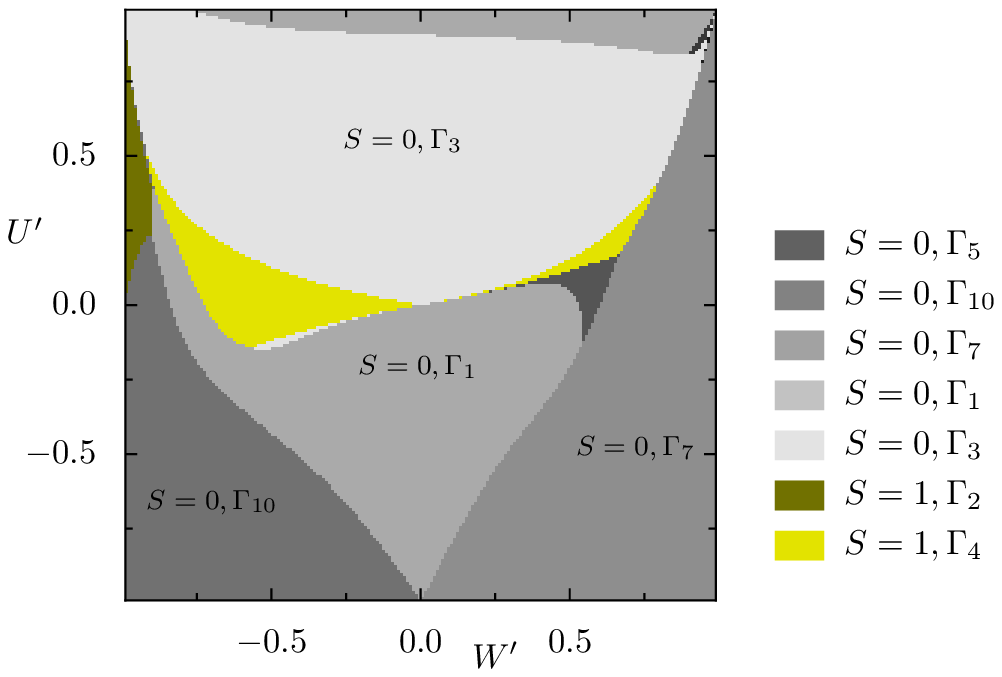}
  \hfil
  \includegraphics[width=\figurewidth]{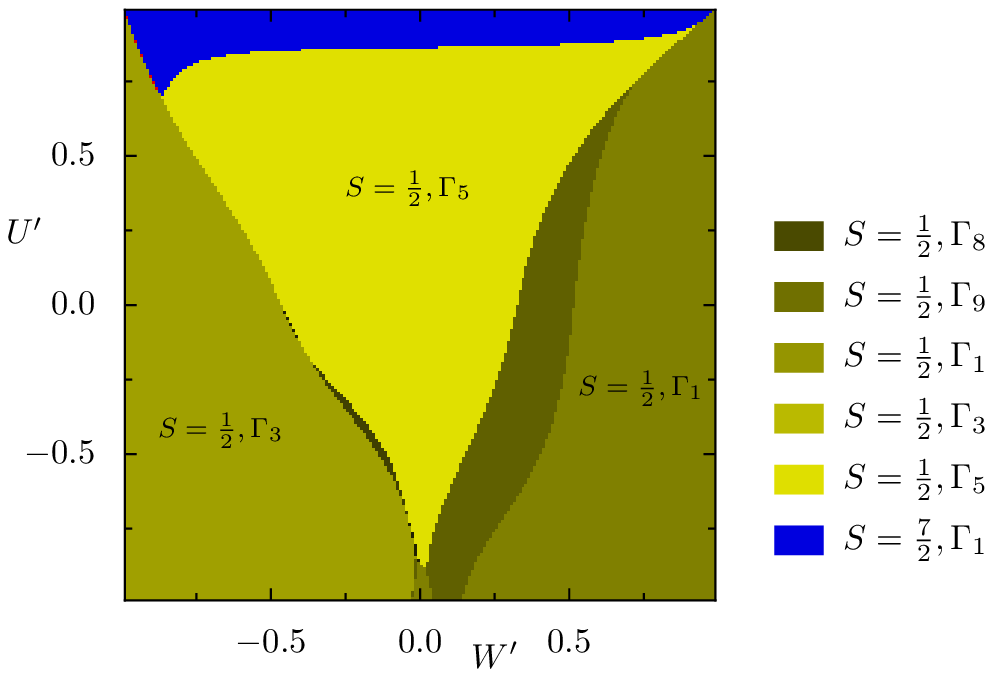}
  \\[3mm]
  \sidecaption
  \includegraphics[width=\figurewidth]{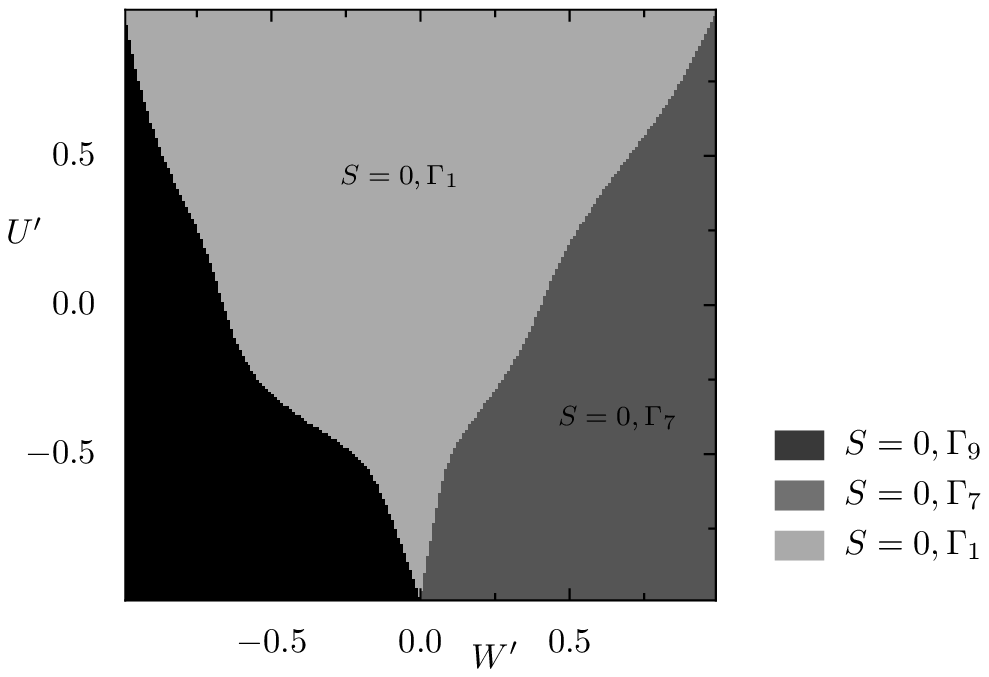}
  \hspace{7mm}
  \caption
  {Canonical GSPDs in dependence on the on-site ($U$)
and nearest-neighbor coulomb interaction ($W$) for the occupation numbers $n=2\ldots8$. Colored
areas stand for ground states of constant quantum numbers, which are listed on the right of each
figure. Because no magnetic field is applied, all states are degenerated with respect to $S_z$.
Degenerations may be read off by multiplying the spin degeneracy by the dimension of the irreducible
representation $\Gamma_i$.
}
  \label{fig:canonical_extended_groundstates_UW}
\end{figure}

 \begin{figure}[p]
 \includegraphics[width=\figurewidth]{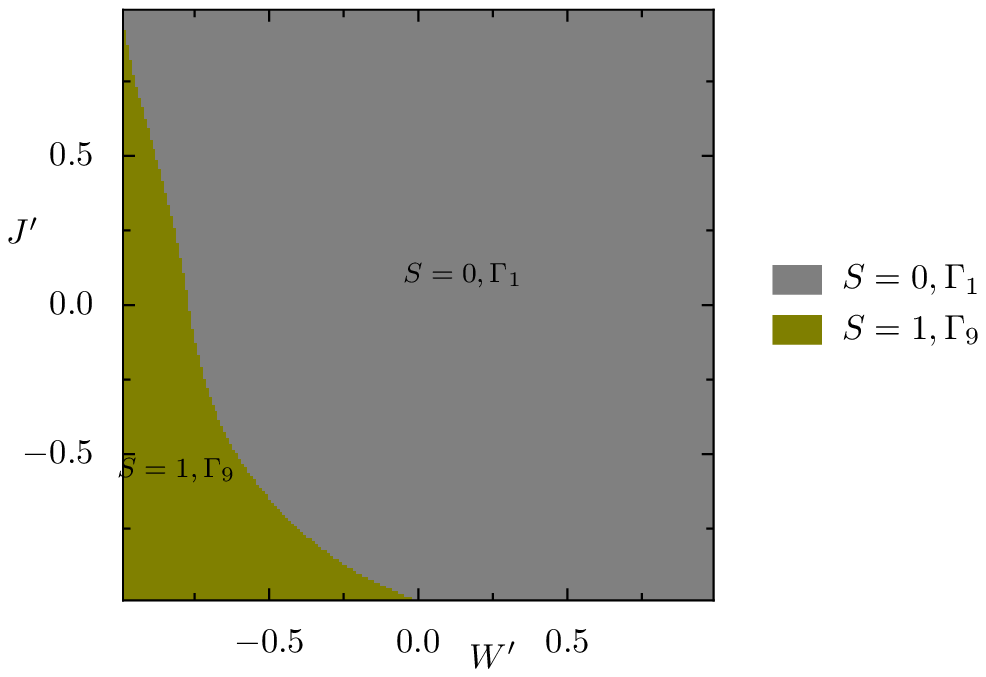}
  \hfil
   \includegraphics[width=\figurewidth]{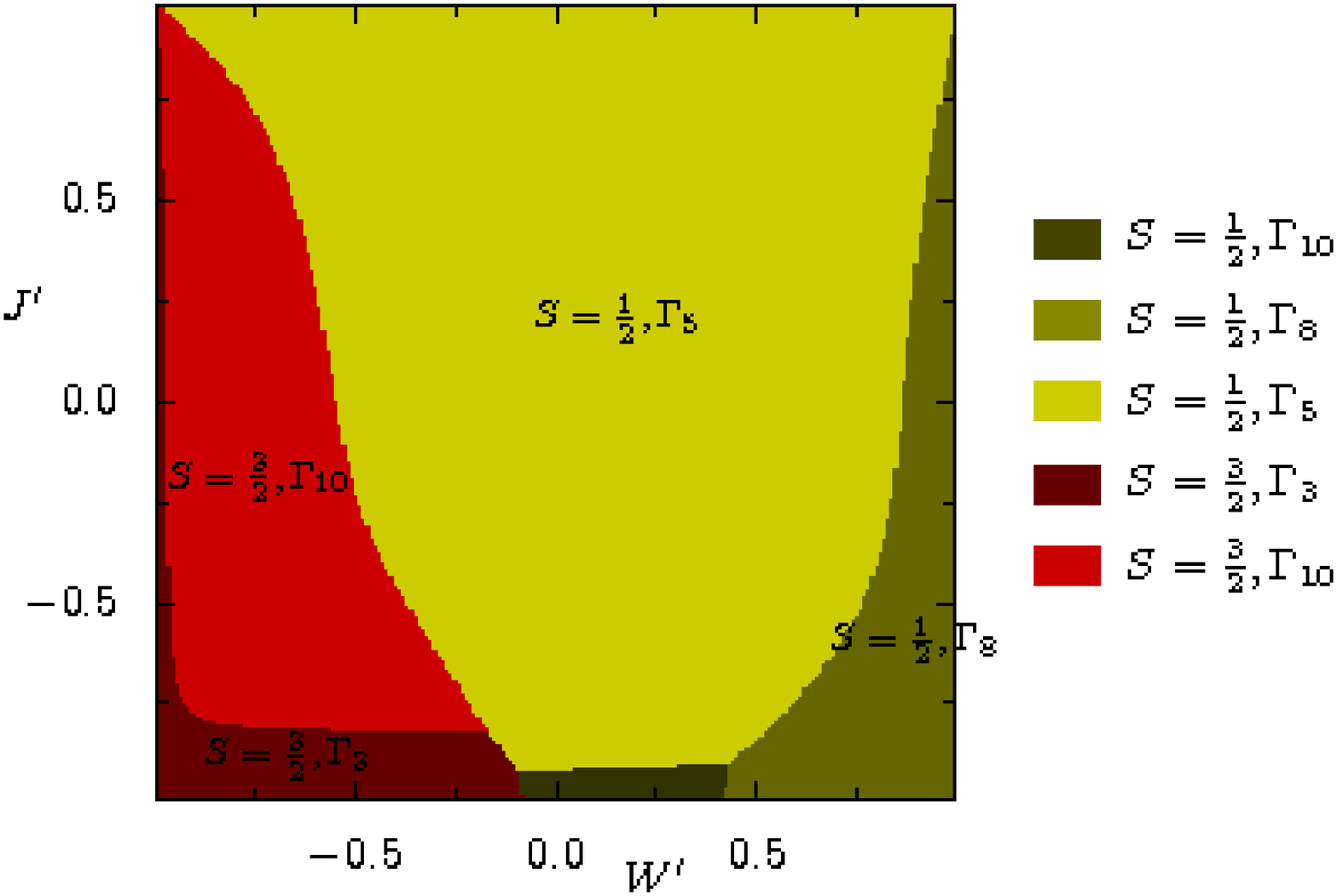}
   \\[3mm]
   \includegraphics[width=\figurewidth]{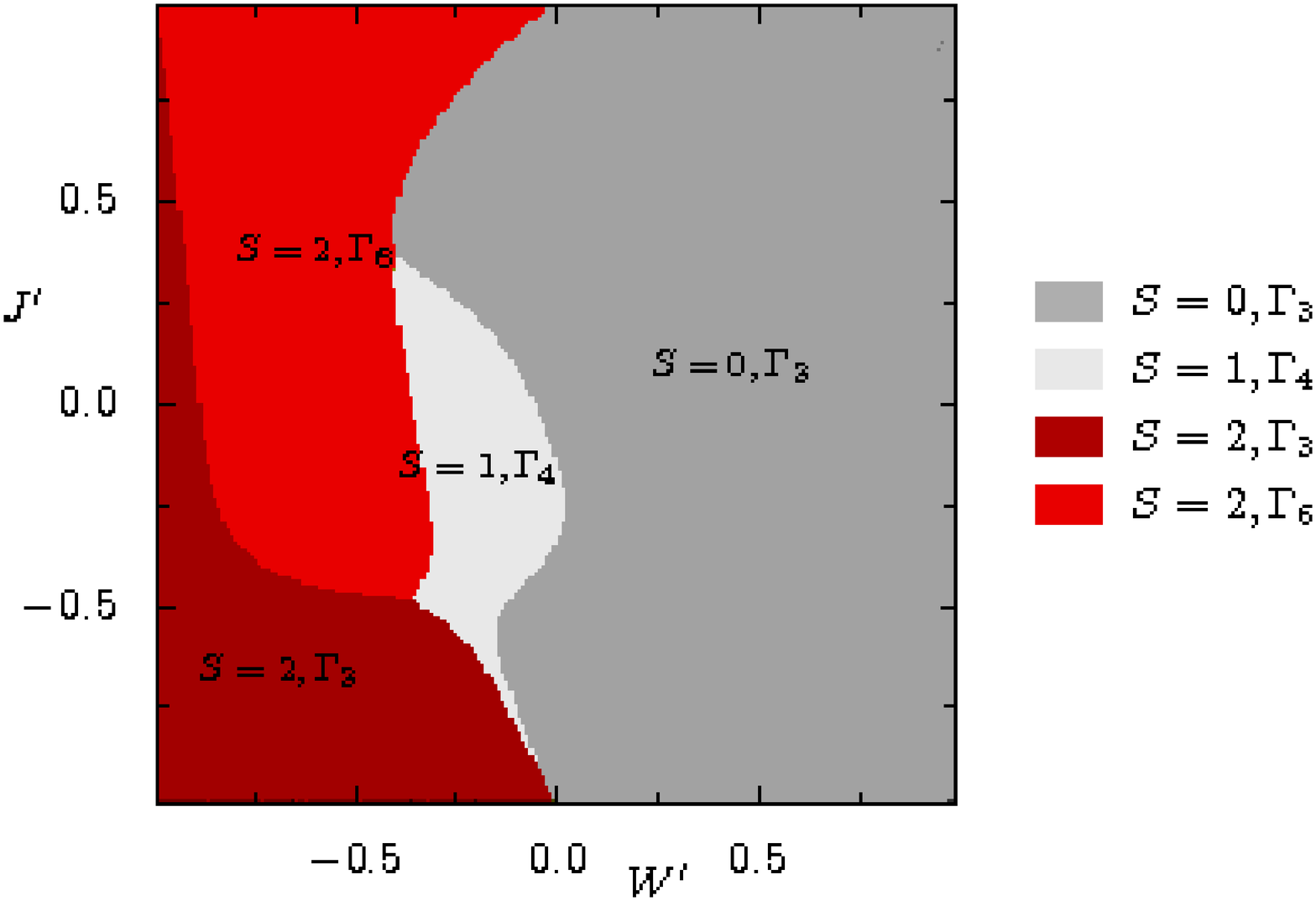}
   \hfil
   \includegraphics[width=\figurewidth]{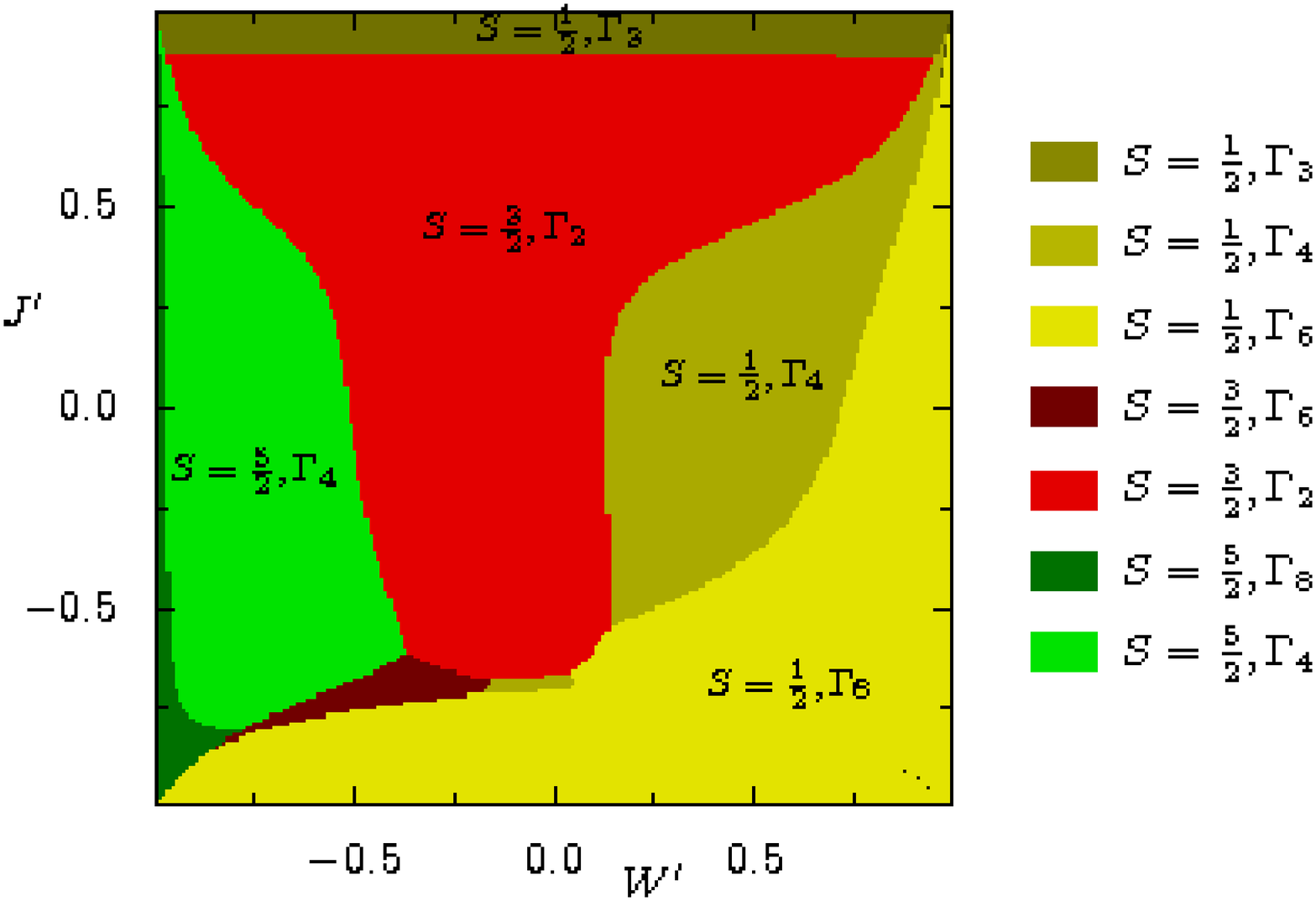}
   \\[3mm]
   \includegraphics[width=\figurewidth]{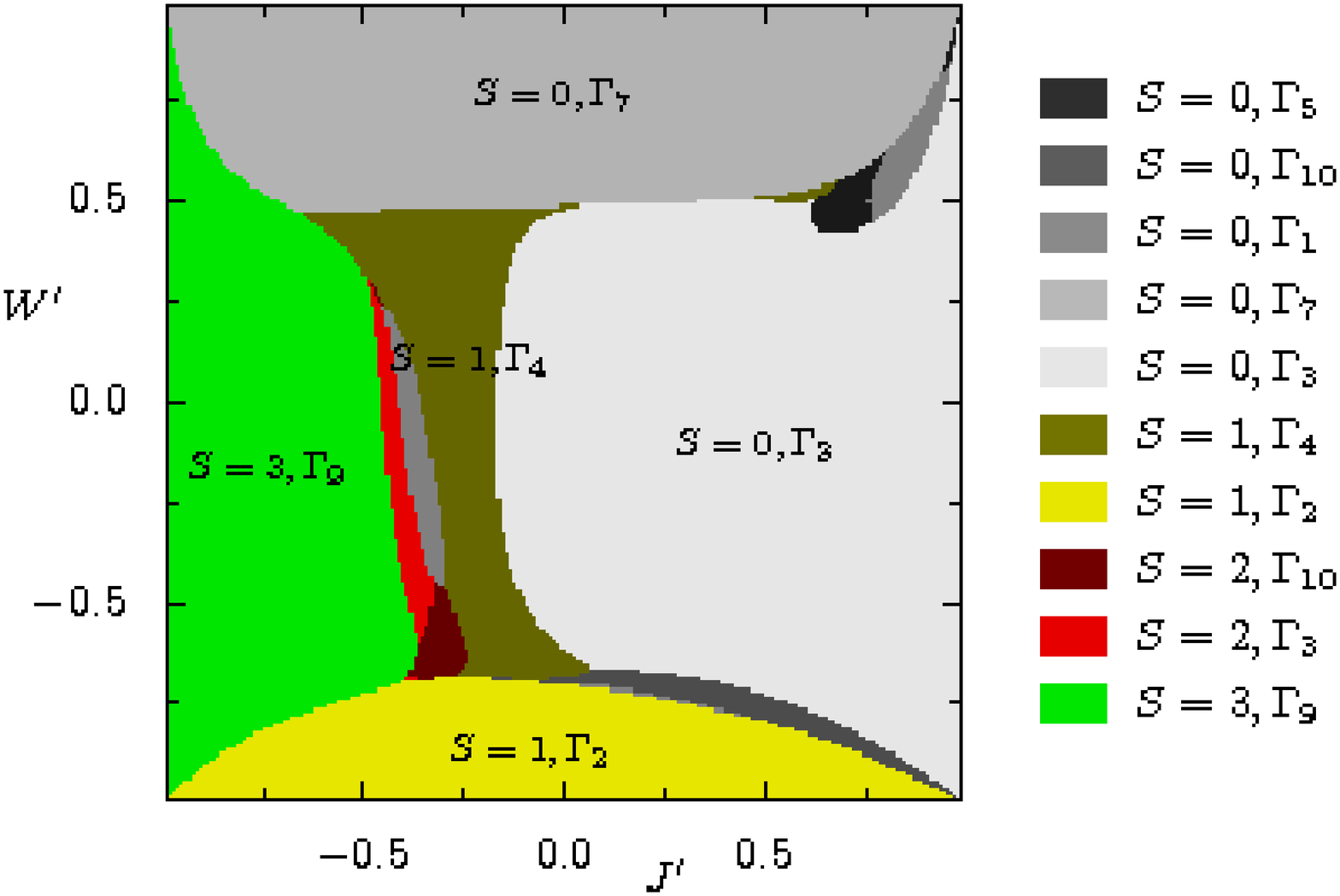}
   \hfil
   \includegraphics[width=\figurewidth]{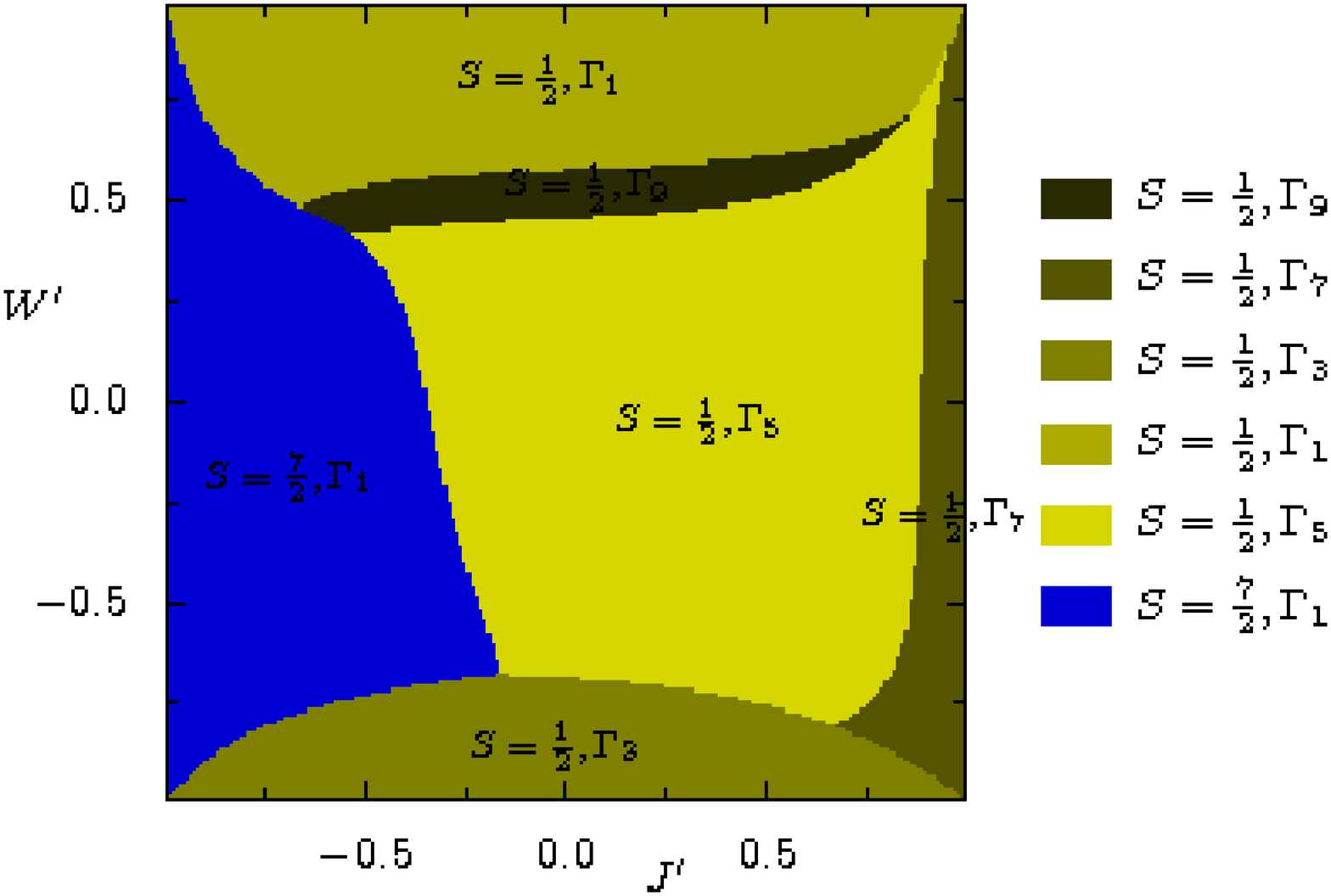}
   \\[3mm]
   \sidecaption
   \includegraphics[width=\figurewidth]{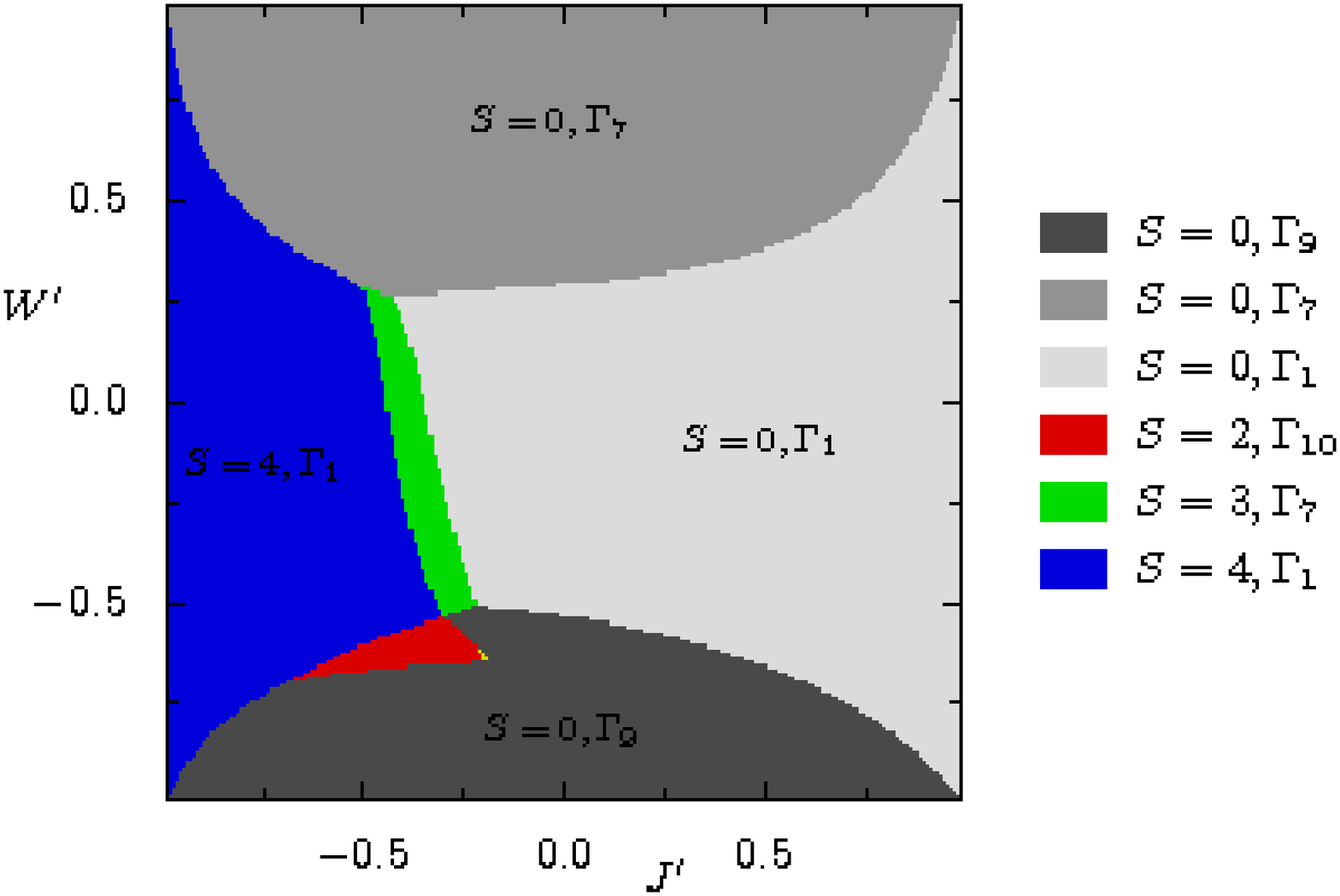}
 \hspace{7mm}
 \caption
   {Canonical GSPDs for the parameters of the extended model at $U=6t$ for
 occupation numbers $n=2\ldots8$. The dependence on the nearest-neighbor coulomb
 interaction $W$ and the exchange interaction $J$ is shown. Colored areas stand for ground
 states of constant quantum numbers, which are listed on the right of each figure. Because
 no magnetic field is applied, all states are degenerated with respect to $S_z$. 
 Degenerations may be read off by multiplying the spin degeneracy by the dimension of 
 the irreducible representation $\Gamma_i$.
 }
  \label{fig:canonical_extended_groundstates_WJ}
 \end{figure}

\begin{figure}
  \setlength{\figurewidth}{45mm}
  \includegraphics[width=\figurewidth]{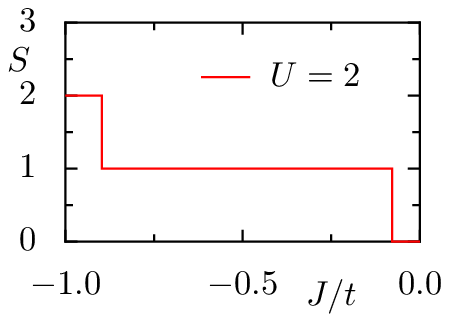}
  \hfil
  \includegraphics[width=\figurewidth]{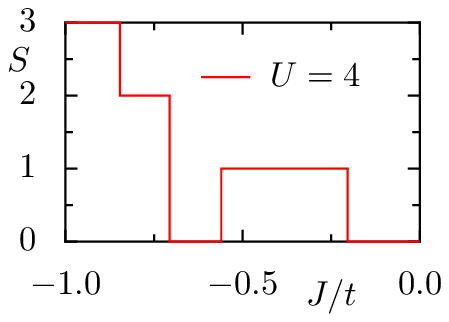}
  \hfil
  \includegraphics[width=\figurewidth]{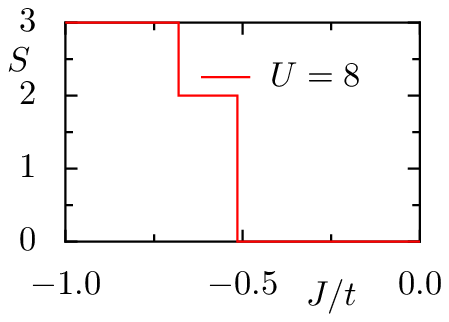}
  \caption{
    Ground state spin $S$ in dependence on the exchange parameter $J$ for the canonical case $n=6$
    at three different on-site correlations $U/t=2,4,8$. Note the intermediate region around $U=4t$,
    where the spin does not increase monotonically with lower $J$, what could have been expected and
    is retrieved for the other cases.
  }
  \label{fig:canonical_extended_crosssection_J}
\end{figure}
	
The additional parameters of the extended model introduce a wide variety of new level crossings,
which might be understood by the help of GSPDs. 
In Fig. \ref{fig:canonical_extended_groundstates_UJ} the dependence on the on-site correlation $U$
and the spin exchange interaction $J$ is shown for occupation numbers $n=2$ to $n=8$. 
The cases $n<2$ are trivial and therefore omitted. According to the Hamiltonian
\eqref{eqn:hamiltonian_exchange}, one has to distinguish between a
ferromagnetic ($J/t<0$) and an antiferromagnetic ($J/t>0$) interaction. 
In the first case, the additional term favors spin alignment. For extremely strong
ferromagnetic interaction, the total spin is increased to its maximum for all $U$ at every
occupation number. In some cases, e.g. $n=4$ or $n=6$ even a very small $J/t<0$ introduces a
change to $S=1$ for small positive on-site correlations, whereas for $n=5$ a small positive $J$
lowers the spin from $S=\frac{3}{2}$ to $\frac{1}{2}$. For the three occupation numbers $n=4,5,6$
the exchange parameter therefore has a big influence on the ground state spin. Contrary to the
case $n=5$, an antiferromagnetic interaction has no drastic effects in general, since the
correlation parameter $U$ favors antiferromagnetic alignment, too. This is especially visible in the
case of the repulsive model, where only a few spatial transitions are visible.
For $n=7$ the  afore mentioned Nagaoka state \cite{Nagaoka65,Tasaki89} 
is obtained for $U>39.642t$ for the pure Hubbard  model. 
The n=7 panel of Fig. \ref{fig:canonical_extended_groundstates_UJ} shows 
the influence of the additional exchange interaction
on the Nagaoka 
state (blue): Whereas the maximum spin state
 is stabilized for $J/t<0$, it is completely destroyed for approximately $J>0.06t$ even
with $U\rightarrow\infty$.

This influence of the exchange term holds in general, although there are some regimes, where the effects
are more subtile. Fig. \ref{fig:canonical_extended_crosssection_J} shows the ground state spin $S$
for different on-site correlations over a relatively small region from $J=-t$ to $J=0t$, where
for lower values an increasing tendency for ferromagnetic alignment is expected. Indeed, this is
found for both weakly and strongly correlated systems at $U=2t$ and $U=8t$, respectively. Contrary,
for an intermediate region around $U=4t$ the spin does not depend monotone on $J$ showing the complex nature of the interaction.

The GSPDs for the nearest-neighbor coulomb interaction $W$ are shown in
Fig. \ref{fig:canonical_extended_groundstates_UW}. Interestingly enough, there are differences in
the effects of both coulomb interactions, most visible in the right column, where the total spin
increases with high positive values of $U$ but not with $W$. The situation is different for
negative $U$, where no change in total spin is observable, nevertheless spatial transitions take
place.
For the even occupation numbers shown in the left column, the total spin is mostly zero. The only
exceptions are for $n=4$ a region for small $W$ and positive $U$ and for $n=6$ at small positive
$U$ and negative $W$. Consequently, adding a small nearest-neighbor Coulomb interaction rarely
introduces ground state changes, notably in the stability of the Nagaoka state at $n=7$.
 The picture of the extended model is completed by the GSPDs in dependence on
 the two extended parameters $W$ and $J$ for fixed correlation 
strength $U=4t$ shown in
 Fig. \ref{fig:canonical_extended_groundstates_WJ}. The images show the rich
 interplay of both interactions.
Although it is hard to detect general rules, one may roughly say that  nearest-neighbor Coulomb  interaction  acts likewise 
 as the on-site correlation $U$  does. The latter is deduced from the comparison
 to Fig. \ref{fig:canonical_extended_groundstates_UJ}, 
where the pictures show similar structure.
\section{Cluster gas}
\subsection{Cluster gas approximation}

In contrast to the last chapter, where the results for fixed occupation number,
were given, the present one is devoted to the so called ``cluster gas''.
what  is an ensemble of non-interacting identical clusters coupled to a reservoir of electrons, thus introducing the chemical potential  $\mu$.
It serves as an approximation to the
full extended lattice, where every second bond is replaced by the indirect hopping via the particle
bath, thus allowing for fluctuations of the occupation numbers on each cluster and therefore modeling the effect of doping.
\subsection{Pure Hubbard model}
\label{sec:grandcanonical_extended}

\begin{figure}
  \includegraphics[width=45mm]{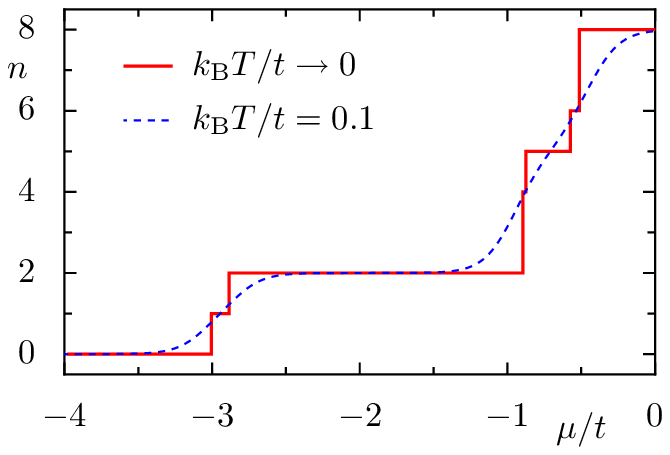}
  \hfil
  \includegraphics[width=45mm]{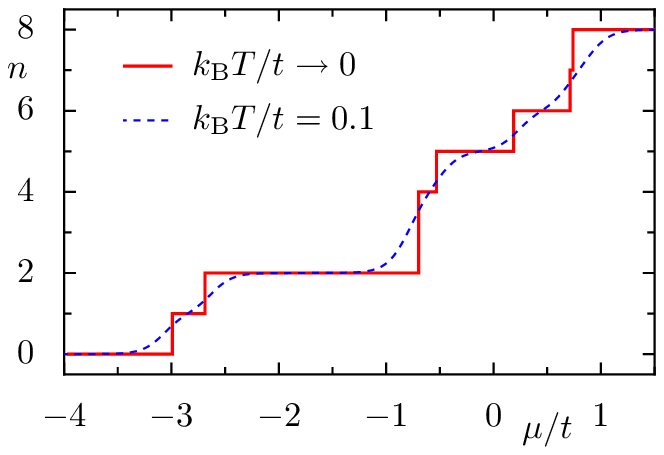}
  \hfil
  \includegraphics[width=45mm]{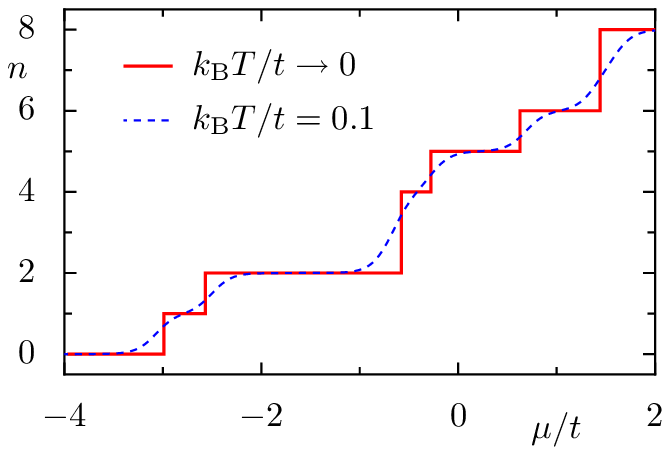}
  \caption
  {The occupation number in dependence of the chemical potential  $\mu$ in the cubic cluster gas
for three different correlation strengths  $U/t=1,4,8$ (f.l.t.r.). The two pictures on the left visualize the
weakly correlated region, whereas the right one shows a strongly correlated system. The dotted curves for a
finite temperature show the usual smearing of the steps.
}
  \label{fig:grandcanonical_simple_n_mu}
\end{figure}

\setlength{\figureheight}{55mm}
\begin{figure}
  \includegraphics[height=\figureheight]{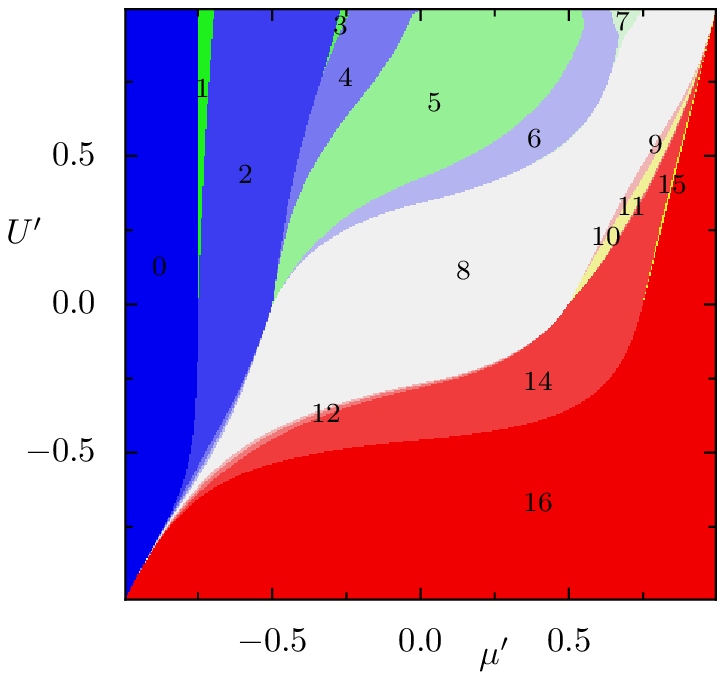}
  \includegraphics[height=\figureheight]{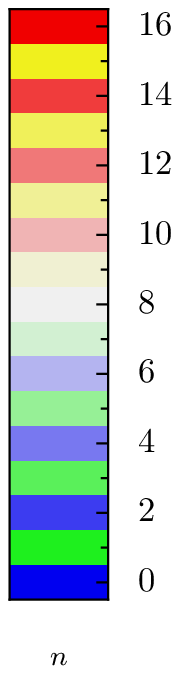}
  \hfil
  \includegraphics[height=\figureheight]{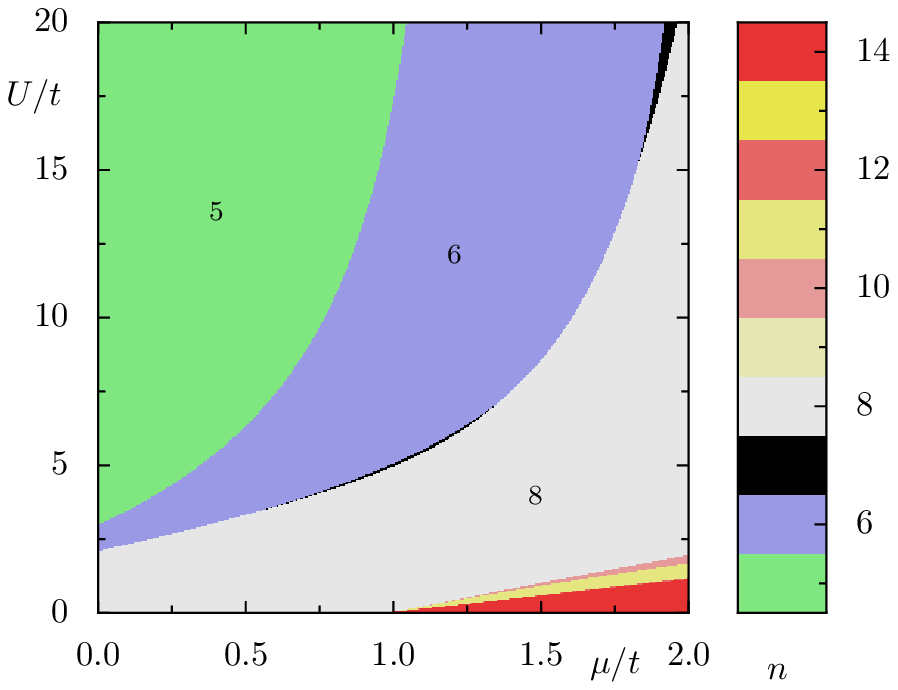}
  \caption
  {The occupation number of the ground state in dependence of the chemical potential  $\mu$ and the
correlation parameter  $U$ in the cubic cluster gas. The left panel shows the complete parameter
space using primed values, whereas the right one is a magnification with unscaled parameters, to
show the small area, where the black marked region $n=7$ is the ground state.}
  \label{fig:grandcanonical_simple_n_muU}
\end{figure}

In a first step, the influence of the chemical potential on the pure Hubbard model will be analyzed. The occupation
number  $n$ in dependence on $\mu$ for different values of the on-site correlation is shown
in Fig.  \ref{fig:grandcanonical_simple_n_mu}. The step functions show similar behavior as in other
small clusters \cite{Schumann08}, e.g. there are steps higher than one. These steps
correspond to degeneration points, where ground states differing in their 
electron occupation by more than one have the same energy. 

\begin{table}
	\tabsidecaption
  \begin{tabular}{@{}lll@{}}
		\hline
		Occupation $n$ & Correlation $U_{\text{TP}}/t$ & Chem. potential $\mu_{\text{TP}}/t$ \\ 
		\hline
		2,3,4 & 13.29885 & -0.4993926 \\
		6,7,8 & 3.372781 &  0.5218219 \\
		6,7,8 & 7.106821 &  1.3549741 \\
		6,7,8 & 15.32480 &  1.8342337 \\
		\hline
  \end{tabular}
  \caption{
  Ground state triple points of the cubic cluster gas for $U/t > 0$ and $n\leq8$. 
  The related	parameters have been calculated to the printed accuracy.
  }
  \label{tbl:grandcanonical_simple_tripelpunkte_U_mu}
\end{table} 

Fig.  \ref{fig:grandcanonical_simple_n_muU} shows an overview of this essential feature.
It generalizes the results to cover the complete parameter space of both the correlation parameter
and the chemical potential in a GSPD using the scaling
functions  \eqref{eqn:scaling_function}. Comparing the two figures, it is convincing that this
non-linear scaling does not change the qualitative picture of the phase diagram and may be used in
the following.
The above mentioned steps reoccur as the boundaries between the colored areas in both images. The steps higher than one may be found at borders of two regions, which are not consecutive in the occupation numbers displayed in the legend to the right.
Additionally, there exist points, where degeneration lines intersect, e.g. three different ground
states coexist. Those ground state triple points may be
calculated with high accuracy. The results for the repulsive model are printed in
Table  \ref{tbl:grandcanonical_simple_tripelpunkte_U_mu} and may be easily checked versus
Fig.  \ref{fig:grandcanonical_simple_n_muU}. They may mainly serve as a reference calculation for other theoretical analysis, since it is very unlikely to have those special conditions in any real experiment. 
In an week interacting extended system we have to expect the interplay of three quantum phase transitions in the vicinity of the triple points. 
Regarding the degeneration lines, we see, that they  are present through a wide parameter range and are therefore reachable by experiments, e.g. using electron or hole doping or pressure.
\begin{figure}
	\sidecaption
  \includegraphics[width=54mm]{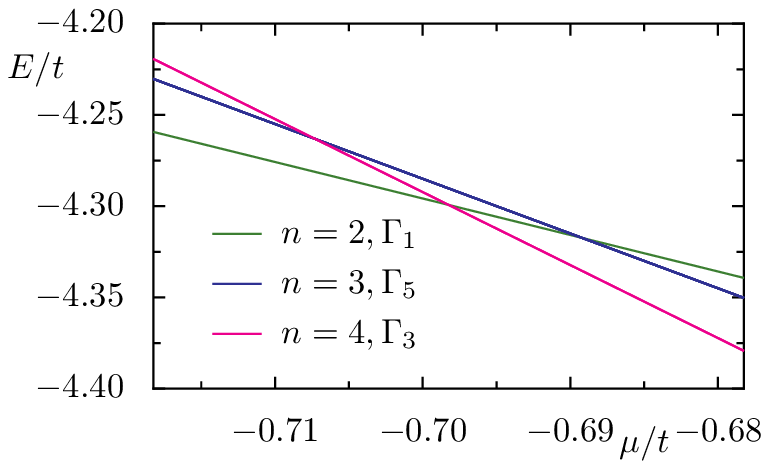}
  \hfil
  \includegraphics[width=54mm]{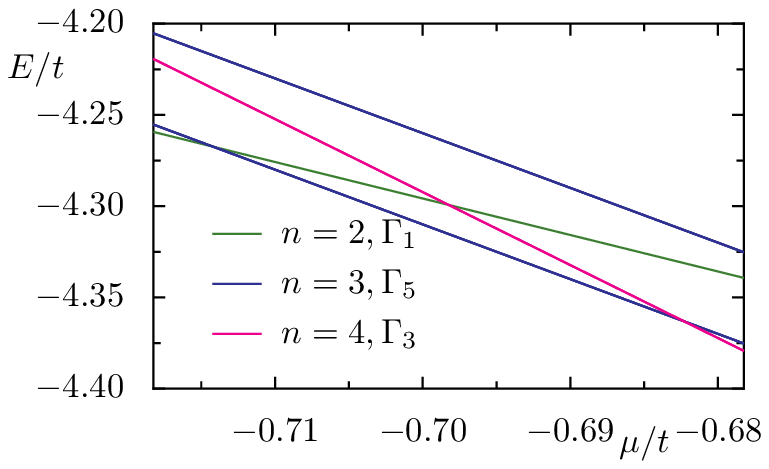}
  \caption
  {Lowest energy levels in the vicinity of a degeneration point in dependence on $\mu$ at $U=4t$. 
  A magnetic field $h/t=0.5$ has been added in the right panel.}
  \label{fig:grandcanonical_simple_E_mu}
\end{figure}

Since the degeneration points -- and therefore especially the highly degenerated ones -- are
considered to be the key to the complexity of the phase diagram, the stability with respect to the
remaining parameters is of major interest. Fig.  \ref{fig:grandcanonical_simple_E_mu} shows the
destruction of such a point by applying a small magnetic field. The
states with even occupation are not magnetic, while the state with $n=3$ is a doublet analogous to
the canonical cases in Fig.  \ref{fig:canonical_simple_Sz_Uh}. On the other hand, new degeneration
points may be introduced by applying an external magnetic field, which can be seen in
Fig.  \ref{fig:grandcanonical_simple_E_mu}, too, where the  $n=2,4$ degeneracy is lifted by the field.

\setlength{\figurewidth}{72mm}
\begin{figure}
  \includegraphics[width=\figurewidth]{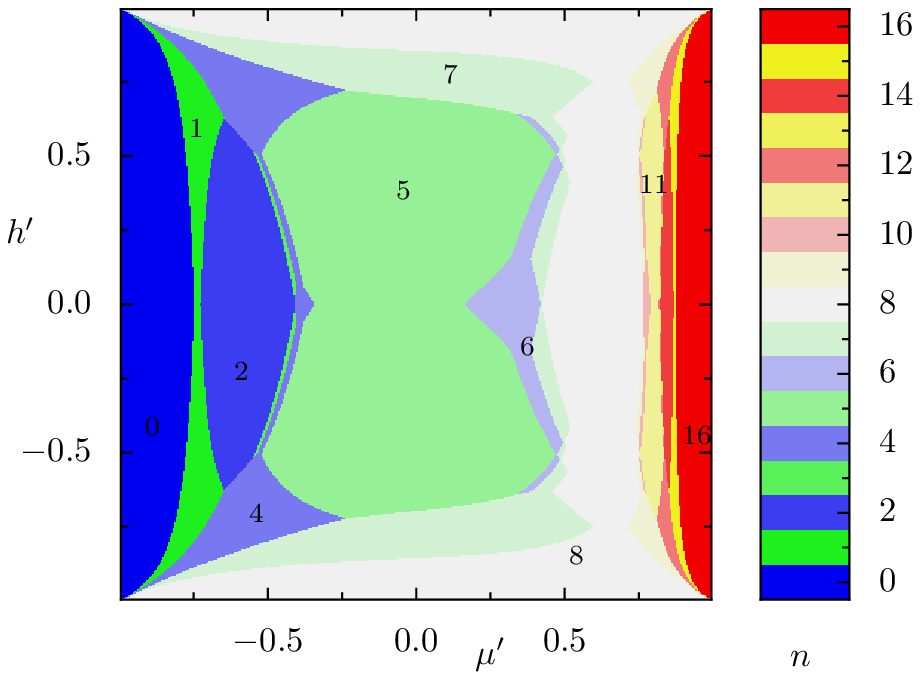}
  \hfil
  \includegraphics[width=\figurewidth]{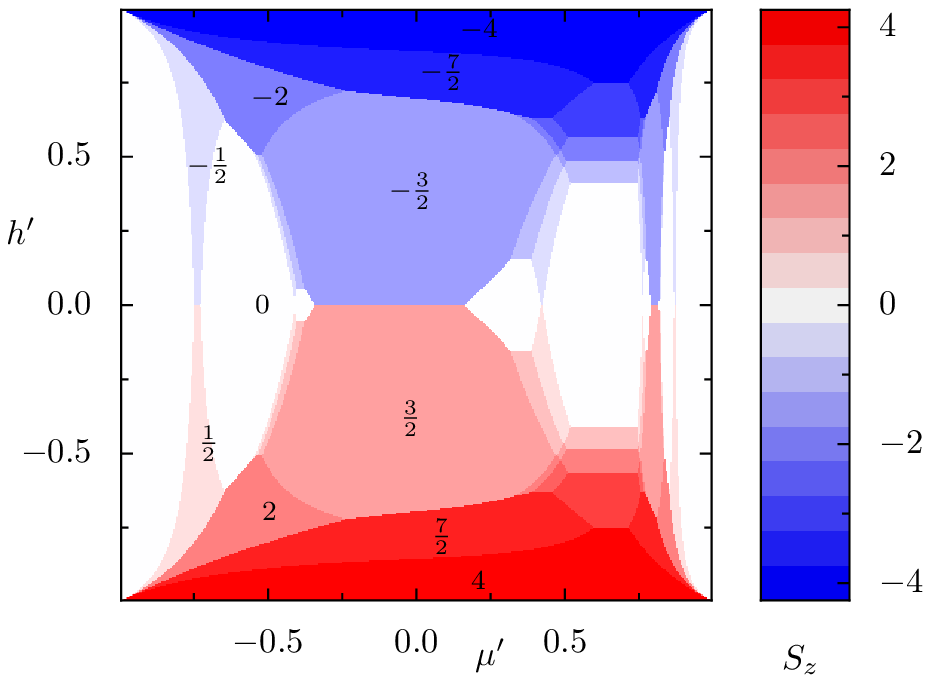}
  \\[3mm]
  \sidecaption
  \includegraphics[width=\figurewidth]{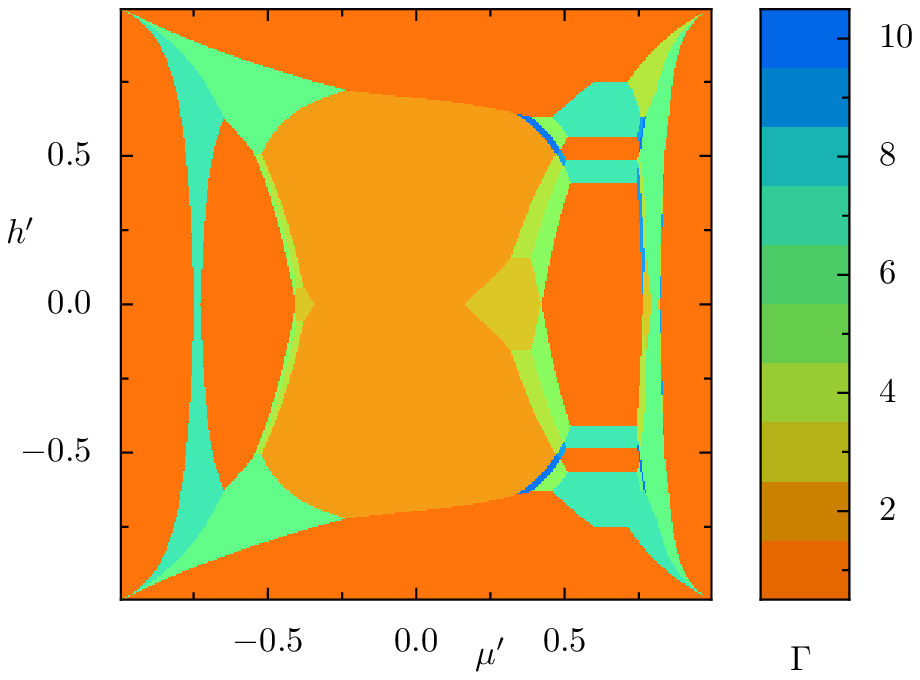}
  \hspace{3mm}
  \caption
  {Occupation number  $n$, spin projection  $S_z$ and spatial symmetry $\Gamma$ of the ground states
in dependence on the chemical potential  $\mu$ and the magnetic field  $h$ at fixed correlation
strength $U=4t$. The meaning of the colors are visualized in the legend right to each figure. Since
all pictures cover the same parameter space, the areas are related, such that $n$, $S_z$ and
$\Gamma$ may be extracted for any given point. The size of the different areas may not be
compared directly, because of the non-linear transformed parameters.}
  \label{fig:grandcanonical_simple_Szn_muh}
\end{figure}

The influence of the magnetic field is studied comprehensively by means of the GSPDs shown in
Fig.  \ref{fig:grandcanonical_simple_Szn_muh}, which show a multitude of new ground state
degeneration lines and triple points. 
In contrast to the canonical cases, the system does not pass all possible
spin projections with increasing magnetic field, but rather exhibits large jumps, e.g. at $\mu'
\approx -0.6t$ from $S_z=0$ to $|S_z|=2$. This is accompanied with a change in the occupation number
from $\mean{n}=2$ to $\mean{n}=4$ and in the spatial symmetry from $\Gamma_1$ to $\Gamma_6$, what
may be deduced from the figures on the left. The system switches abruptly from two electrons with
antiparallel spins to a configuration, where four spins are aligned.
This example shows the rich interplay of charge and spin degrees of freedom under the influence of
a magnetic field, which is a completely new feature of the cluster gas and was not present in the
canonical case in Fig.  \ref{fig:canonical_simple_Sz_Uh}. At small magnetic fields,
the right figure also displays an antiferromagnetic ground state, with mostly minimal spin
projection. The only exception is the case $\mean{n}=5$ around $h=\mu=0t$, where four spins seem to
be aligned, while the remaining one is antiparallel. This is in accordance with the related canonical results.


\begin{figure}
  \includegraphics[width=68mm]{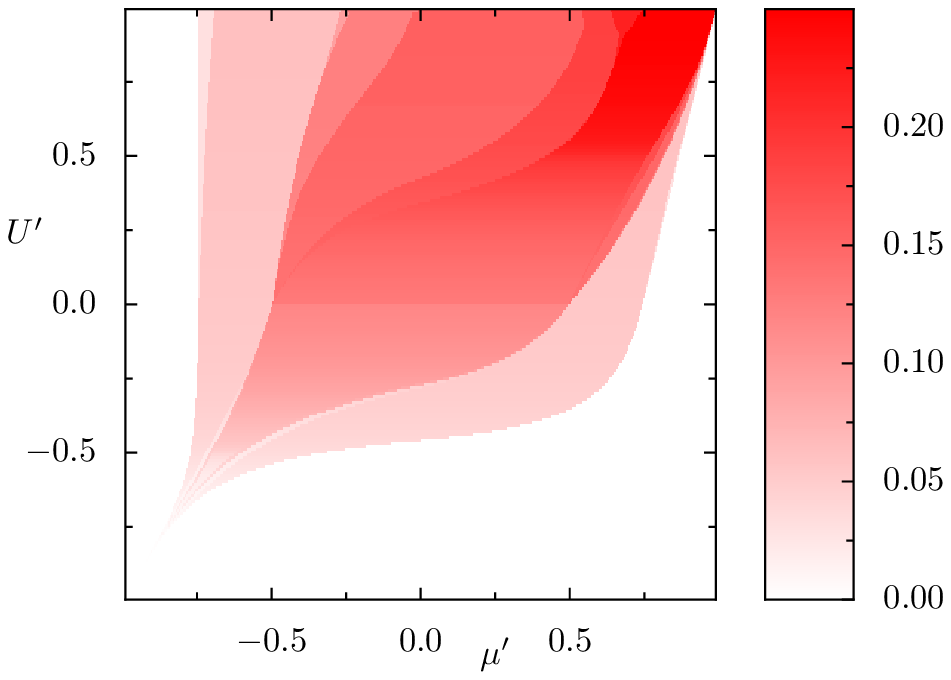}  a)
  \hfil
  \includegraphics[width=68mm]{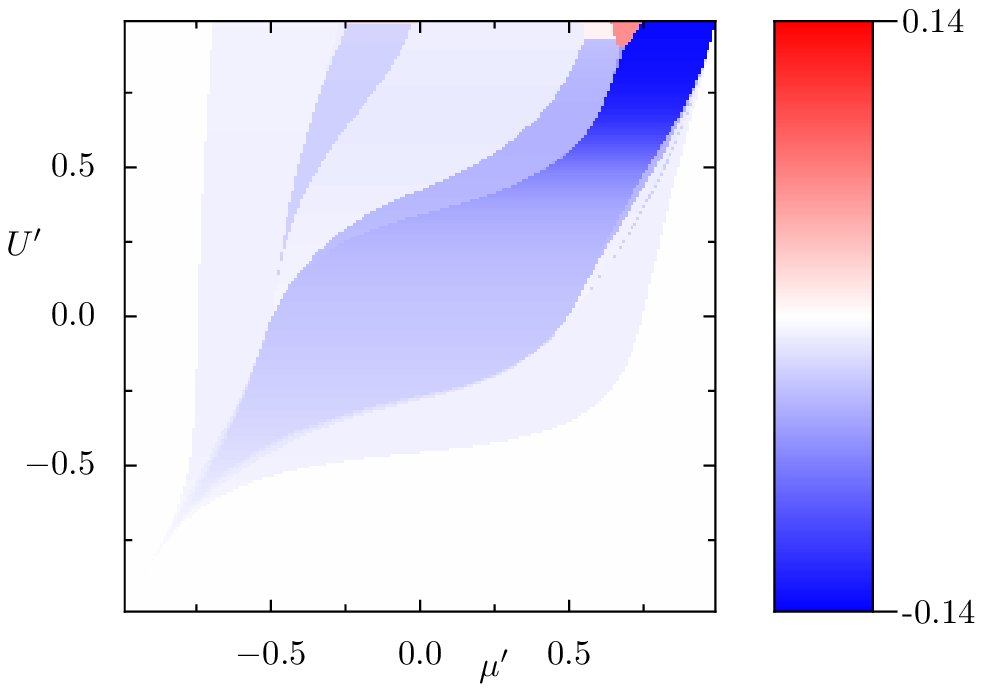}  b)
  \\[3mm]
  \includegraphics[width=68mm]{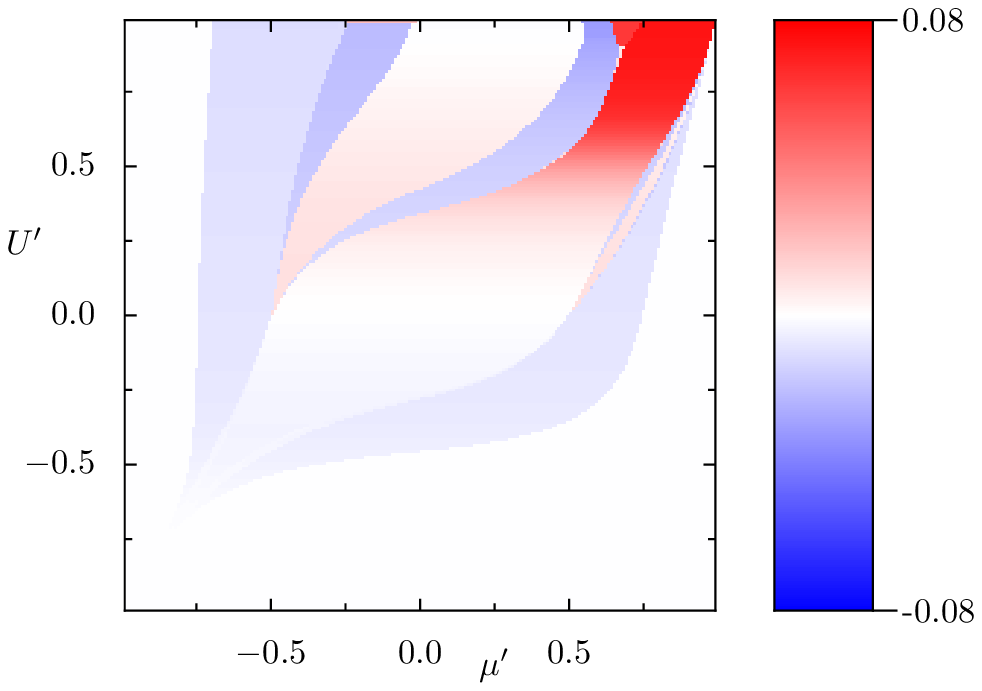}  c)
  \hfil
  \includegraphics[width=68mm]{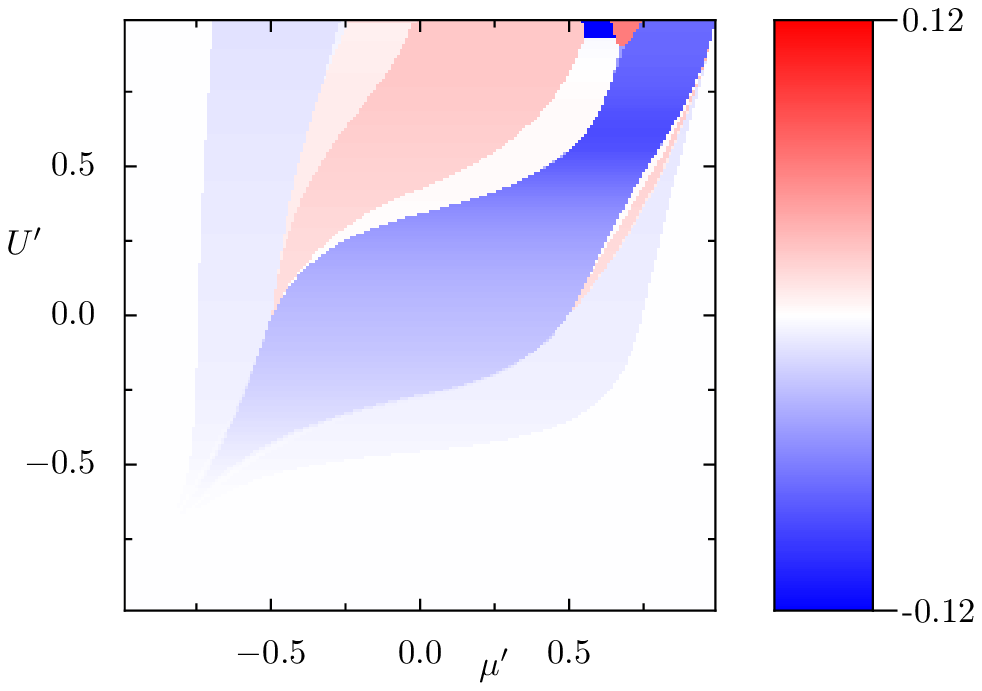}  d)
  
  \caption{The four different spin correlation functions of the cubic cluster gas in dependence on
the correlation parameter  $U$ and the chemical potential  $\mu$. The values of the functions are
visualized by colors, which might be associated using the legend right to each figure. The four
functions shown are the same as in Ref. \cite{Callaway87}: Picture a) shows the mean local spin
square $\mean{S_{z,i}^2}$, whereas the other cases correlate spins on different
sites $i$ and $j$, with increasing distance to each other: $\mean{S_{z,i}S_{z,j}}$. There are three
distinct possibilities on the cube: b) nearest-neighbor, c) face diagonal and d) space diagonal.}
  \label{fig:grandcanonical_simple_spincorrelation}
\end{figure}

The spatial order of the local spins can be discussed in great detail using spin correlation
functions, what was first done for the canonical case by Callaway et al. \cite{Callaway87}.
The grand canonical case is shown in Fig.  \ref{fig:grandcanonical_simple_spincorrelation}.
The degeneration lines of Fig.  \ref{fig:grandcanonical_simple_n_muU} are once again visible, since
the same parameter range is covered and the spin configuration depends heavily on the number of electrons.
Consequently, the mean occupation numbers shown in Fig.  \ref{fig:grandcanonical_simple_n_muU} may be used to explain
features of the correlation functions.

The key feature is visible in the top right picture, where the nearest-neighbor correlation is
shown: Only the region with $\mean{n}=7$ exhibits a positive spin correlation denoting a tendency
for a
ferromagnetic ground state, which is in accordance with Nagaoka's theorem \cite{Nagaoka65,Tasaki89},
since $\mean{n}=7$ is the occupation number, where exactly one electron is missing to half filling.
Moreover, the calculation shows, that the Nagaoka state is not only stable for infinite on-site
correlation  $U$, but for the whole region $\mean{n}=7$ with $U>39.642t$, where the transition to
 the $S=\frac{1}{2}$ state takes place. An interesting
conjecture may be drawn regarding the spatial change of the state $\mean{n}=6$ at $U=61.313t$,
which occurs in the vicinity of the Nagaoka state. The switch of the irreducible representation from
$\Gamma_5$ to $\Gamma_1$ may be induced by the ferromagnetic $n=7$ state, which has
$\Gamma_1$-symmetry, too.

All other regions have a tendency for antiparallel spins, leading to antiferromagnetism in the
extended lattice. The next pictures support this point of view, since especially for half
filling $\mean{n}=8$ parallel alignment is favored across the face diagonal in c), but once again
reversed across the space diagonal in d), which is in accordance with the view of an alternating
spin configuration. Contrary, in the case of $\mean{n}=7$, all correlation functions are positive,
supporting a parallel order. The picture is less clear for the smaller occupation numbers, since the
spatial configuration of the spins seems to be
more complex and a connection between neighboring spins can not be drawn that easily. This is also
supported by the overall smaller absolute value of the correlation functions, showing the weaker
correlation directly.

\begin{figure}
  \includegraphics[width=45mm]{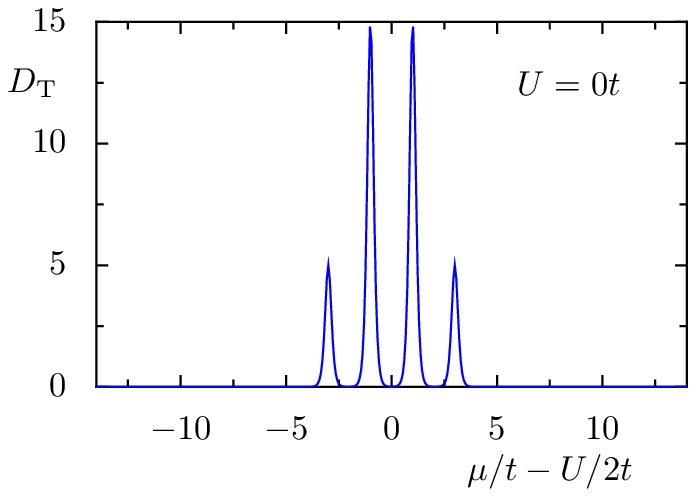}
  \hfil
  \includegraphics[width=45mm]{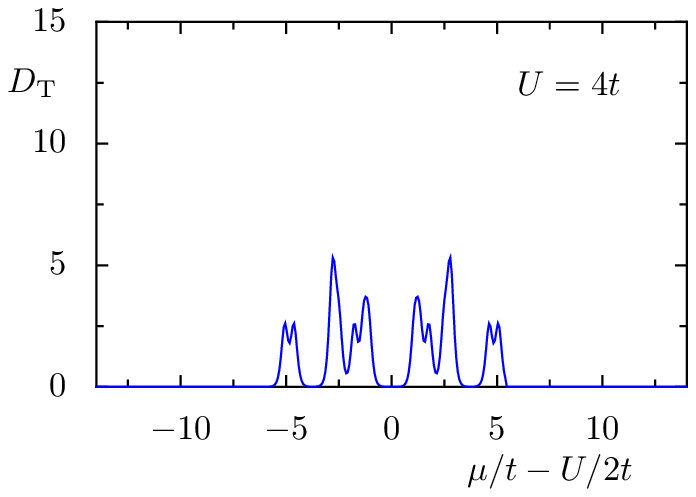}
  \hfil  
  \includegraphics[width=45mm]{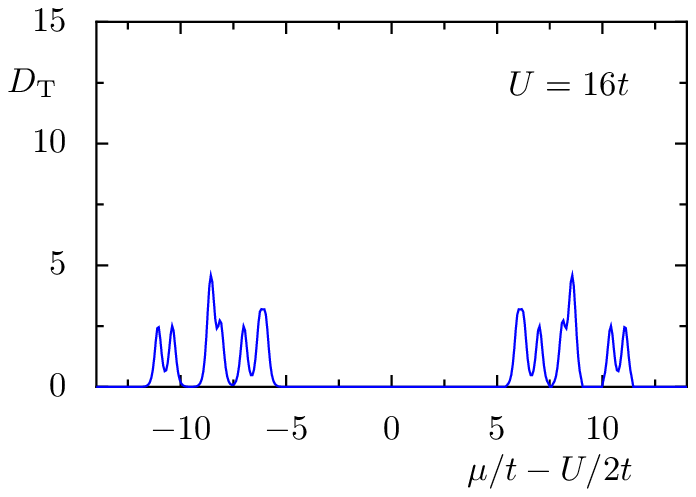}
  \caption
  {Thermodynamical density of states for the cubic cluster gas in dependence on the chemical
potential for different
correlation parameters at the temperature $\kb T=0.1t$. The function is shown for
three increasingly strong correlations $U/t=0,4,16$. The axis of the chemical
potential  $\mu$
has been shifted by $U/2t$ to emphasize the particle-hole-symmetry.}
  \label{fig:grandcanonical_simple_TDOS_mu}
\end{figure}

To complete the picture of the cubic cluster gas, this section ends with some further
characteristic functions, which were also computed for e.g. the square \cite{Schumann06}.
The spectral function $J_{\cminus{i\sigma}\cplus{i\sigma}}$ of the cubic cluster has already
been reported in Ref. \cite{Callaway91} and has been confirmed. Additionally, the thermodynamic
density of states shown in Fig.  \ref{fig:grandcanonical_simple_TDOS_mu} is defined as
\begin{align}
  D_{\text{T}}(\mu) := \frac{\partial n(\mu)}{\partial \mu}.
\end{align}
With increasing on-site correlation the initial four-peak structure of the uncorrelated 
model transforms into a more complex picture: Two groups of peaks separate with an increasing
gap inbetween. The separation takes place at the transition region around $U=6t$.
This is similar to the spectral function and in qualitative accordance with the common picture of
the Hubbard-Mott transition.

\subsection{Extended Hubbard model}

In the same manner as in the canonical case, additional interactions may be added for the cluster
gas. To compare the results to the canonical case, the same terms are considered.
It is evident, that the complete phase diagram is extremely complex and only a small part will
be discussed here. 

\begin{figure}[htbp]
  \includegraphics[width=73mm]{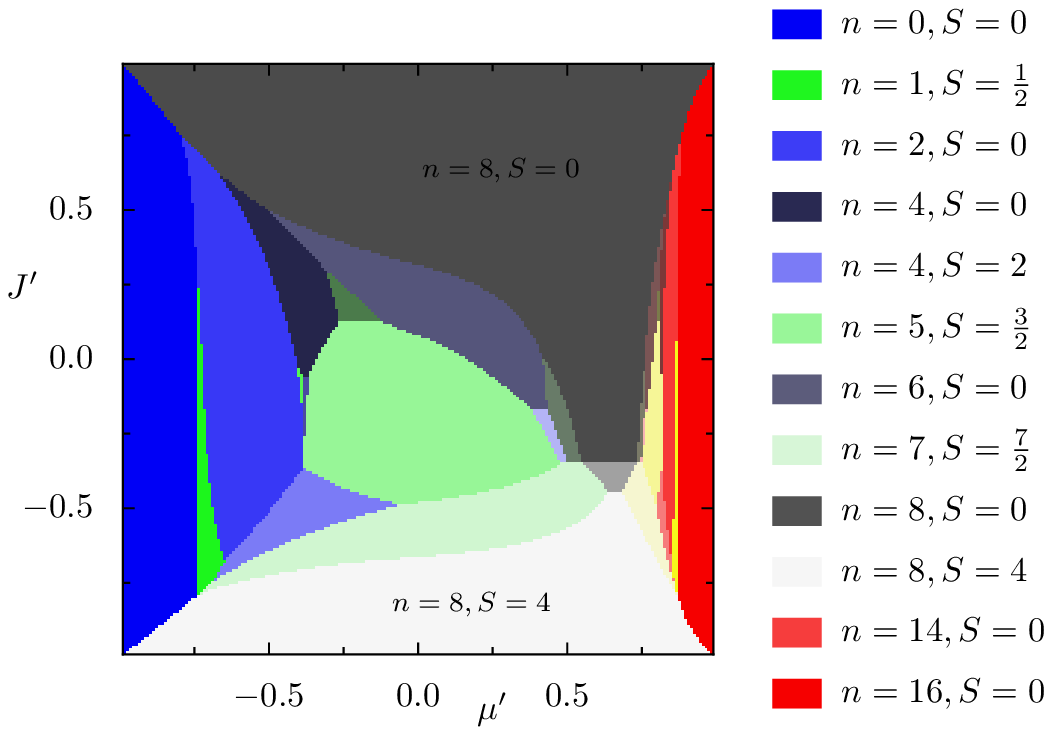}
  \hfil
  \includegraphics[width=73mm]{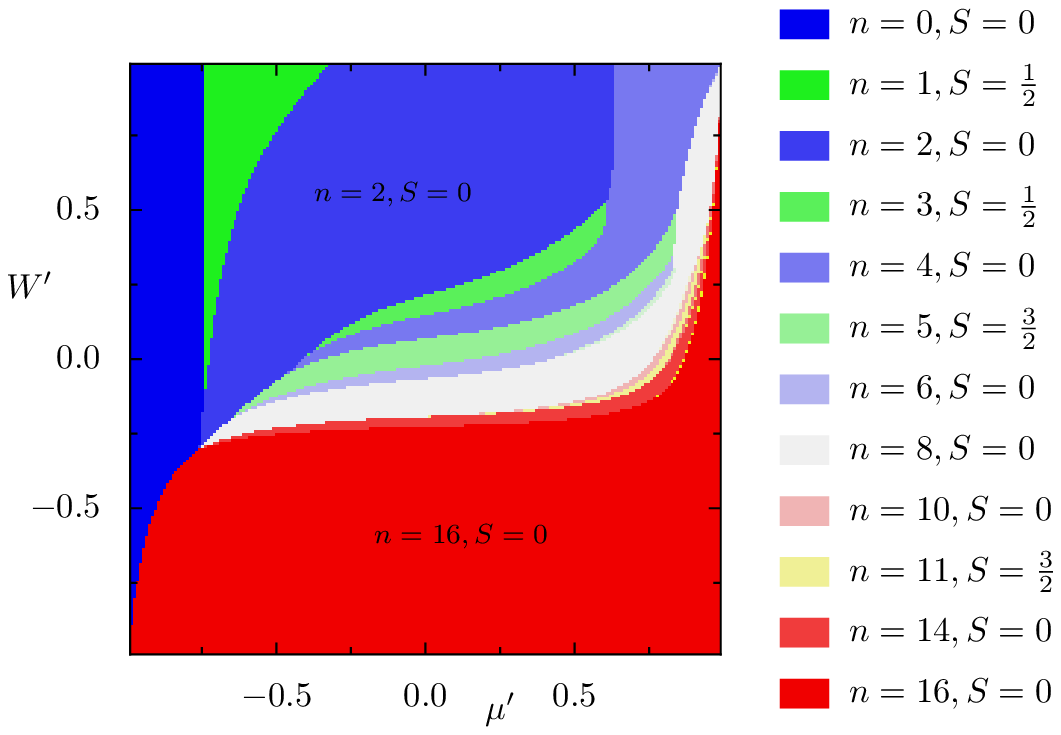}
  \caption{GSPDs of the extended model of the cubic cluster gas. The
correlation parameter $U=4t$ is fixed in both figures. The groundstates are pictured as
colored areas in dependence on the chemical potential $\mu$ and both additional parameters $J$ and
$W$, respectively. The quantum numbers are shown only for large areas in the legend right to the
figures.}
  \label{fig:grandcanonical_extended_groundstate}
\end{figure}

The influence of a ferromagnetic ($J<0t$) and antiferromagnetic ($J>0t$) exchange
interaction is
shown on the left panel of Fig. \ref{fig:grandcanonical_extended_groundstate}. The multitude of
degeneration points increases drastically. Qualitatively, the same features as in the canonical
case of Fig. \ref{fig:canonical_extended_groundstates_UJ} are visible, in that a ferromagnetic
exchange leads to charge ordering and the opposing antiferromagnetic one does not have those
qualitative impact, but stabilizes the tendency for antiparallel spin ordering. Nonetheless, both
characteristics lead to a half-filled state for extreme values showing the tendency for an ordered
state in general. The degeneration points discussed in section \ref{sec:grandcanonical_extended}
are mostly visible throughout the whole range of $J$, although the state $n=7$ is stable around
$J\approx-t$. New degeneration points, e.g. between $n=0$ and $n=2$ are introduced for positive $J$.

The nearest-neighbor Coulomb interaction is shown of the right side of the same figure. The
qualitative picture is quite similar to Fig. \ref{fig:grandcanonical_simple_n_muU}, the GSPD in 
dependence on $U$ and $\mu$. This is reasonable considering both $U$ and $W$ denote Coulomb
interactions, only differing in their spatial properties. More quantitatively, it has been shown
for the triangle, the tetrahedron and the square, that both parameters occur in fixed combinations
in most eigenvalues of the problem \cite{Schumann08,Schumann09}. The basic degeneration points of
section \ref{sec:grandcanonical_extended} are stabilized for high absolute values of $W$ and new points are introduced, where $n$ steps  e.g. from 4 to 8 at $W=t$.
 
\section{Discussion}
The comparison of our semi-analytical treatment  of the Hubbard model on a cubic cluster with the former results of our analytical solutions on smaller clusters 
makes clear, that the complexity is increased considerably by increasing
the cluster size or by adding additional interactions to the Hamiltonians. 
For eight sites we were still able to get the exact spectrum and eigenfunctions  for the complete parameter region from $-\infty$ to
$+\infty$ for all parameters, thus providing a reference
for other studies, where the (extended) Hubbard model is employed. 

The detailed reproduction of the results by Callaway et
al. is  a result, which provides us with confidence in our symbolic code.
It was a prerequisite for the treatment of the Hubbard model
extended by additional terms, i.e.  an external magnetic
field $h$ coupled to the spins, and the nearest neighbor Coulomb and exchange interaction, $W$ and $J$, respectively. 
The sometimes complex effects of increasing $h$ have also been found in other small clusters showing the
competition between aligning the spins and minimizing their kinetic energy. The situation becomes
even more difficult with the additional interactions: Although the tendency for ferromagnetic
alignment caused by $J/t<0$ has been recovered for most parameter regions, there are exceptions,
where lowering $J$ also lowers the total spin, which would have rather been expected for positive
$J$. The influence of the nearest neighbor Coulomb interaction $W$ is comparable to the
on-site correlation $U$ and does mostly introduce spatial transitions in the GSPDs. It is
questionable, whether the enormous amount of spatial changes may be seen in bulk systems, but it
has been recovered that the difference between odd and even electron occupation is of qualitative character, which has been reported for all other small clusters, too.

With regard to extended systems, the cubic cluster gas is a model for the simple
cubic lattice. The approximation will be not too bad, if one assumes a weak inter-cluster hopping,
which is than replaced by the indirect exchange via the particle reservoir. 
Besides the highly non-trivial analytical closed-form solution for two electrons, 
the most important results reported in this paper are the ground state level crossings being
visible in the various GSPDs. They may give rise to quantum phase transition in
extended systems. With respect to the cluster gas, the degeneration points of states with occupation
numbers differing by two or more are of central interest, e.g. the jump from $n=6$ to $n=8$ visible
in Fig. \ref{fig:grandcanonical_simple_n_mu}. This is especially important, since this degeneration
line is present at moderate on-site correlation and lies in the weakly underdoped region, thus being
reachable by experiments. Utilizing doping, the mean electron occupation
of the cluster gas may be fixed to $\mean{n}=7$ and the result will be a mixture of clusters with
six and eight electrons, if $U$ lies between $3.37t$ and $7.10t$ or is above $15.32t$. On can
presume, that something similar to the stripe structure of 2D-systems may be observed.

The alignment of the spins has been studied using the spin correlation functions and it has been
found, that the half-filled case shows a tendency for antiparallel alignment, as expected.
If the on-site correlation is high enough, introducing one hole alters the system completely, such
that the $n=7$ spins are parallel aligned, which accounts for a ferromagnetic ground state in the
extended system, showing that Nagaoka's result is valid down to $U=39.642t$. 
As a consequence one has to expect a ferromagnetic groundstate for a mean
electron occupation $\mean{n}=\frac{7}{8}$ and large on-site correlation $U>79.284t$ if
the cubic lattice is approximated by an array of week interacting cubic clusters with periodic boundary conditions. 
For $n=6$ we have a  singlet ground state with $\Gamma_1$ symmetry for strong electron correlation $U>61.313t$  , which may  be considered as the two-hole counterpart of the Nagaoka state,
apart from the fact that the Nagaoka state does not depend on $U$, whereas the $n=6$ ground state
does. Moreover, the limit value $\lim_{U->\infty                                                } {E(n=6,S=0,\Gamma_1)/t}=-2\sqrt{4+\sqrt{3}}$ 
may serve as an excellent proving tool for qualifying numerical codes.

Regarding the influence of the two additional parameters, the nearest-neighbor Coulomb
interaction $W$ and the spin exchange interaction $J$, it has been found, that the
effect of $W$ is comparable to the on-site correlation $U$, which confirms the finding that the combination of $U$ and $W$ may be replaced by an effective
on-site correlation parameter in most cases, what was shown analytically for smaller clusters. 
Contrary, the effect of $J$ is more important, since even small values may induce a
change of the ground state spin in some areas. An striking example is the vanishing of
the Nagaoka ground-state in the presence of  a small anti-ferromagnetic exchange.
Additionally, the simple picture of $J$ emphasizing
the tendency for (anti-)parallel spin alignment does not always hold, since there are parameter
regions, where the contrary has been observed.

The qualitative consequences of the degeneration points onto extended systems was discussed in detail
in \cite{Schumann08} for smaller clusters. Since our results in general fit well to
the given scenario, we abstain from discussing it again. One remarkable new feature
is the existence of two separated regions of the correlation parameter $U$ where
a $6-8$ degeneration happens. 
It is clear that we may expect the most interesting effects in the vicinity  of these
degeneration points, i.e. for $n=7$ or a hole density of 0.125 \%.

Regarding frustration, our results fortify the view, that it is impossible to
classify a cluster as frustrated from geometry alone. It is
inevitable to consider it together with the electron occupation and the 
correlation strength. Typical frustrated situations will cover only small areas
in the $U$-$\mu$-GSPDs or even be absent. On the contrary the area of unfrustrated configurations 
will grow at the costs of the area of neighboring frustrated configurations. 
This reflects the fact, that frustration increases the ground state energy, thus only a small
amount of chemical energy $\Delta\mu(\mean{N})$ is needed to suppress it.

Probably, it will be hard, to realize an experimental situation, where strongly correlated electrons reside
on a cubic cluster or simple cubic lattice.
Most probably seems to be a situation were d- or f-atoms are embedded in organic molecules.
An other possibility would be the realization of a compound, were an effective simple cubic model
remains after projecting out other parts of the full Hamiltonian, in analogy to the reduction
of Hamiltonian of the copper oxides planes to an effective Hubbard model on a 2-d simple cubic lattice.

A final conclusion from our work on small clusters will be in order. 
We have meanwhile the complete eigensystem 
for all clusters, where the correlation independent symmetries are enough to break 
down the Hamilton matrix to blocks lower than 5. 
Sorry to say the cube does not belong to that set. 
Nevertheless, the big number of symmetry operators made it possible to block-diagonalize
the model analytically, with block sizes allowing a diagonalization to arbitrary numerical 
accuracy by help of symbolic computer facilities.
This exact solution allowed us to get an impression of the unforeseen complexity of the model.
We therefore do not intend to go to bigger cluster sizes or further additional terms in the
Hamiltonian, since with increasing size of the final blocks we will loose the benefit
from our analytical treatment, therefore, ab initio numerical algorithms will 
be more appropriate for such tasks. We see the main benefit from our analytical/exact
cluster solutions in its reference function for other methods dealing with
strongly correlated electrons and as toy models for problems intimately connected to small clusters, e.g. entanglement, quantum dot arrays or electron transport through single molecules.
\begin{acknowledgement}
We are much indebted to  J. Richter for his lively interest in the subject and valuable discussions.
\end{acknowledgement}


\newpage
\appendix
\section{Analytical solution for N=2}
\label{appendix1}
In the following we give the eigensystem for the
electron occupation $N=2$.
The original Hamilton matrix in that space is of dimension 120, which 
is symmetry reduced to 61 subspaces with maximum dimension 4. 
Thus, it is easy to get the eigensystem in its explicite dependency on the correlation parameters 
Unfortunately the lengthy form of some explicite eigenvalues and eigenvectors 
prevents printing them. Here we give the closed form expression of the groundstate for anti-ferromagnetic or small ferromagnetic exchange
 (gray area in Figs. \ref{fig:canonical_extended_groundstates_UJ} and \ref{fig:canonical_extended_groundstates_UW})
belonging to the non-degenerate
eigenstate with $\Gamma_1$ and $s=0$ :
\beq
E_{16-GS}&=& -\frac{A_{14}}{8}-\frac{\sqrt{A_1}}{2}-\frac{\sqrt{A_2}}{2}
\eeq
 \beq
A_1 &=& \frac{A_{14}^2}{8}-\frac{A_9}{4 \sqrt{A_2}}-\frac{A_{10}}{3 2^{2/3} \sqrt[3]{A_3}}+\frac{2 A_{12}}{3}-\frac{\sqrt[3]{A_3}}{6 \sqrt[3]{2}} \nonumber \\
A_2 &=& \frac{A_{14}^2}{16}+\frac{A_{10}}{3 2^{2/3} \sqrt[3]{A_3}}+\frac{A_{12}}{3}+\frac{\sqrt[3]{A_3}}{6 \sqrt[3]{2}} \nonumber \\
A_3 &=& A_7+\sqrt{A_4} \nonumber \\
A_4 &=& A_5^2-4 A_6^3 \nonumber \\
A_5 &=& -2 A_{12}^3+1728 A_{11} A_{12}-36 A_{13} A_{14} A_{12}+864 A_{13}^2+324 A_{11} A_{14}^2 \nonumber \\
A_6 &=& A_{12}^2+288 A_{11}+12 A_{13} A_{14} \nonumber \\
A_7 &=& -2 A_{12}^3+1728 A_{11} A_{12}-36 A_{13} A_{14} A_{12}+864 A_{13}^2+324 A_{11} A_{14}^2 \nonumber \\
A_8 &=& A_{12}^2+288 A_{11}+12 A_{13} A_{14} \nonumber \\
A_9 &=& -\frac{A_{14}^3}{8}-A_{12} A_{14}+16 A_{13} \nonumber \\
A_{10} &=& 324 J^2 t^2+9 J^2 U^2+624 J t^2 U-1728 J t^2 W-48 J U^2 W
 \nonumber \\ &&
+13312 t^4+336 t^2 U^2-1664 t^2 U W+2304 t^2 W^2+64 U^2 W^2 \nonumber \\
A_{11} &=& 3 J t^2 U+24 t^4-8 t^2 U W \nonumber \\
A_{12} &=& 3 J U+80 t^2-8 U W \nonumber \\
A_{13} &=& 9 J t^2-14 t^2 U-24 t^2 W \nonumber \\
A_{14} &=& 3 J-2 U-8 W \nonumber 
 \eeq
and the  groundstate for strong ferromagnetic exchange
(dark yellow area in Fig. \ref{fig:canonical_extended_groundstates_UJ}) is
ninefold degenerated, since it belongs  to $\Gamma_9$ and $s=1$. The related energy is
\beq
E_{10-GS}&=& \frac{1}{3} \left(\frac{J}{2}+4 W\right)-\frac{2 \sqrt{A_1} \cos \left(B_1\right)}{\sqrt{3}}
\eeq
 \beq
 B_{1} &=& \frac{1}{3} \cos ^{-1}\left(-\frac{(J+8 W) \left(J^2+16 J W-144 t^2+64 W^2\right)}{24 \sqrt{3} A_1^{3/2}}\right) \nonumber \\
 A_{1} &=& \frac{1}{12} (J+8 W)^2+16 t^2 \nonumber 
\punkt
\eeq
Therefore we decided to give the analytical form of the 
symmetry reduced Hamilton matrices together with the basis, 
utilized for the calculation. 
This form contains the full information in a more condensed form,
since the coefficients of the basis  are independent on the model parameters up to this stage of diagonalization.
The headline to every subspace gives the eigenvalues of the $U,W,J$-independent symmetry operators, i.e. the eigenvalues 
of $\opS$ together with the spin-projection 
$\opS_z$ in z-direction $m_s$. Furthermore, the spatial
symmetry is indicated by $\Gamma_{i,j}$, where the first index labels the irreducible
representation of the point group and the second numbers the partner.
The notation is based on Ref. \cite{CornwellBook}.
Next we give the basis, where we have chosen the same notation as in Ref. \cite{Schumann02}
and finally the related Hamilton block matrix.\\
\subsection*{Subspace No. 1 with {\boldmath $N=2$, $s=2$, $m_s=-1$, $\Gamma_{4-1}$}}
\beq
\ket{\Psi_{1-1}} &=& \frac{1}{2 \sqrt{2}}\left ( {\scriptstyle \ket{000000dd}} - {\scriptstyle \ket{0000dd00}} - {\scriptstyle \ket{000d000d}} + {\scriptstyle \ket{00d000d0}}\right . \nonumber \\                   \nonumber 
                                          && \hspace{0.5cm} \left . -{\scriptstyle \ket{00dd0000}} + {\scriptstyle \ket{0d000d00}} - {\scriptstyle \ket{d000d000}} + {\scriptstyle \ket{dd000000}} \nonumber \right )
\nonumber \\ 
\ket{\Psi_{1-2}} &=& \frac{1}{2 \sqrt{2}}\left ( {\scriptstyle \ket{00000d0d}} - {\scriptstyle \ket{0000d0d0}} - {\scriptstyle \ket{000dd000}} + {\scriptstyle \ket{00d00d00}}\right . \nonumber \\                     \nonumber 
                                          && \hspace{0.5cm} \left .  + {\scriptstyle \ket{0d0000d0}} - {\scriptstyle \ket{0d0d0000}} - {\scriptstyle \ket{d000000d}} + {\scriptstyle \ket{d0d00000}} \nonumber \right )
\nonumber \\ 
 {\bf H}_{1}&=&
 \left ( 
\begin{array}{cc}
\frac{1}{2} (J+8 W)&
2 t \\ 
2 t&
0
\end{array} 
 \right ) 
\nonumber 
\nonumber \eeq 
\subsection*{Subspace No. 2 with {\boldmath $N=2$, $s=2$, $m_s=-1$, $\Gamma_{4-2}$}}
\beq
\ket{\Psi_{2-1}} &=& -\frac{1}{2 \sqrt{2}}\left ( {\scriptstyle \ket{00000d0d}} + {\scriptstyle \ket{0000d0d0}} - {\scriptstyle \ket{000d00d0}} - {\scriptstyle \ket{00d0000d}}\right . \nonumber \\                     \nonumber 
                                           && \hspace{0.5cm} \left .  + {\scriptstyle \ket{0d00d000}} - {\scriptstyle \ket{0d0d0000}} + {\scriptstyle \ket{d0000d00}} - {\scriptstyle \ket{d0d00000}} \nonumber \right )
\nonumber \\ 
\ket{\Psi_{2-2}} &=& -\frac{1}{2 \sqrt{2}}\left ( {\scriptstyle \ket{00000dd0}} + {\scriptstyle \ket{0000d00d}} - {\scriptstyle \ket{000d000d}} - {\scriptstyle \ket{00d000d0}}\right . \nonumber \\                     \nonumber 
                                           && \hspace{0.5cm} \left .  + {\scriptstyle \ket{0d000d00}} - {\scriptstyle \ket{0dd00000}} + {\scriptstyle \ket{d000d000}} - {\scriptstyle \ket{d00d0000}} \nonumber \right )
\nonumber \\ 
 {\bf H}_{2}&=&
 \left ( 
\begin{array}{cc}
0&
2 t \\ 
2 t&
\frac{1}{2} (J+8 W)
\end{array} 
 \right ) 
\nonumber 
\nonumber \eeq 
\subsection*{Subspace No. 3 with {\boldmath $N=2$, $s=2$, $m_s=-1$, $\Gamma_{4-3}$}}
\beq
\ket{\Psi_{3-1}} &=& \frac{1}{2 \sqrt{2}}\left ( {\scriptstyle \ket{000000dd}} + {\scriptstyle \ket{00000dd0}} - {\scriptstyle \ket{0000d00d}} + {\scriptstyle \ket{0000dd00}}\right . \nonumber \\                     \nonumber 
                                          && \hspace{0.5cm} \left .  + {\scriptstyle \ket{00dd0000}} + {\scriptstyle \ket{0dd00000}} - {\scriptstyle \ket{d00d0000}} + {\scriptstyle \ket{dd000000}} \nonumber \right )
\nonumber \\ 
\ket{\Psi_{3-2}} &=& -\frac{1}{2 \sqrt{2}}\left ( {\scriptstyle \ket{000d00d0}} - {\scriptstyle \ket{000dd000}} - {\scriptstyle \ket{00d0000d}} + {\scriptstyle \ket{00d00d00}}\right . \nonumber \\                   \nonumber 
                                           && \hspace{0.5cm} \left . -{\scriptstyle \ket{0d0000d0}} + {\scriptstyle \ket{0d00d000}} + {\scriptstyle \ket{d000000d}} - {\scriptstyle \ket{d0000d00}} \nonumber \right )
\nonumber \\ 
 {\bf H}_{3}&=&
 \left ( 
\begin{array}{cc}
\frac{1}{2} (J+8 W)&
2 t \\ 
2 t&
0
\end{array} 
 \right ) 
\nonumber 
\nonumber \eeq 
\subsection*{Subspace No. 4 with {\boldmath $N=2$, $s=2$, $m_s=-1$, $\Gamma_{5-1}$}}
\beq
\ket{\Psi_{4-1}} &=& \frac{1}{2 \sqrt{2}}\left ( {\scriptstyle \ket{000000dd}} - {\scriptstyle \ket{00000dd0}} + {\scriptstyle \ket{0000d00d}} + {\scriptstyle \ket{0000dd00}}\right . \nonumber \\                     \nonumber 
                                          && \hspace{0.5cm} \left .  + {\scriptstyle \ket{00dd0000}} - {\scriptstyle \ket{0dd00000}} + {\scriptstyle \ket{d00d0000}} + {\scriptstyle \ket{dd000000}} \nonumber \right )
\nonumber \\ 
\ket{\Psi_{4-2}} &=& -\frac{1}{2 \sqrt{2}}\left ( {\scriptstyle \ket{000d00d0}} + {\scriptstyle \ket{000dd000}} - {\scriptstyle \ket{00d0000d}} - {\scriptstyle \ket{00d00d00}}\right . \nonumber \\                     \nonumber 
                                           && \hspace{0.5cm} \left .  + {\scriptstyle \ket{0d0000d0}} + {\scriptstyle \ket{0d00d000}} - {\scriptstyle \ket{d000000d}} - {\scriptstyle \ket{d0000d00}} \nonumber \right )
\nonumber \\ 
 {\bf H}_{4}&=&
 \left ( 
\begin{array}{cc}
\frac{1}{2} (J+8 W)&
2 t \\ 
2 t&
0
\end{array} 
 \right ) 
\nonumber 
\nonumber \eeq 
\subsection*{Subspace No. 5 with {\boldmath $N=2$, $s=2$, $m_s=-1$, $\Gamma_{5-2}$}}
\beq
\ket{\Psi_{5-1}} &=& \frac{1}{2 \sqrt{2}}\left ( {\scriptstyle \ket{000000dd}} - {\scriptstyle \ket{0000dd00}} + {\scriptstyle \ket{000d000d}} - {\scriptstyle \ket{00d000d0}}\right . \nonumber \\                   \nonumber 
                                          && \hspace{0.5cm} \left . -{\scriptstyle \ket{00dd0000}} - {\scriptstyle \ket{0d000d00}} + {\scriptstyle \ket{d000d000}} + {\scriptstyle \ket{dd000000}} \nonumber \right )
\nonumber \\ 
\ket{\Psi_{5-2}} &=& \frac{1}{2 \sqrt{2}}\left ( {\scriptstyle \ket{00000d0d}} - {\scriptstyle \ket{0000d0d0}} + {\scriptstyle \ket{000dd000}} - {\scriptstyle \ket{00d00d00}}\right . \nonumber \\                   \nonumber 
                                          && \hspace{0.5cm} \left . -{\scriptstyle \ket{0d0000d0}} - {\scriptstyle \ket{0d0d0000}} + {\scriptstyle \ket{d000000d}} + {\scriptstyle \ket{d0d00000}} \nonumber \right )
\nonumber \\ 
 {\bf H}_{5}&=&
 \left ( 
\begin{array}{cc}
\frac{1}{2} (J+8 W)&
2 t \\ 
2 t&
0
\end{array} 
 \right ) 
\nonumber 
\nonumber \eeq 
\subsection*{Subspace No. 6 with {\boldmath $N=2$, $s=2$, $m_s=-1$, $\Gamma_{5-3}$}}
\beq
\ket{\Psi_{6-1}} &=& -\frac{1}{2 \sqrt{2}}\left ( {\scriptstyle \ket{00000d0d}} + {\scriptstyle \ket{0000d0d0}} + {\scriptstyle \ket{000d00d0}} + {\scriptstyle \ket{00d0000d}}\right . \nonumber \\                   \nonumber 
                                           && \hspace{0.5cm} \left . -{\scriptstyle \ket{0d00d000}} - {\scriptstyle \ket{0d0d0000}} - {\scriptstyle \ket{d0000d00}} - {\scriptstyle \ket{d0d00000}} \nonumber \right )
\nonumber \\ 
\ket{\Psi_{6-2}} &=& -\frac{1}{2 \sqrt{2}}\left ( {\scriptstyle \ket{00000dd0}} + {\scriptstyle \ket{0000d00d}} + {\scriptstyle \ket{000d000d}} + {\scriptstyle \ket{00d000d0}}\right . \nonumber \\                   \nonumber 
                                           && \hspace{0.5cm} \left . -{\scriptstyle \ket{0d000d00}} - {\scriptstyle \ket{0dd00000}} - {\scriptstyle \ket{d000d000}} - {\scriptstyle \ket{d00d0000}} \nonumber \right )
\nonumber \\ 
 {\bf H}_{6}&=&
 \left ( 
\begin{array}{cc}
0&
2 t \\ 
2 t&
\frac{1}{2} (J+8 W)
\end{array} 
 \right ) 
\nonumber 
\nonumber \eeq 
\subsection*{Subspace No. 7 with {\boldmath $N=2$, $s=2$, $m_s=-1$, $\Gamma_{7}$}}
\beq
\ket{\Psi_{7-1}} &=& -\frac{1}{2 \sqrt{3}}\left ( {\scriptstyle \ket{000000dd}} - {\scriptstyle \ket{00000dd0}} + {\scriptstyle \ket{0000d00d}} + {\scriptstyle \ket{0000dd00}} + {\scriptstyle \ket{000d000d}} - {\scriptstyle \ket{00d000d0}}\right . \nonumber \\                   \nonumber 
                                           && \hspace{0.5cm} \left . -{\scriptstyle \ket{00dd0000}} + {\scriptstyle \ket{0d000d00}} + {\scriptstyle \ket{0dd00000}} - {\scriptstyle \ket{d000d000}} - {\scriptstyle \ket{d00d0000}} - {\scriptstyle \ket{dd000000}} \nonumber \right )
\nonumber \\ 
\ket{\Psi_{7-2}} &=& -\frac{1}{2}\left ( {\scriptstyle \ket{000d0d00}} - {\scriptstyle \ket{00d0d000}} + {\scriptstyle \ket{0d00000d}} - {\scriptstyle \ket{d00000d0}} \nonumber \right ) \nonumber 
\nonumber \\ 
 {\bf H}_{7}&=&
 \left ( 
\begin{array}{cc}
\frac{1}{2} (J+8 W)&
0 \\ 
0&
0
\end{array} 
 \right ) 
\nonumber 
\nonumber \eeq 
\subsection*{Subspace No. 8 with {\boldmath $N=2$, $s=2$, $m_s=-1$, $\Gamma_{8-1}$}}
\beq
\ket{\Psi_{8-1}} &=& -\frac{1}{2 \sqrt{6}}\left ( {\scriptstyle \ket{000000dd}} - {\scriptstyle \ket{00000dd0}} + {\scriptstyle \ket{0000d00d}} + {\scriptstyle \ket{0000dd00}}\right . \nonumber \\                   \nonumber 
                                           && \hspace{0.5cm} \left . -{\scriptstyle \ket{00dd0000}} + {\scriptstyle \ket{0dd00000}} - {\scriptstyle \ket{d00d0000}} - {\scriptstyle \ket{dd000000}} \nonumber \right )
\\ 
 && +\frac{1}{\sqrt{6}}\left ( {\scriptstyle \ket{000d000d}} - {\scriptstyle \ket{00d000d0}} + {\scriptstyle \ket{0d000d00}} - {\scriptstyle \ket{d000d000}} \nonumber \right )
\nonumber \\ 
 {\bf H}_{8}&=&
 \left ( 
\begin{array}{c}
\frac{1}{2} (J+8 W)
\end{array} 
 \right ) 
\nonumber 
\nonumber \eeq 
\subsection*{Subspace No. 9 with {\boldmath $N=2$, $s=2$, $m_s=-1$, $\Gamma_{8-2}$}}
\beq
\ket{\Psi_{9-1}} &=& -\frac{1}{2 \sqrt{2}}\left ( {\scriptstyle \ket{000000dd}} + {\scriptstyle \ket{00000dd0}} - {\scriptstyle \ket{0000d00d}} + {\scriptstyle \ket{0000dd00}}\right . \nonumber \\                   \nonumber 
                                           && \hspace{0.5cm} \left . -{\scriptstyle \ket{00dd0000}} - {\scriptstyle \ket{0dd00000}} + {\scriptstyle \ket{d00d0000}} - {\scriptstyle \ket{dd000000}} \nonumber \right )
\nonumber \\ 
 {\bf H}_{9}&=&
 \left ( 
\begin{array}{c}
\frac{1}{2} (J+8 W)
\end{array} 
 \right ) 
\nonumber 
\nonumber \eeq 
\subsection*{Subspace No. 10 with {\boldmath $N=2$, $s=2$, $m_s=-1$, $\Gamma_{9-1}$}}
\beq
\ket{\Psi_{10-1}} &=& \frac{1}{2 \sqrt{2}}\left ( {\scriptstyle \ket{00000d0d}} + {\scriptstyle \ket{0000d0d0}} - {\scriptstyle \ket{000dd000}} - {\scriptstyle \ket{00d00d00}}\right . \nonumber \\                     \nonumber 
                                           && \hspace{0.5cm} \left .  + {\scriptstyle \ket{0d0000d0}} + {\scriptstyle \ket{0d0d0000}} + {\scriptstyle \ket{d000000d}} + {\scriptstyle \ket{d0d00000}} \nonumber \right )
\nonumber \\ 
\ket{\Psi_{10-2}} &=& \frac{1}{2}\left ( {\scriptstyle \ket{00000dd0}} + {\scriptstyle \ket{0000d00d}} + {\scriptstyle \ket{0dd00000}} + {\scriptstyle \ket{d00d0000}} \nonumber \right ) \nonumber 
\nonumber \\ 
\ket{\Psi_{10-3}} &=& -\frac{1}{2}\left ( {\scriptstyle \ket{000d0d00}} + {\scriptstyle \ket{00d0d000}} - {\scriptstyle \ket{0d00000d}} - {\scriptstyle \ket{d00000d0}} \nonumber \right ) \nonumber 
\nonumber \\ 
 {\bf H}_{10}&=&
 \left ( 
\begin{array}{ccc}
0&
2 \sqrt{2} t&
2 \sqrt{2} t \\ 
2 \sqrt{2} t&
\frac{1}{2} (J+8 W)&
0 \\ 
2 \sqrt{2} t&
0&
0
\end{array} 
 \right ) 
\nonumber 
\nonumber \eeq 
\subsection*{Subspace No. 11 with {\boldmath $N=2$, $s=2$, $m_s=-1$, $\Gamma_{9-1}$}}
\beq
\ket{\Psi_{11-1}} &=& -\frac{1}{2}\left ( {\scriptstyle \ket{000000dd}} - {\scriptstyle \ket{0000dd00}} + {\scriptstyle \ket{00dd0000}} - {\scriptstyle \ket{dd000000}} \nonumber \right ) \nonumber 
\nonumber \\ 
\ket{\Psi_{11-2}} &=& -\frac{1}{2 \sqrt{2}}\left ( {\scriptstyle \ket{00000d0d}} - {\scriptstyle \ket{0000d0d0}} - {\scriptstyle \ket{000d00d0}} + {\scriptstyle \ket{00d0000d}}\right . \nonumber \\                     \nonumber 
                                            && \hspace{0.5cm} \left .  + {\scriptstyle \ket{0d00d000}} + {\scriptstyle \ket{0d0d0000}} - {\scriptstyle \ket{d0000d00}} - {\scriptstyle \ket{d0d00000}} \nonumber \right )
\nonumber \\ 
\ket{\Psi_{11-3}} &=& \frac{1}{2}\left ( {\scriptstyle \ket{000d0d00}} - {\scriptstyle \ket{00d0d000}} - {\scriptstyle \ket{0d00000d}} + {\scriptstyle \ket{d00000d0}} \nonumber \right ) \nonumber 
\nonumber \\ 
 {\bf H}_{11}&=&
 \left ( 
\begin{array}{ccc}
\frac{1}{2} (J+8 W)&
2 \sqrt{2} t&
0 \\ 
2 \sqrt{2} t&
0&
2 \sqrt{2} t \\ 
0&
2 \sqrt{2} t&
0
\end{array} 
 \right ) 
\nonumber 
\nonumber \eeq 
\subsection*{Subspace No. 12 with {\boldmath $N=2$, $s=2$, $m_s=-1$, $\Gamma_{9-3}$}}
\beq
\ket{\Psi_{12-1}} &=& \frac{1}{2}\left ( {\scriptstyle \ket{000d000d}} + {\scriptstyle \ket{00d000d0}} + {\scriptstyle \ket{0d000d00}} + {\scriptstyle \ket{d000d000}} \nonumber \right ) \nonumber 
\nonumber \\ 
\ket{\Psi_{12-2}} &=& \frac{1}{2 \sqrt{2}}\left ( {\scriptstyle \ket{000d00d0}} + {\scriptstyle \ket{000dd000}} + {\scriptstyle \ket{00d0000d}} + {\scriptstyle \ket{00d00d00}}\right . \nonumber \\                     \nonumber 
                                           && \hspace{0.5cm} \left .  + {\scriptstyle \ket{0d0000d0}} + {\scriptstyle \ket{0d00d000}} + {\scriptstyle \ket{d000000d}} + {\scriptstyle \ket{d0000d00}} \nonumber \right )
\nonumber \\ 
\ket{\Psi_{12-3}} &=& \frac{1}{2}\left ( {\scriptstyle \ket{000d0d00}} + {\scriptstyle \ket{00d0d000}} + {\scriptstyle \ket{0d00000d}} + {\scriptstyle \ket{d00000d0}} \nonumber \right ) \nonumber 
\nonumber \\ 
 {\bf H}_{12}&=&
 \left ( 
\begin{array}{ccc}
\frac{1}{2} (J+8 W)&
2 \sqrt{2} t&
0 \\ 
2 \sqrt{2} t&
0&
2 \sqrt{2} t \\ 
0&
2 \sqrt{2} t&
0
\end{array} 
 \right ) 
\nonumber 
\nonumber \eeq 
\subsection*{Subspace No. 13 with {\boldmath $N=2$, $s=2$, $m_s=-1$, $\Gamma_{10-1}$}}
\beq
\ket{\Psi_{13-1}} &=& \frac{1}{2 \sqrt{2}}\left ( {\scriptstyle \ket{000d00d0}} - {\scriptstyle \ket{000dd000}} + {\scriptstyle \ket{00d0000d}} - {\scriptstyle \ket{00d00d00}}\right . \nonumber \\                   \nonumber 
                                           && \hspace{0.5cm} \left . -{\scriptstyle \ket{0d0000d0}} + {\scriptstyle \ket{0d00d000}} - {\scriptstyle \ket{d000000d}} + {\scriptstyle \ket{d0000d00}} \nonumber \right )
\nonumber \\ 
 {\bf H}_{13}&=&
 \left ( 
\begin{array}{c}
0
\end{array} 
 \right ) 
\nonumber 
\nonumber \eeq 
\subsection*{Subspace No. 14 with {\boldmath $N=2$, $s=2$, $m_s=-1$, $\Gamma_{10-2}$}}
\beq
\ket{\Psi_{14-1}} &=& \frac{1}{2 \sqrt{2}}\left ( {\scriptstyle \ket{00000d0d}} + {\scriptstyle \ket{0000d0d0}} + {\scriptstyle \ket{000dd000}} + {\scriptstyle \ket{00d00d00}}\right . \nonumber \\                   \nonumber 
                                           && \hspace{0.5cm} \left . -{\scriptstyle \ket{0d0000d0}} + {\scriptstyle \ket{0d0d0000}} - {\scriptstyle \ket{d000000d}} + {\scriptstyle \ket{d0d00000}} \nonumber \right )
\nonumber \\ 
 {\bf H}_{14}&=&
 \left ( 
\begin{array}{c}
0
\end{array} 
 \right ) 
\nonumber 
\nonumber \eeq 
\subsection*{Subspace No. 15 with {\boldmath $N=2$, $s=2$, $m_s=-1$, $\Gamma_{10-3}$}}
\beq
\ket{\Psi_{15-1}} &=& -\frac{1}{2 \sqrt{2}}\left ( {\scriptstyle \ket{00000d0d}} - {\scriptstyle \ket{0000d0d0}} + {\scriptstyle \ket{000d00d0}} - {\scriptstyle \ket{00d0000d}}\right . \nonumber \\                   \nonumber 
                                            && \hspace{0.5cm} \left . -{\scriptstyle \ket{0d00d000}} + {\scriptstyle \ket{0d0d0000}} + {\scriptstyle \ket{d0000d00}} - {\scriptstyle \ket{d0d00000}} \nonumber \right )
\nonumber \\ 
 {\bf H}_{15}&=&
 \left ( 
\begin{array}{c}
0
\end{array} 
 \right ) 
\nonumber 
\nonumber \eeq 
\subsection*{Subspace No. 16 with {\boldmath $N=2$, $s=0$, $m_s=0$, $\Gamma_{1}$}}
\beq
\ket{\Psi_{16-1}} &=& \frac{1}{2 \sqrt{2}}\left ( {\scriptstyle \ket{00000002}} + {\scriptstyle \ket{00000020}} + {\scriptstyle \ket{00000200}} + {\scriptstyle \ket{00002000}}\right . \nonumber \\                     \nonumber 
                                           && \hspace{0.5cm} \left .  + {\scriptstyle \ket{00020000}} + {\scriptstyle \ket{00200000}} + {\scriptstyle \ket{02000000}} + {\scriptstyle \ket{20000000}} \nonumber \right )
\nonumber \\ 
\ket{\Psi_{16-2}} &=& \frac{1}{2 \sqrt{6}}\left ( {\scriptstyle \ket{000000du}} - {\scriptstyle \ket{000000ud}} + {\scriptstyle \ket{00000du0}} - {\scriptstyle \ket{00000ud0}} + {\scriptstyle \ket{0000d00u}} + {\scriptstyle \ket{0000du00}}\right . \nonumber \\                      \nonumber 
                                           && \hspace{0.5cm} \left . -{\scriptstyle \ket{0000u00d}} - {\scriptstyle \ket{0000ud00}} + {\scriptstyle \ket{000d000u}} - {\scriptstyle \ket{000u000d}} + {\scriptstyle \ket{00d000u0}} + {\scriptstyle \ket{00du0000}}\right . \nonumber \\ 
                                           && \hspace{0.5cm} \left . -{\scriptstyle \ket{00u000d0}} - {\scriptstyle \ket{00ud0000}} + {\scriptstyle \ket{0d000u00}} + {\scriptstyle \ket{0du00000}} - {\scriptstyle \ket{0u000d00}} - {\scriptstyle \ket{0ud00000}}\right . \nonumber \\ 
                                           && \hspace{0.5cm} \left .  + {\scriptstyle \ket{d000u000}} + {\scriptstyle \ket{d00u0000}} + {\scriptstyle \ket{du000000}} - {\scriptstyle \ket{u000d000}} - {\scriptstyle \ket{u00d0000}} - {\scriptstyle \ket{ud000000}} \nonumber \right )
\nonumber \\ 
\ket{\Psi_{16-3}} &=& \frac{1}{2 \sqrt{6}}\left ( {\scriptstyle \ket{00000d0u}} - {\scriptstyle \ket{00000u0d}} + {\scriptstyle \ket{0000d0u0}} - {\scriptstyle \ket{0000u0d0}} + {\scriptstyle \ket{000d00u0}} + {\scriptstyle \ket{000du000}}\right . \nonumber \\                        \nonumber 
                                           && \hspace{0.5cm} \left . -{\scriptstyle \ket{000u00d0}} - {\scriptstyle \ket{000ud000}} + {\scriptstyle \ket{00d0000u}} + {\scriptstyle \ket{00d00u00}} - {\scriptstyle \ket{00u0000d}} - {\scriptstyle \ket{00u00d00}}\right . \nonumber \\ 
                                           && \hspace{0.5cm} \left .  + {\scriptstyle \ket{0d0000u0}} + {\scriptstyle \ket{0d00u000}} + {\scriptstyle \ket{0d0u0000}} - {\scriptstyle \ket{0u0000d0}} - {\scriptstyle \ket{0u00d000}} - {\scriptstyle \ket{0u0d0000}}\right . \nonumber \\ 
                                           && \hspace{0.5cm} \left .  + {\scriptstyle \ket{d000000u}} + {\scriptstyle \ket{d0000u00}} + {\scriptstyle \ket{d0u00000}} - {\scriptstyle \ket{u000000d}} - {\scriptstyle \ket{u0000d00}} - {\scriptstyle \ket{u0d00000}} \nonumber \right )
\nonumber \\ 
\ket{\Psi_{16-4}} &=& \frac{1}{2 \sqrt{2}}\left ( {\scriptstyle \ket{000d0u00}} - {\scriptstyle \ket{000u0d00}} + {\scriptstyle \ket{00d0u000}} - {\scriptstyle \ket{00u0d000}}\right . \nonumber \\                     \nonumber 
                                           && \hspace{0.5cm} \left .  + {\scriptstyle \ket{0d00000u}} - {\scriptstyle \ket{0u00000d}} + {\scriptstyle \ket{d00000u0}} - {\scriptstyle \ket{u00000d0}} \nonumber \right )
\nonumber \\ 
 {\bf H}_{16}&=&
 \left ( 
\begin{array}{cccc}
U&
-2 \sqrt{3} t&
0&
0 \\ 
-2 \sqrt{3} t&
4 W-\frac{3 J}{2}&
4 t&
0 \\ 
0&
4 t&
0&
2 \sqrt{3} t \\ 
0&
0&
2 \sqrt{3} t&
0
\end{array} 
 \right ) 
\nonumber 
\nonumber \eeq 
\subsection*{Subspace No. 17 with {\boldmath $N=2$, $s=0$, $m_s=0$, $\Gamma_{3-1}$}}
\beq
\ket{\Psi_{17-1}} &=& \frac{1}{4}\left ( {\scriptstyle \ket{000000du}} - {\scriptstyle \ket{000000ud}} - {\scriptstyle \ket{00000du0}} + {\scriptstyle \ket{00000ud0}} - {\scriptstyle \ket{0000d00u}} + {\scriptstyle \ket{0000du00}}\right . \nonumber \\                        \nonumber 
                                  && \hspace{0.5cm} \left .  + {\scriptstyle \ket{0000u00d}} - {\scriptstyle \ket{0000ud00}} + {\scriptstyle \ket{00du0000}} - {\scriptstyle \ket{00ud0000}} - {\scriptstyle \ket{0du00000}} + {\scriptstyle \ket{0ud00000}}\right . \nonumber \\ 
                                  && \hspace{0.5cm} \left . -{\scriptstyle \ket{d00u0000}} + {\scriptstyle \ket{du000000}} + {\scriptstyle \ket{u00d0000}} - {\scriptstyle \ket{ud000000}} \nonumber \right )
\nonumber \\ 
\ket{\Psi_{17-2}} &=& \frac{1}{4}\left ( {\scriptstyle \ket{000d00u0}} - {\scriptstyle \ket{000du000}} - {\scriptstyle \ket{000u00d0}} + {\scriptstyle \ket{000ud000}} + {\scriptstyle \ket{00d0000u}} - {\scriptstyle \ket{00d00u00}}\right . \nonumber \\                      \nonumber 
                                  && \hspace{0.5cm} \left . -{\scriptstyle \ket{00u0000d}} + {\scriptstyle \ket{00u00d00}} - {\scriptstyle \ket{0d0000u0}} + {\scriptstyle \ket{0d00u000}} + {\scriptstyle \ket{0u0000d0}} - {\scriptstyle \ket{0u00d000}}\right . \nonumber \\ 
                                  && \hspace{0.5cm} \left . -{\scriptstyle \ket{d000000u}} + {\scriptstyle \ket{d0000u00}} + {\scriptstyle \ket{u000000d}} - {\scriptstyle \ket{u0000d00}} \nonumber \right )
\nonumber \\ 
 {\bf H}_{17}&=&
 \left ( 
\begin{array}{cc}
4 W-\frac{3 J}{2}&
2 t \\ 
2 t&
0
\end{array} 
 \right ) 
\nonumber 
\nonumber \eeq 
\subsection*{Subspace No. 18 with {\boldmath $N=2$, $s=0$, $m_s=0$, $\Gamma_{3-2}$}}
\beq
\ket{\Psi_{18-1}} &=& \frac{1}{4 \sqrt{3}}\left ( {\scriptstyle \ket{000000du}} - {\scriptstyle \ket{000000ud}} + {\scriptstyle \ket{00000du0}} - {\scriptstyle \ket{00000ud0}} + {\scriptstyle \ket{0000d00u}} + {\scriptstyle \ket{0000du00}}\right . \nonumber \\                      \nonumber 
                                           && \hspace{0.5cm} \left . -{\scriptstyle \ket{0000u00d}} - {\scriptstyle \ket{0000ud00}} + {\scriptstyle \ket{00du0000}} - {\scriptstyle \ket{00ud0000}} + {\scriptstyle \ket{0du00000}} - {\scriptstyle \ket{0ud00000}}\right . \nonumber \\ 
                                           && \hspace{0.5cm} \left .  + {\scriptstyle \ket{d00u0000}} + {\scriptstyle \ket{du000000}} - {\scriptstyle \ket{u00d0000}} - {\scriptstyle \ket{ud000000}} \nonumber \right )
\\ 
 &&  -\frac{1}{2 \sqrt{3}}\left ( {\scriptstyle \ket{000d000u}} - {\scriptstyle \ket{000u000d}} + {\scriptstyle \ket{00d000u0}} - {\scriptstyle \ket{00u000d0}}\right . \nonumber \\ 
                           && \hspace{0.5cm} \left .  + {\scriptstyle \ket{0d000u00}} - {\scriptstyle \ket{0u000d00}} + {\scriptstyle \ket{d000u000}} - {\scriptstyle \ket{u000d000}} \nonumber \right )
\nonumber \\ 
\ket{\Psi_{18-2}} &=& \frac{1}{2 \sqrt{3}}\left ( {\scriptstyle \ket{00000d0u}} - {\scriptstyle \ket{00000u0d}} + {\scriptstyle \ket{0000d0u0}} - {\scriptstyle \ket{0000u0d0}}\right . \nonumber \\                     \nonumber 
                                           && \hspace{0.5cm} \left .  + {\scriptstyle \ket{0d0u0000}} - {\scriptstyle \ket{0u0d0000}} + {\scriptstyle \ket{d0u00000}} - {\scriptstyle \ket{u0d00000}} \nonumber \right )
\\ 
 &&  -\frac{1}{4 \sqrt{3}}\left ( {\scriptstyle \ket{000d00u0}} + {\scriptstyle \ket{000du000}} - {\scriptstyle \ket{000u00d0}} - {\scriptstyle \ket{000ud000}} + {\scriptstyle \ket{00d0000u}} + {\scriptstyle \ket{00d00u00}}\right . \nonumber \\ 
                           && \hspace{0.5cm} \left . -{\scriptstyle \ket{00u0000d}} - {\scriptstyle \ket{00u00d00}} + {\scriptstyle \ket{0d0000u0}} + {\scriptstyle \ket{0d00u000}} - {\scriptstyle \ket{0u0000d0}} - {\scriptstyle \ket{0u00d000}}\right . \nonumber \\ 
                           && \hspace{0.5cm} \left .  + {\scriptstyle \ket{d000000u}} + {\scriptstyle \ket{d0000u00}} - {\scriptstyle \ket{u000000d}} - {\scriptstyle \ket{u0000d00}} \nonumber \right )
\nonumber \\ 
 {\bf H}_{18}&=&
 \left ( 
\begin{array}{cc}
4 W-\frac{3 J}{2}&
2 t \\ 
2 t&
0
\end{array} 
 \right ) 
\nonumber 
\nonumber \eeq 
\subsection*{Subspace No. 19 with {\boldmath $N=2$, $s=0$, $m_s=0$, $\Gamma_{5-1}$}}
\beq
\ket{\Psi_{19-1}} &=& -\frac{1}{2 \sqrt{2}}\left ( {\scriptstyle \ket{00000002}} - {\scriptstyle \ket{00000020}} + {\scriptstyle \ket{00000200}} - {\scriptstyle \ket{00002000}}\right . \nonumber \\                     \nonumber 
                                            && \hspace{0.5cm} \left .  + {\scriptstyle \ket{00020000}} - {\scriptstyle \ket{00200000}} + {\scriptstyle \ket{02000000}} - {\scriptstyle \ket{20000000}} \nonumber \right )
\nonumber \\ 
\ket{\Psi_{19-2}} &=& -\frac{1}{2 \sqrt{2}}\left ( {\scriptstyle \ket{00000d0u}} - {\scriptstyle \ket{00000u0d}} - {\scriptstyle \ket{0000d0u0}} + {\scriptstyle \ket{0000u0d0}}\right . \nonumber \\                     \nonumber 
                                            && \hspace{0.5cm} \left .  + {\scriptstyle \ket{0d0u0000}} - {\scriptstyle \ket{0u0d0000}} - {\scriptstyle \ket{d0u00000}} + {\scriptstyle \ket{u0d00000}} \nonumber \right )
\nonumber \\ 
\ket{\Psi_{19-3}} &=& -\frac{1}{2 \sqrt{2}}\left ( {\scriptstyle \ket{000d000u}} - {\scriptstyle \ket{000u000d}} - {\scriptstyle \ket{00d000u0}} + {\scriptstyle \ket{00u000d0}}\right . \nonumber \\                     \nonumber 
                                            && \hspace{0.5cm} \left .  + {\scriptstyle \ket{0d000u00}} - {\scriptstyle \ket{0u000d00}} - {\scriptstyle \ket{d000u000}} + {\scriptstyle \ket{u000d000}} \nonumber \right )
\nonumber \\ 
\ket{\Psi_{19-4}} &=& -\frac{1}{2 \sqrt{2}}\left ( {\scriptstyle \ket{000d0u00}} - {\scriptstyle \ket{000u0d00}} - {\scriptstyle \ket{00d0u000}} + {\scriptstyle \ket{00u0d000}}\right . \nonumber \\                     \nonumber 
                                            && \hspace{0.5cm} \left .  + {\scriptstyle \ket{0d00000u}} - {\scriptstyle \ket{0u00000d}} - {\scriptstyle \ket{d00000u0}} + {\scriptstyle \ket{u00000d0}} \nonumber \right )
\nonumber \\ 
 {\bf H}_{19}&=&
 \left ( 
\begin{array}{cccc}
U&
0&
-2 t&
0 \\ 
0&
0&
0&
2 t \\ 
-2 t&
0&
4 W-\frac{3 J}{2}&
0 \\ 
0&
2 t&
0&
0
\end{array} 
 \right ) 
\nonumber 
\nonumber \eeq 
\subsection*{Subspace No. 20 with {\boldmath $N=2$, $s=0$, $m_s=0$, $\Gamma_{5-2}$}}
\beq
\ket{\Psi_{20-1}} &=& -\frac{1}{2 \sqrt{2}}\left ( {\scriptstyle \ket{00000002}} - {\scriptstyle \ket{00000020}} - {\scriptstyle \ket{00000200}} + {\scriptstyle \ket{00002000}}\right . \nonumber \\                   \nonumber 
                                            && \hspace{0.5cm} \left . -{\scriptstyle \ket{00020000}} + {\scriptstyle \ket{00200000}} + {\scriptstyle \ket{02000000}} - {\scriptstyle \ket{20000000}} \nonumber \right )
\nonumber \\ 
\ket{\Psi_{20-2}} &=& \frac{1}{2 \sqrt{2}}\left ( {\scriptstyle \ket{00000du0}} - {\scriptstyle \ket{00000ud0}} - {\scriptstyle \ket{0000d00u}} + {\scriptstyle \ket{0000u00d}}\right . \nonumber \\                   \nonumber 
                                           && \hspace{0.5cm} \left . -{\scriptstyle \ket{0du00000}} + {\scriptstyle \ket{0ud00000}} + {\scriptstyle \ket{d00u0000}} - {\scriptstyle \ket{u00d0000}} \nonumber \right )
\nonumber \\ 
\ket{\Psi_{20-3}} &=& \frac{1}{2 \sqrt{2}}\left ( {\scriptstyle \ket{000d00u0}} - {\scriptstyle \ket{000u00d0}} - {\scriptstyle \ket{00d0000u}} + {\scriptstyle \ket{00u0000d}}\right . \nonumber \\                   \nonumber 
                                           && \hspace{0.5cm} \left . -{\scriptstyle \ket{0d00u000}} + {\scriptstyle \ket{0u00d000}} + {\scriptstyle \ket{d0000u00}} - {\scriptstyle \ket{u0000d00}} \nonumber \right )
\nonumber \\ 
\ket{\Psi_{20-4}} &=& \frac{1}{2 \sqrt{2}}\left ( {\scriptstyle \ket{000d0u00}} - {\scriptstyle \ket{000u0d00}} - {\scriptstyle \ket{00d0u000}} + {\scriptstyle \ket{00u0d000}}\right . \nonumber \\                   \nonumber 
                                           && \hspace{0.5cm} \left . -{\scriptstyle \ket{0d00000u}} + {\scriptstyle \ket{0u00000d}} + {\scriptstyle \ket{d00000u0}} - {\scriptstyle \ket{u00000d0}} \nonumber \right )
\nonumber \\ 
 {\bf H}_{20}&=&
 \left ( 
\begin{array}{cccc}
U&
-2 t&
0&
0 \\ 
-2 t&
4 W-\frac{3 J}{2}&
0&
0 \\ 
0&
0&
0&
2 t \\ 
0&
0&
2 t&
0
\end{array} 
 \right ) 
\nonumber 
\nonumber \eeq 
\subsection*{Subspace No. 21 with {\boldmath $N=2$, $s=0$, $m_s=0$, $\Gamma_{5-3}$}}
\beq
\ket{\Psi_{21-1}} &=& \frac{1}{2 \sqrt{2}}\left ( {\scriptstyle \ket{00000002}} + {\scriptstyle \ket{00000020}} - {\scriptstyle \ket{00000200}} - {\scriptstyle \ket{00002000}}\right . \nonumber \\                   \nonumber 
                                           && \hspace{0.5cm} \left . -{\scriptstyle \ket{00020000}} - {\scriptstyle \ket{00200000}} + {\scriptstyle \ket{02000000}} + {\scriptstyle \ket{20000000}} \nonumber \right )
\nonumber \\ 
\ket{\Psi_{21-2}} &=& \frac{1}{2 \sqrt{2}}\left ( {\scriptstyle \ket{000000du}} - {\scriptstyle \ket{000000ud}} - {\scriptstyle \ket{0000du00}} + {\scriptstyle \ket{0000ud00}}\right . \nonumber \\                   \nonumber 
                                           && \hspace{0.5cm} \left . -{\scriptstyle \ket{00du0000}} + {\scriptstyle \ket{00ud0000}} + {\scriptstyle \ket{du000000}} - {\scriptstyle \ket{ud000000}} \nonumber \right )
\nonumber \\ 
\ket{\Psi_{21-3}} &=& -\frac{1}{2 \sqrt{2}}\left ( {\scriptstyle \ket{000d0u00}} - {\scriptstyle \ket{000u0d00}} + {\scriptstyle \ket{00d0u000}} - {\scriptstyle \ket{00u0d000}}\right . \nonumber \\                   \nonumber 
                                            && \hspace{0.5cm} \left . -{\scriptstyle \ket{0d00000u}} + {\scriptstyle \ket{0u00000d}} - {\scriptstyle \ket{d00000u0}} + {\scriptstyle \ket{u00000d0}} \nonumber \right )
\nonumber \\ 
\ket{\Psi_{21-4}} &=& -\frac{1}{2 \sqrt{2}}\left ( {\scriptstyle \ket{000du000}} - {\scriptstyle \ket{000ud000}} + {\scriptstyle \ket{00d00u00}} - {\scriptstyle \ket{00u00d00}}\right . \nonumber \\                   \nonumber 
                                            && \hspace{0.5cm} \left . -{\scriptstyle \ket{0d0000u0}} + {\scriptstyle \ket{0u0000d0}} - {\scriptstyle \ket{d000000u}} + {\scriptstyle \ket{u000000d}} \nonumber \right )
\nonumber \\ 
 {\bf H}_{21}&=&
 \left ( 
\begin{array}{cccc}
U&
-2 t&
0&
0 \\ 
-2 t&
4 W-\frac{3 J}{2}&
0&
0 \\ 
0&
0&
0&
2 t \\ 
0&
0&
2 t&
0
\end{array} 
 \right ) 
\nonumber 
\nonumber \eeq 
\subsection*{Subspace No. 22 with {\boldmath $N=2$, $s=0$, $m_s=0$, $\Gamma_{7}$}}
\beq
\ket{\Psi_{22-1}} &=& \frac{1}{2 \sqrt{6}}\left ( {\scriptstyle \ket{00000d0u}} - {\scriptstyle \ket{00000u0d}} - {\scriptstyle \ket{0000d0u0}} + {\scriptstyle \ket{0000u0d0}} - {\scriptstyle \ket{000d00u0}} - {\scriptstyle \ket{000du000}}\right . \nonumber \\                        \nonumber 
                                           && \hspace{0.5cm} \left .  + {\scriptstyle \ket{000u00d0}} + {\scriptstyle \ket{000ud000}} + {\scriptstyle \ket{00d0000u}} + {\scriptstyle \ket{00d00u00}} - {\scriptstyle \ket{00u0000d}} - {\scriptstyle \ket{00u00d00}}\right . \nonumber \\ 
                                           && \hspace{0.5cm} \left . -{\scriptstyle \ket{0d0000u0}} - {\scriptstyle \ket{0d00u000}} - {\scriptstyle \ket{0d0u0000}} + {\scriptstyle \ket{0u0000d0}} + {\scriptstyle \ket{0u00d000}} + {\scriptstyle \ket{0u0d0000}}\right . \nonumber \\ 
                                           && \hspace{0.5cm} \left .  + {\scriptstyle \ket{d000000u}} + {\scriptstyle \ket{d0000u00}} + {\scriptstyle \ket{d0u00000}} - {\scriptstyle \ket{u000000d}} - {\scriptstyle \ket{u0000d00}} - {\scriptstyle \ket{u0d00000}} \nonumber \right )
\nonumber \\ 
 {\bf H}_{22}&=&
 \left ( 
\begin{array}{c}
0
\end{array} 
 \right ) 
\nonumber 
\nonumber \eeq 
\subsection*{Subspace No. 23 with {\boldmath $N=2$, $s=0$, $m_s=0$, $\Gamma_{8-1}$}}
\beq
\ket{\Psi_{23-1}} &=& \frac{1}{2 \sqrt{3}}\left ( {\scriptstyle \ket{00000d0u}} - {\scriptstyle \ket{00000u0d}} - {\scriptstyle \ket{0000d0u0}} + {\scriptstyle \ket{0000u0d0}}\right . \nonumber \\                   \nonumber 
                                           && \hspace{0.5cm} \left . -{\scriptstyle \ket{0d0u0000}} + {\scriptstyle \ket{0u0d0000}} + {\scriptstyle \ket{d0u00000}} - {\scriptstyle \ket{u0d00000}} \nonumber \right )
\\ 
 && +\frac{1}{4 \sqrt{3}}\left ( {\scriptstyle \ket{000d00u0}} + {\scriptstyle \ket{000du000}} - {\scriptstyle \ket{000u00d0}} - {\scriptstyle \ket{000ud000}} - {\scriptstyle \ket{00d0000u}} - {\scriptstyle \ket{00d00u00}}\right . \nonumber \\ 
                          && \hspace{0.5cm} \left .  + {\scriptstyle \ket{00u0000d}} + {\scriptstyle \ket{00u00d00}} + {\scriptstyle \ket{0d0000u0}} + {\scriptstyle \ket{0d00u000}} - {\scriptstyle \ket{0u0000d0}} - {\scriptstyle \ket{0u00d000}}\right . \nonumber \\ 
                          && \hspace{0.5cm} \left . -{\scriptstyle \ket{d000000u}} - {\scriptstyle \ket{d0000u00}} + {\scriptstyle \ket{u000000d}} + {\scriptstyle \ket{u0000d00}} \nonumber \right )
\nonumber \\ 
 {\bf H}_{23}&=&
 \left ( 
\begin{array}{c}
0
\end{array} 
 \right ) 
\nonumber 
\nonumber \eeq 
\subsection*{Subspace No. 24 with {\boldmath $N=2$, $s=0$, $m_s=0$, $\Gamma_{8-2}$}}
\beq
\ket{\Psi_{24-1}} &=& -\frac{1}{4}\left ( {\scriptstyle \ket{000d00u0}} - {\scriptstyle \ket{000du000}} - {\scriptstyle \ket{000u00d0}} + {\scriptstyle \ket{000ud000}} - {\scriptstyle \ket{00d0000u}} + {\scriptstyle \ket{00d00u00}}\right . \nonumber \\                        \nonumber 
                                   && \hspace{0.5cm} \left .  + {\scriptstyle \ket{00u0000d}} - {\scriptstyle \ket{00u00d00}} - {\scriptstyle \ket{0d0000u0}} + {\scriptstyle \ket{0d00u000}} + {\scriptstyle \ket{0u0000d0}} - {\scriptstyle \ket{0u00d000}}\right . \nonumber \\ 
                                   && \hspace{0.5cm} \left .  + {\scriptstyle \ket{d000000u}} - {\scriptstyle \ket{d0000u00}} - {\scriptstyle \ket{u000000d}} + {\scriptstyle \ket{u0000d00}} \nonumber \right )
\nonumber \\ 
 {\bf H}_{24}&=&
 \left ( 
\begin{array}{c}
0
\end{array} 
 \right ) 
\nonumber 
\nonumber \eeq 
\subsection*{Subspace No. 25 with {\boldmath $N=2$, $s=0$, $m_s=0$, $\Gamma_{9-1}$}}
\beq
\ket{\Psi_{25-1}} &=& -\frac{1}{2 \sqrt{2}}\left ( {\scriptstyle \ket{00000002}} + {\scriptstyle \ket{00000020}} - {\scriptstyle \ket{00000200}} - {\scriptstyle \ket{00002000}}\right . \nonumber \\                     \nonumber 
                                            && \hspace{0.5cm} \left .  + {\scriptstyle \ket{00020000}} + {\scriptstyle \ket{00200000}} - {\scriptstyle \ket{02000000}} - {\scriptstyle \ket{20000000}} \nonumber \right )
\nonumber \\ 
\ket{\Psi_{25-2}} &=& -\frac{1}{4}\left ( {\scriptstyle \ket{000000du}} - {\scriptstyle \ket{000000ud}} - {\scriptstyle \ket{0000du00}} + {\scriptstyle \ket{0000ud00}} + {\scriptstyle \ket{000d000u}} - {\scriptstyle \ket{000u000d}}\right . \nonumber \\                        \nonumber 
                                   && \hspace{0.5cm} \left .  + {\scriptstyle \ket{00d000u0}} + {\scriptstyle \ket{00du0000}} - {\scriptstyle \ket{00u000d0}} - {\scriptstyle \ket{00ud0000}} - {\scriptstyle \ket{0d000u00}} + {\scriptstyle \ket{0u000d00}}\right . \nonumber \\ 
                                   && \hspace{0.5cm} \left . -{\scriptstyle \ket{d000u000}} - {\scriptstyle \ket{du000000}} + {\scriptstyle \ket{u000d000}} + {\scriptstyle \ket{ud000000}} \nonumber \right )
\nonumber \\ 
\ket{\Psi_{25-3}} &=& -\frac{1}{2 \sqrt{2}}\left ( {\scriptstyle \ket{000d00u0}} - {\scriptstyle \ket{000u00d0}} + {\scriptstyle \ket{00d0000u}} - {\scriptstyle \ket{00u0000d}}\right . \nonumber \\                   \nonumber 
                                            && \hspace{0.5cm} \left . -{\scriptstyle \ket{0d00u000}} + {\scriptstyle \ket{0u00d000}} - {\scriptstyle \ket{d0000u00}} + {\scriptstyle \ket{u0000d00}} \nonumber \right )
\nonumber \\ 
 {\bf H}_{25}&=&
 \left ( 
\begin{array}{ccc}
U&
-2 \sqrt{2} t&
0 \\ 
-2 \sqrt{2} t&
4 W-\frac{3 J}{2}&
2 \sqrt{2} t \\ 
0&
2 \sqrt{2} t&
0
\end{array} 
 \right ) 
\nonumber 
\nonumber \eeq 
\subsection*{Subspace No. 26 with {\boldmath $N=2$, $s=0$, $m_s=0$, $\Gamma_{9-1}$}}
\beq
\ket{\Psi_{26-1}} &=& -\frac{1}{2 \sqrt{2}}\left ( {\scriptstyle \ket{00000002}} - {\scriptstyle \ket{00000020}} - {\scriptstyle \ket{00000200}} + {\scriptstyle \ket{00002000}}\right . \nonumber \\                     \nonumber 
                                            && \hspace{0.5cm} \left .  + {\scriptstyle \ket{00020000}} - {\scriptstyle \ket{00200000}} - {\scriptstyle \ket{02000000}} + {\scriptstyle \ket{20000000}} \nonumber \right )
\nonumber \\ 
\ket{\Psi_{26-2}} &=& -\frac{1}{4}\left ( {\scriptstyle \ket{00000du0}} - {\scriptstyle \ket{00000ud0}} - {\scriptstyle \ket{0000d00u}} + {\scriptstyle \ket{0000u00d}} - {\scriptstyle \ket{000d000u}} + {\scriptstyle \ket{000u000d}}\right . \nonumber \\                        \nonumber 
                                   && \hspace{0.5cm} \left .  + {\scriptstyle \ket{00d000u0}} - {\scriptstyle \ket{00u000d0}} + {\scriptstyle \ket{0d000u00}} + {\scriptstyle \ket{0du00000}} - {\scriptstyle \ket{0u000d00}} - {\scriptstyle \ket{0ud00000}}\right . \nonumber \\ 
                                   && \hspace{0.5cm} \left . -{\scriptstyle \ket{d000u000}} - {\scriptstyle \ket{d00u0000}} + {\scriptstyle \ket{u000d000}} + {\scriptstyle \ket{u00d0000}} \nonumber \right )
\nonumber \\ 
\ket{\Psi_{26-3}} &=& \frac{1}{2 \sqrt{2}}\left ( {\scriptstyle \ket{000du000}} - {\scriptstyle \ket{000ud000}} - {\scriptstyle \ket{00d00u00}} + {\scriptstyle \ket{00u00d00}}\right . \nonumber \\                   \nonumber 
                                           && \hspace{0.5cm} \left . -{\scriptstyle \ket{0d0000u0}} + {\scriptstyle \ket{0u0000d0}} + {\scriptstyle \ket{d000000u}} - {\scriptstyle \ket{u000000d}} \nonumber \right )
\nonumber \\ 
 {\bf H}_{26}&=&
 \left ( 
\begin{array}{ccc}
U&
2 \sqrt{2} t&
0 \\ 
2 \sqrt{2} t&
4 W-\frac{3 J}{2}&
2 \sqrt{2} t \\ 
0&
2 \sqrt{2} t&
0
\end{array} 
 \right ) 
\nonumber 
\nonumber \eeq 
\subsection*{Subspace No. 27 with {\boldmath $N=2$, $s=0$, $m_s=0$, $\Gamma_{9-3}$}}
\beq
\ket{\Psi_{27-1}} &=& -\frac{1}{2 \sqrt{2}}\left ( {\scriptstyle \ket{00000002}} + {\scriptstyle \ket{00000020}} + {\scriptstyle \ket{00000200}} + {\scriptstyle \ket{00002000}}\right . \nonumber \\                   \nonumber 
                                            && \hspace{0.5cm} \left . -{\scriptstyle \ket{00020000}} - {\scriptstyle \ket{00200000}} - {\scriptstyle \ket{02000000}} - {\scriptstyle \ket{20000000}} \nonumber \right )
\nonumber \\ 
\ket{\Psi_{27-2}} &=& -\frac{1}{4}\left ( {\scriptstyle \ket{000000du}} - {\scriptstyle \ket{000000ud}} + {\scriptstyle \ket{00000du0}} - {\scriptstyle \ket{00000ud0}} + {\scriptstyle \ket{0000d00u}} + {\scriptstyle \ket{0000du00}}\right . \nonumber \\                      \nonumber 
                                   && \hspace{0.5cm} \left . -{\scriptstyle \ket{0000u00d}} - {\scriptstyle \ket{0000ud00}} - {\scriptstyle \ket{00du0000}} + {\scriptstyle \ket{00ud0000}} - {\scriptstyle \ket{0du00000}} + {\scriptstyle \ket{0ud00000}}\right . \nonumber \\ 
                                   && \hspace{0.5cm} \left . -{\scriptstyle \ket{d00u0000}} - {\scriptstyle \ket{du000000}} + {\scriptstyle \ket{u00d0000}} + {\scriptstyle \ket{ud000000}} \nonumber \right )
\nonumber \\ 
\ket{\Psi_{27-3}} &=& -\frac{1}{2 \sqrt{2}}\left ( {\scriptstyle \ket{00000d0u}} - {\scriptstyle \ket{00000u0d}} + {\scriptstyle \ket{0000d0u0}} - {\scriptstyle \ket{0000u0d0}}\right . \nonumber \\                   \nonumber 
                                            && \hspace{0.5cm} \left . -{\scriptstyle \ket{0d0u0000}} + {\scriptstyle \ket{0u0d0000}} - {\scriptstyle \ket{d0u00000}} + {\scriptstyle \ket{u0d00000}} \nonumber \right )
\nonumber \\ 
 {\bf H}_{27}&=&
 \left ( 
\begin{array}{ccc}
U&
-2 \sqrt{2} t&
0 \\ 
-2 \sqrt{2} t&
4 W-\frac{3 J}{2}&
2 \sqrt{2} t \\ 
0&
2 \sqrt{2} t&
0
\end{array} 
 \right ) 
\nonumber 
\nonumber \eeq 
\subsection*{Subspace No. 28 with {\boldmath $N=2$, $s=0$, $m_s=0$, $\Gamma_{10-1}$}}
\beq
\ket{\Psi_{28-1}} &=& -\frac{1}{4}\left ( {\scriptstyle \ket{000000du}} - {\scriptstyle \ket{000000ud}} - {\scriptstyle \ket{00000du0}} + {\scriptstyle \ket{00000ud0}} - {\scriptstyle \ket{0000d00u}} + {\scriptstyle \ket{0000du00}}\right . \nonumber \\                        \nonumber 
                                   && \hspace{0.5cm} \left .  + {\scriptstyle \ket{0000u00d}} - {\scriptstyle \ket{0000ud00}} - {\scriptstyle \ket{00du0000}} + {\scriptstyle \ket{00ud0000}} + {\scriptstyle \ket{0du00000}} - {\scriptstyle \ket{0ud00000}}\right . \nonumber \\ 
                                   && \hspace{0.5cm} \left .  + {\scriptstyle \ket{d00u0000}} - {\scriptstyle \ket{du000000}} - {\scriptstyle \ket{u00d0000}} + {\scriptstyle \ket{ud000000}} \nonumber \right )
\nonumber \\ 
 {\bf H}_{28}&=&
 \left ( 
\begin{array}{c}
4 W-\frac{3 J}{2}
\end{array} 
 \right ) 
\nonumber 
\nonumber \eeq 
\subsection*{Subspace No. 29 with {\boldmath $N=2$, $s=0$, $m_s=0$, $\Gamma_{10-2}$}}
\beq
\ket{\Psi_{29-1}} &=& -\frac{1}{4}\left ( {\scriptstyle \ket{000000du}} - {\scriptstyle \ket{000000ud}} - {\scriptstyle \ket{0000du00}} + {\scriptstyle \ket{0000ud00}} - {\scriptstyle \ket{000d000u}} + {\scriptstyle \ket{000u000d}}\right . \nonumber \\                      \nonumber 
                                   && \hspace{0.5cm} \left . -{\scriptstyle \ket{00d000u0}} + {\scriptstyle \ket{00du0000}} + {\scriptstyle \ket{00u000d0}} - {\scriptstyle \ket{00ud0000}} + {\scriptstyle \ket{0d000u00}} - {\scriptstyle \ket{0u000d00}}\right . \nonumber \\ 
                                   && \hspace{0.5cm} \left .  + {\scriptstyle \ket{d000u000}} - {\scriptstyle \ket{du000000}} - {\scriptstyle \ket{u000d000}} + {\scriptstyle \ket{ud000000}} \nonumber \right )
\nonumber \\ 
 {\bf H}_{29}&=&
 \left ( 
\begin{array}{c}
4 W-\frac{3 J}{2}
\end{array} 
 \right ) 
\nonumber 
\nonumber \eeq 
\subsection*{Subspace No. 30 with {\boldmath $N=2$, $s=0$, $m_s=0$, $\Gamma_{10-3}$}}
\beq
\ket{\Psi_{30-1}} &=& -\frac{1}{4}\left ( {\scriptstyle \ket{00000du0}} - {\scriptstyle \ket{00000ud0}} - {\scriptstyle \ket{0000d00u}} + {\scriptstyle \ket{0000u00d}} + {\scriptstyle \ket{000d000u}} - {\scriptstyle \ket{000u000d}}\right . \nonumber \\                      \nonumber 
                                   && \hspace{0.5cm} \left . -{\scriptstyle \ket{00d000u0}} + {\scriptstyle \ket{00u000d0}} - {\scriptstyle \ket{0d000u00}} + {\scriptstyle \ket{0du00000}} + {\scriptstyle \ket{0u000d00}} - {\scriptstyle \ket{0ud00000}}\right . \nonumber \\ 
                                   && \hspace{0.5cm} \left .  + {\scriptstyle \ket{d000u000}} - {\scriptstyle \ket{d00u0000}} - {\scriptstyle \ket{u000d000}} + {\scriptstyle \ket{u00d0000}} \nonumber \right )
\nonumber \\ 
 {\bf H}_{30}&=&
 \left ( 
\begin{array}{c}
4 W-\frac{3 J}{2}
\end{array} 
 \right ) 
\nonumber 
\nonumber \eeq 
\subsection*{Subspace No. 31 with {\boldmath $N=2$, $s=0$, $m_s=0$, $\Gamma_{7}$}}
\beq
\ket{\Psi_{31-1}} &=& \frac{1}{2 \sqrt{2}}\left ( {\scriptstyle \ket{00000002}} - {\scriptstyle \ket{00000020}} + {\scriptstyle \ket{00000200}} - {\scriptstyle \ket{00002000}}\right . \nonumber \\                   \nonumber 
                                           && \hspace{0.5cm} \left . -{\scriptstyle \ket{00020000}} + {\scriptstyle \ket{00200000}} - {\scriptstyle \ket{02000000}} + {\scriptstyle \ket{20000000}} \nonumber \right )
\nonumber \\ 
 {\bf H}_{31}&=&
 \left ( 
\begin{array}{c}
U
\end{array} 
 \right ) 
\nonumber 
\nonumber \eeq 
\subsection*{Subspace No. 32 with {\boldmath $N=2$, $s=2$, $m_s=0$, $\Gamma_{4-1}$}}
\beq
\ket{\Psi_{32-1}} &=& \frac{1}{4}\left ( {\scriptstyle \ket{000000du}} + {\scriptstyle \ket{000000ud}} - {\scriptstyle \ket{0000du00}} - {\scriptstyle \ket{0000ud00}} - {\scriptstyle \ket{000d000u}} - {\scriptstyle \ket{000u000d}}\right . \nonumber \\                        \nonumber 
                                  && \hspace{0.5cm} \left .  + {\scriptstyle \ket{00d000u0}} - {\scriptstyle \ket{00du0000}} + {\scriptstyle \ket{00u000d0}} - {\scriptstyle \ket{00ud0000}} + {\scriptstyle \ket{0d000u00}} + {\scriptstyle \ket{0u000d00}}\right . \nonumber \\ 
                                  && \hspace{0.5cm} \left . -{\scriptstyle \ket{d000u000}} + {\scriptstyle \ket{du000000}} - {\scriptstyle \ket{u000d000}} + {\scriptstyle \ket{ud000000}} \nonumber \right )
\nonumber \\ 
\ket{\Psi_{32-2}} &=& \frac{1}{4}\left ( {\scriptstyle \ket{00000d0u}} + {\scriptstyle \ket{00000u0d}} - {\scriptstyle \ket{0000d0u0}} - {\scriptstyle \ket{0000u0d0}} - {\scriptstyle \ket{000du000}} - {\scriptstyle \ket{000ud000}}\right . \nonumber \\                        \nonumber 
                                  && \hspace{0.5cm} \left .  + {\scriptstyle \ket{00d00u00}} + {\scriptstyle \ket{00u00d00}} + {\scriptstyle \ket{0d0000u0}} - {\scriptstyle \ket{0d0u0000}} + {\scriptstyle \ket{0u0000d0}} - {\scriptstyle \ket{0u0d0000}}\right . \nonumber \\ 
                                  && \hspace{0.5cm} \left . -{\scriptstyle \ket{d000000u}} + {\scriptstyle \ket{d0u00000}} - {\scriptstyle \ket{u000000d}} + {\scriptstyle \ket{u0d00000}} \nonumber \right )
\nonumber \\ 
 {\bf H}_{32}&=&
 \left ( 
\begin{array}{cc}
\frac{1}{2} (J+8 W)&
2 t \\ 
2 t&
0
\end{array} 
 \right ) 
\nonumber 
\nonumber \eeq 
\subsection*{Subspace No. 33 with {\boldmath $N=2$, $s=2$, $m_s=0$, $\Gamma_{4-2}$}}
\beq
\ket{\Psi_{33-1}} &=& -\frac{1}{4}\left ( {\scriptstyle \ket{00000d0u}} + {\scriptstyle \ket{00000u0d}} + {\scriptstyle \ket{0000d0u0}} + {\scriptstyle \ket{0000u0d0}} - {\scriptstyle \ket{000d00u0}} - {\scriptstyle \ket{000u00d0}}\right . \nonumber \\                      \nonumber 
                                   && \hspace{0.5cm} \left . -{\scriptstyle \ket{00d0000u}} - {\scriptstyle \ket{00u0000d}} + {\scriptstyle \ket{0d00u000}} - {\scriptstyle \ket{0d0u0000}} + {\scriptstyle \ket{0u00d000}} - {\scriptstyle \ket{0u0d0000}}\right . \nonumber \\ 
                                   && \hspace{0.5cm} \left .  + {\scriptstyle \ket{d0000u00}} - {\scriptstyle \ket{d0u00000}} + {\scriptstyle \ket{u0000d00}} - {\scriptstyle \ket{u0d00000}} \nonumber \right )
\nonumber \\ 
\ket{\Psi_{33-2}} &=& -\frac{1}{4}\left ( {\scriptstyle \ket{00000du0}} + {\scriptstyle \ket{00000ud0}} + {\scriptstyle \ket{0000d00u}} + {\scriptstyle \ket{0000u00d}} - {\scriptstyle \ket{000d000u}} - {\scriptstyle \ket{000u000d}}\right . \nonumber \\                      \nonumber 
                                   && \hspace{0.5cm} \left . -{\scriptstyle \ket{00d000u0}} - {\scriptstyle \ket{00u000d0}} + {\scriptstyle \ket{0d000u00}} - {\scriptstyle \ket{0du00000}} + {\scriptstyle \ket{0u000d00}} - {\scriptstyle \ket{0ud00000}}\right . \nonumber \\ 
                                   && \hspace{0.5cm} \left .  + {\scriptstyle \ket{d000u000}} - {\scriptstyle \ket{d00u0000}} + {\scriptstyle \ket{u000d000}} - {\scriptstyle \ket{u00d0000}} \nonumber \right )
\nonumber \\ 
 {\bf H}_{33}&=&
 \left ( 
\begin{array}{cc}
0&
2 t \\ 
2 t&
\frac{1}{2} (J+8 W)
\end{array} 
 \right ) 
\nonumber 
\nonumber \eeq 
\subsection*{Subspace No. 34 with {\boldmath $N=2$, $s=2$, $m_s=0$, $\Gamma_{4-3}$}}
\beq
\ket{\Psi_{34-1}} &=& \frac{1}{4}\left ( {\scriptstyle \ket{000000du}} + {\scriptstyle \ket{000000ud}} + {\scriptstyle \ket{00000du0}} + {\scriptstyle \ket{00000ud0}} - {\scriptstyle \ket{0000d00u}} + {\scriptstyle \ket{0000du00}}\right . \nonumber \\                      \nonumber 
                                  && \hspace{0.5cm} \left . -{\scriptstyle \ket{0000u00d}} + {\scriptstyle \ket{0000ud00}} + {\scriptstyle \ket{00du0000}} + {\scriptstyle \ket{00ud0000}} + {\scriptstyle \ket{0du00000}} + {\scriptstyle \ket{0ud00000}}\right . \nonumber \\ 
                                  && \hspace{0.5cm} \left . -{\scriptstyle \ket{d00u0000}} + {\scriptstyle \ket{du000000}} - {\scriptstyle \ket{u00d0000}} + {\scriptstyle \ket{ud000000}} \nonumber \right )
\nonumber \\ 
\ket{\Psi_{34-2}} &=& -\frac{1}{4}\left ( {\scriptstyle \ket{000d00u0}} - {\scriptstyle \ket{000du000}} + {\scriptstyle \ket{000u00d0}} - {\scriptstyle \ket{000ud000}} - {\scriptstyle \ket{00d0000u}} + {\scriptstyle \ket{00d00u00}}\right . \nonumber \\                      \nonumber 
                                   && \hspace{0.5cm} \left . -{\scriptstyle \ket{00u0000d}} + {\scriptstyle \ket{00u00d00}} - {\scriptstyle \ket{0d0000u0}} + {\scriptstyle \ket{0d00u000}} - {\scriptstyle \ket{0u0000d0}} + {\scriptstyle \ket{0u00d000}}\right . \nonumber \\ 
                                   && \hspace{0.5cm} \left .  + {\scriptstyle \ket{d000000u}} - {\scriptstyle \ket{d0000u00}} + {\scriptstyle \ket{u000000d}} - {\scriptstyle \ket{u0000d00}} \nonumber \right )
\nonumber \\ 
 {\bf H}_{34}&=&
 \left ( 
\begin{array}{cc}
\frac{1}{2} (J+8 W)&
2 t \\ 
2 t&
0
\end{array} 
 \right ) 
\nonumber 
\nonumber \eeq 
\subsection*{Subspace No. 35 with {\boldmath $N=2$, $s=2$, $m_s=0$, $\Gamma_{5-1}$}}
\beq
\ket{\Psi_{35-1}} &=& \frac{1}{4}\left ( {\scriptstyle \ket{000000du}} + {\scriptstyle \ket{000000ud}} - {\scriptstyle \ket{00000du0}} - {\scriptstyle \ket{00000ud0}} + {\scriptstyle \ket{0000d00u}} + {\scriptstyle \ket{0000du00}}\right . \nonumber \\                        \nonumber 
                                  && \hspace{0.5cm} \left .  + {\scriptstyle \ket{0000u00d}} + {\scriptstyle \ket{0000ud00}} + {\scriptstyle \ket{00du0000}} + {\scriptstyle \ket{00ud0000}} - {\scriptstyle \ket{0du00000}} - {\scriptstyle \ket{0ud00000}}\right . \nonumber \\ 
                                  && \hspace{0.5cm} \left .  + {\scriptstyle \ket{d00u0000}} + {\scriptstyle \ket{du000000}} + {\scriptstyle \ket{u00d0000}} + {\scriptstyle \ket{ud000000}} \nonumber \right )
\nonumber \\ 
\ket{\Psi_{35-2}} &=& -\frac{1}{4}\left ( {\scriptstyle \ket{000d00u0}} + {\scriptstyle \ket{000du000}} + {\scriptstyle \ket{000u00d0}} + {\scriptstyle \ket{000ud000}} - {\scriptstyle \ket{00d0000u}} - {\scriptstyle \ket{00d00u00}}\right . \nonumber \\                      \nonumber 
                                   && \hspace{0.5cm} \left . -{\scriptstyle \ket{00u0000d}} - {\scriptstyle \ket{00u00d00}} + {\scriptstyle \ket{0d0000u0}} + {\scriptstyle \ket{0d00u000}} + {\scriptstyle \ket{0u0000d0}} + {\scriptstyle \ket{0u00d000}}\right . \nonumber \\ 
                                   && \hspace{0.5cm} \left . -{\scriptstyle \ket{d000000u}} - {\scriptstyle \ket{d0000u00}} - {\scriptstyle \ket{u000000d}} - {\scriptstyle \ket{u0000d00}} \nonumber \right )
\nonumber \\ 
 {\bf H}_{35}&=&
 \left ( 
\begin{array}{cc}
\frac{1}{2} (J+8 W)&
2 t \\ 
2 t&
0
\end{array} 
 \right ) 
\nonumber 
\nonumber \eeq 
\subsection*{Subspace No. 36 with {\boldmath $N=2$, $s=2$, $m_s=0$, $\Gamma_{5-2}$}}
\beq
\ket{\Psi_{36-1}} &=& \frac{1}{4}\left ( {\scriptstyle \ket{000000du}} + {\scriptstyle \ket{000000ud}} - {\scriptstyle \ket{0000du00}} - {\scriptstyle \ket{0000ud00}} + {\scriptstyle \ket{000d000u}} + {\scriptstyle \ket{000u000d}}\right . \nonumber \\                      \nonumber 
                                  && \hspace{0.5cm} \left . -{\scriptstyle \ket{00d000u0}} - {\scriptstyle \ket{00du0000}} - {\scriptstyle \ket{00u000d0}} - {\scriptstyle \ket{00ud0000}} - {\scriptstyle \ket{0d000u00}} - {\scriptstyle \ket{0u000d00}}\right . \nonumber \\ 
                                  && \hspace{0.5cm} \left .  + {\scriptstyle \ket{d000u000}} + {\scriptstyle \ket{du000000}} + {\scriptstyle \ket{u000d000}} + {\scriptstyle \ket{ud000000}} \nonumber \right )
\nonumber \\ 
\ket{\Psi_{36-2}} &=& \frac{1}{4}\left ( {\scriptstyle \ket{00000d0u}} + {\scriptstyle \ket{00000u0d}} - {\scriptstyle \ket{0000d0u0}} - {\scriptstyle \ket{0000u0d0}} + {\scriptstyle \ket{000du000}} + {\scriptstyle \ket{000ud000}}\right . \nonumber \\                      \nonumber 
                                  && \hspace{0.5cm} \left . -{\scriptstyle \ket{00d00u00}} - {\scriptstyle \ket{00u00d00}} - {\scriptstyle \ket{0d0000u0}} - {\scriptstyle \ket{0d0u0000}} - {\scriptstyle \ket{0u0000d0}} - {\scriptstyle \ket{0u0d0000}}\right . \nonumber \\ 
                                  && \hspace{0.5cm} \left .  + {\scriptstyle \ket{d000000u}} + {\scriptstyle \ket{d0u00000}} + {\scriptstyle \ket{u000000d}} + {\scriptstyle \ket{u0d00000}} \nonumber \right )
\nonumber \\ 
 {\bf H}_{36}&=&
 \left ( 
\begin{array}{cc}
\frac{1}{2} (J+8 W)&
2 t \\ 
2 t&
0
\end{array} 
 \right ) 
\nonumber 
\nonumber \eeq 
\subsection*{Subspace No. 37 with {\boldmath $N=2$, $s=2$, $m_s=0$, $\Gamma_{5-3}$}}
\beq
\ket{\Psi_{37-1}} &=& -\frac{1}{4}\left ( {\scriptstyle \ket{00000d0u}} + {\scriptstyle \ket{00000u0d}} + {\scriptstyle \ket{0000d0u0}} + {\scriptstyle \ket{0000u0d0}} + {\scriptstyle \ket{000d00u0}} + {\scriptstyle \ket{000u00d0}}\right . \nonumber \\                        \nonumber 
                                   && \hspace{0.5cm} \left .  + {\scriptstyle \ket{00d0000u}} + {\scriptstyle \ket{00u0000d}} - {\scriptstyle \ket{0d00u000}} - {\scriptstyle \ket{0d0u0000}} - {\scriptstyle \ket{0u00d000}} - {\scriptstyle \ket{0u0d0000}}\right . \nonumber \\ 
                                   && \hspace{0.5cm} \left . -{\scriptstyle \ket{d0000u00}} - {\scriptstyle \ket{d0u00000}} - {\scriptstyle \ket{u0000d00}} - {\scriptstyle \ket{u0d00000}} \nonumber \right )
\nonumber \\ 
\ket{\Psi_{37-2}} &=& -\frac{1}{4}\left ( {\scriptstyle \ket{00000du0}} + {\scriptstyle \ket{00000ud0}} + {\scriptstyle \ket{0000d00u}} + {\scriptstyle \ket{0000u00d}} + {\scriptstyle \ket{000d000u}} + {\scriptstyle \ket{000u000d}}\right . \nonumber \\                        \nonumber 
                                   && \hspace{0.5cm} \left .  + {\scriptstyle \ket{00d000u0}} + {\scriptstyle \ket{00u000d0}} - {\scriptstyle \ket{0d000u00}} - {\scriptstyle \ket{0du00000}} - {\scriptstyle \ket{0u000d00}} - {\scriptstyle \ket{0ud00000}}\right . \nonumber \\ 
                                   && \hspace{0.5cm} \left . -{\scriptstyle \ket{d000u000}} - {\scriptstyle \ket{d00u0000}} - {\scriptstyle \ket{u000d000}} - {\scriptstyle \ket{u00d0000}} \nonumber \right )
\nonumber \\ 
 {\bf H}_{37}&=&
 \left ( 
\begin{array}{cc}
0&
2 t \\ 
2 t&
\frac{1}{2} (J+8 W)
\end{array} 
 \right ) 
\nonumber 
\nonumber \eeq 
\subsection*{Subspace No. 38 with {\boldmath $N=2$, $s=2$, $m_s=0$, $\Gamma_{7}$}}
\beq
\ket{\Psi_{38-1}} &=& -\frac{1}{2 \sqrt{6}}\left ( {\scriptstyle \ket{000000du}} + {\scriptstyle \ket{000000ud}} - {\scriptstyle \ket{00000du0}} - {\scriptstyle \ket{00000ud0}} + {\scriptstyle \ket{0000d00u}} + {\scriptstyle \ket{0000du00}}\right . \nonumber \\                        \nonumber 
                                            && \hspace{0.5cm} \left .  + {\scriptstyle \ket{0000u00d}} + {\scriptstyle \ket{0000ud00}} + {\scriptstyle \ket{000d000u}} + {\scriptstyle \ket{000u000d}} - {\scriptstyle \ket{00d000u0}} - {\scriptstyle \ket{00du0000}}\right . \nonumber \\ 
                                            && \hspace{0.5cm} \left . -{\scriptstyle \ket{00u000d0}} - {\scriptstyle \ket{00ud0000}} + {\scriptstyle \ket{0d000u00}} + {\scriptstyle \ket{0du00000}} + {\scriptstyle \ket{0u000d00}} + {\scriptstyle \ket{0ud00000}}\right . \nonumber \\ 
                                            && \hspace{0.5cm} \left . -{\scriptstyle \ket{d000u000}} - {\scriptstyle \ket{d00u0000}} - {\scriptstyle \ket{du000000}} - {\scriptstyle \ket{u000d000}} - {\scriptstyle \ket{u00d0000}} - {\scriptstyle \ket{ud000000}} \nonumber \right )
\nonumber \\ 
\ket{\Psi_{38-2}} &=& -\frac{1}{2 \sqrt{2}}\left ( {\scriptstyle \ket{000d0u00}} + {\scriptstyle \ket{000u0d00}} - {\scriptstyle \ket{00d0u000}} - {\scriptstyle \ket{00u0d000}}\right . \nonumber \\                     \nonumber 
                                            && \hspace{0.5cm} \left .  + {\scriptstyle \ket{0d00000u}} + {\scriptstyle \ket{0u00000d}} - {\scriptstyle \ket{d00000u0}} - {\scriptstyle \ket{u00000d0}} \nonumber \right )
\nonumber \\ 
 {\bf H}_{38}&=&
 \left ( 
\begin{array}{cc}
\frac{1}{2} (J+8 W)&
0 \\ 
0&
0
\end{array} 
 \right ) 
\nonumber 
\nonumber \eeq 
\subsection*{Subspace No. 39 with {\boldmath $N=2$, $s=2$, $m_s=0$, $\Gamma_{8-1}$}}
\beq
\ket{\Psi_{39-1}} &=& -\frac{1}{4 \sqrt{3}}\left ( {\scriptstyle \ket{000000du}} + {\scriptstyle \ket{000000ud}} - {\scriptstyle \ket{00000du0}} - {\scriptstyle \ket{00000ud0}} + {\scriptstyle \ket{0000d00u}} + {\scriptstyle \ket{0000du00}}\right . \nonumber \\                        \nonumber 
                                            && \hspace{0.5cm} \left .  + {\scriptstyle \ket{0000u00d}} + {\scriptstyle \ket{0000ud00}} - {\scriptstyle \ket{00du0000}} - {\scriptstyle \ket{00ud0000}} + {\scriptstyle \ket{0du00000}} + {\scriptstyle \ket{0ud00000}}\right . \nonumber \\ 
                                            && \hspace{0.5cm} \left . -{\scriptstyle \ket{d00u0000}} - {\scriptstyle \ket{du000000}} - {\scriptstyle \ket{u00d0000}} - {\scriptstyle \ket{ud000000}} \nonumber \right )
\\ 
 && +\frac{1}{2 \sqrt{3}}\left ( {\scriptstyle \ket{000d000u}} + {\scriptstyle \ket{000u000d}} - {\scriptstyle \ket{00d000u0}} - {\scriptstyle \ket{00u000d0}}\right . \nonumber \\ 
                          && \hspace{0.5cm} \left .  + {\scriptstyle \ket{0d000u00}} + {\scriptstyle \ket{0u000d00}} - {\scriptstyle \ket{d000u000}} - {\scriptstyle \ket{u000d000}} \nonumber \right )
\nonumber \\ 
 {\bf H}_{39}&=&
 \left ( 
\begin{array}{c}
\frac{1}{2} (J+8 W)
\end{array} 
 \right ) 
\nonumber 
\nonumber \eeq 
\subsection*{Subspace No. 40 with {\boldmath $N=2$, $s=2$, $m_s=0$, $\Gamma_{8-2}$}}
\beq
\ket{\Psi_{40-1}} &=& -\frac{1}{4}\left ( {\scriptstyle \ket{000000du}} + {\scriptstyle \ket{000000ud}} + {\scriptstyle \ket{00000du0}} + {\scriptstyle \ket{00000ud0}} - {\scriptstyle \ket{0000d00u}} + {\scriptstyle \ket{0000du00}}\right . \nonumber \\                      \nonumber 
                                   && \hspace{0.5cm} \left . -{\scriptstyle \ket{0000u00d}} + {\scriptstyle \ket{0000ud00}} - {\scriptstyle \ket{00du0000}} - {\scriptstyle \ket{00ud0000}} - {\scriptstyle \ket{0du00000}} - {\scriptstyle \ket{0ud00000}}\right . \nonumber \\ 
                                   && \hspace{0.5cm} \left .  + {\scriptstyle \ket{d00u0000}} - {\scriptstyle \ket{du000000}} + {\scriptstyle \ket{u00d0000}} - {\scriptstyle \ket{ud000000}} \nonumber \right )
\nonumber \\ 
 {\bf H}_{40}&=&
 \left ( 
\begin{array}{c}
\frac{1}{2} (J+8 W)
\end{array} 
 \right ) 
\nonumber 
\nonumber \eeq 
\subsection*{Subspace No. 41 with {\boldmath $N=2$, $s=2$, $m_s=0$, $\Gamma_{9-1}$}}
\beq
\ket{\Psi_{41-1}} &=& \frac{1}{4}\left ( {\scriptstyle \ket{00000d0u}} + {\scriptstyle \ket{00000u0d}} + {\scriptstyle \ket{0000d0u0}} + {\scriptstyle \ket{0000u0d0}} - {\scriptstyle \ket{000du000}} - {\scriptstyle \ket{000ud000}}\right . \nonumber \\                      \nonumber 
                                  && \hspace{0.5cm} \left . -{\scriptstyle \ket{00d00u00}} - {\scriptstyle \ket{00u00d00}} + {\scriptstyle \ket{0d0000u0}} + {\scriptstyle \ket{0d0u0000}} + {\scriptstyle \ket{0u0000d0}} + {\scriptstyle \ket{0u0d0000}}\right . \nonumber \\ 
                                  && \hspace{0.5cm} \left .  + {\scriptstyle \ket{d000000u}} + {\scriptstyle \ket{d0u00000}} + {\scriptstyle \ket{u000000d}} + {\scriptstyle \ket{u0d00000}} \nonumber \right )
\nonumber \\ 
\ket{\Psi_{41-2}} &=& \frac{1}{2 \sqrt{2}}\left ( {\scriptstyle \ket{00000du0}} + {\scriptstyle \ket{00000ud0}} + {\scriptstyle \ket{0000d00u}} + {\scriptstyle \ket{0000u00d}}\right . \nonumber \\                     \nonumber 
                                           && \hspace{0.5cm} \left .  + {\scriptstyle \ket{0du00000}} + {\scriptstyle \ket{0ud00000}} + {\scriptstyle \ket{d00u0000}} + {\scriptstyle \ket{u00d0000}} \nonumber \right )
\nonumber \\ 
\ket{\Psi_{41-3}} &=& -\frac{1}{2 \sqrt{2}}\left ( {\scriptstyle \ket{000d0u00}} + {\scriptstyle \ket{000u0d00}} + {\scriptstyle \ket{00d0u000}} + {\scriptstyle \ket{00u0d000}}\right . \nonumber \\                   \nonumber 
                                            && \hspace{0.5cm} \left . -{\scriptstyle \ket{0d00000u}} - {\scriptstyle \ket{0u00000d}} - {\scriptstyle \ket{d00000u0}} - {\scriptstyle \ket{u00000d0}} \nonumber \right )
\nonumber \\ 
 {\bf H}_{41}&=&
 \left ( 
\begin{array}{ccc}
0&
2 \sqrt{2} t&
2 \sqrt{2} t \\ 
2 \sqrt{2} t&
\frac{1}{2} (J+8 W)&
0 \\ 
2 \sqrt{2} t&
0&
0
\end{array} 
 \right ) 
\nonumber 
\nonumber \eeq 
\subsection*{Subspace No. 42 with {\boldmath $N=2$, $s=2$, $m_s=0$, $\Gamma_{9-1}$}}
\beq
\ket{\Psi_{42-1}} &=& -\frac{1}{2 \sqrt{2}}\left ( {\scriptstyle \ket{000000du}} + {\scriptstyle \ket{000000ud}} - {\scriptstyle \ket{0000du00}} - {\scriptstyle \ket{0000ud00}}\right . \nonumber \\                     \nonumber 
                                            && \hspace{0.5cm} \left .  + {\scriptstyle \ket{00du0000}} + {\scriptstyle \ket{00ud0000}} - {\scriptstyle \ket{du000000}} - {\scriptstyle \ket{ud000000}} \nonumber \right )
\nonumber \\ 
\ket{\Psi_{42-2}} &=& -\frac{1}{4}\left ( {\scriptstyle \ket{00000d0u}} + {\scriptstyle \ket{00000u0d}} - {\scriptstyle \ket{0000d0u0}} - {\scriptstyle \ket{0000u0d0}} - {\scriptstyle \ket{000d00u0}} - {\scriptstyle \ket{000u00d0}}\right . \nonumber \\                        \nonumber 
                                   && \hspace{0.5cm} \left .  + {\scriptstyle \ket{00d0000u}} + {\scriptstyle \ket{00u0000d}} + {\scriptstyle \ket{0d00u000}} + {\scriptstyle \ket{0d0u0000}} + {\scriptstyle \ket{0u00d000}} + {\scriptstyle \ket{0u0d0000}}\right . \nonumber \\ 
                                   && \hspace{0.5cm} \left . -{\scriptstyle \ket{d0000u00}} - {\scriptstyle \ket{d0u00000}} - {\scriptstyle \ket{u0000d00}} - {\scriptstyle \ket{u0d00000}} \nonumber \right )
\nonumber \\ 
\ket{\Psi_{42-3}} &=& \frac{1}{2 \sqrt{2}}\left ( {\scriptstyle \ket{000d0u00}} + {\scriptstyle \ket{000u0d00}} - {\scriptstyle \ket{00d0u000}} - {\scriptstyle \ket{00u0d000}}\right . \nonumber \\                   \nonumber 
                                           && \hspace{0.5cm} \left . -{\scriptstyle \ket{0d00000u}} - {\scriptstyle \ket{0u00000d}} + {\scriptstyle \ket{d00000u0}} + {\scriptstyle \ket{u00000d0}} \nonumber \right )
\nonumber \\ 
 {\bf H}_{42}&=&
 \left ( 
\begin{array}{ccc}
\frac{1}{2} (J+8 W)&
2 \sqrt{2} t&
0 \\ 
2 \sqrt{2} t&
0&
2 \sqrt{2} t \\ 
0&
2 \sqrt{2} t&
0
\end{array} 
 \right ) 
\nonumber 
\nonumber \eeq 
\subsection*{Subspace No. 43 with {\boldmath $N=2$, $s=2$, $m_s=0$, $\Gamma_{9-3}$}}
\beq
\ket{\Psi_{43-1}} &=& \frac{1}{2 \sqrt{2}}\left ( {\scriptstyle \ket{000d000u}} + {\scriptstyle \ket{000u000d}} + {\scriptstyle \ket{00d000u0}} + {\scriptstyle \ket{00u000d0}}\right . \nonumber \\                     \nonumber 
                                           && \hspace{0.5cm} \left .  + {\scriptstyle \ket{0d000u00}} + {\scriptstyle \ket{0u000d00}} + {\scriptstyle \ket{d000u000}} + {\scriptstyle \ket{u000d000}} \nonumber \right )
\nonumber \\ 
\ket{\Psi_{43-2}} &=& \frac{1}{4}\left ( {\scriptstyle \ket{000d00u0}} + {\scriptstyle \ket{000du000}} + {\scriptstyle \ket{000u00d0}} + {\scriptstyle \ket{000ud000}} + {\scriptstyle \ket{00d0000u}} + {\scriptstyle \ket{00d00u00}}\right . \nonumber \\                        \nonumber 
                                  && \hspace{0.5cm} \left .  + {\scriptstyle \ket{00u0000d}} + {\scriptstyle \ket{00u00d00}} + {\scriptstyle \ket{0d0000u0}} + {\scriptstyle \ket{0d00u000}} + {\scriptstyle \ket{0u0000d0}} + {\scriptstyle \ket{0u00d000}}\right . \nonumber \\ 
                                  && \hspace{0.5cm} \left .  + {\scriptstyle \ket{d000000u}} + {\scriptstyle \ket{d0000u00}} + {\scriptstyle \ket{u000000d}} + {\scriptstyle \ket{u0000d00}} \nonumber \right )
\nonumber \\ 
\ket{\Psi_{43-3}} &=& \frac{1}{2 \sqrt{2}}\left ( {\scriptstyle \ket{000d0u00}} + {\scriptstyle \ket{000u0d00}} + {\scriptstyle \ket{00d0u000}} + {\scriptstyle \ket{00u0d000}}\right . \nonumber \\                     \nonumber 
                                           && \hspace{0.5cm} \left .  + {\scriptstyle \ket{0d00000u}} + {\scriptstyle \ket{0u00000d}} + {\scriptstyle \ket{d00000u0}} + {\scriptstyle \ket{u00000d0}} \nonumber \right )
\nonumber \\ 
 {\bf H}_{43}&=&
 \left ( 
\begin{array}{ccc}
\frac{1}{2} (J+8 W)&
2 \sqrt{2} t&
0 \\ 
2 \sqrt{2} t&
0&
2 \sqrt{2} t \\ 
0&
2 \sqrt{2} t&
0
\end{array} 
 \right ) 
\nonumber 
\nonumber \eeq 
\subsection*{Subspace No. 44 with {\boldmath $N=2$, $s=2$, $m_s=0$, $\Gamma_{10-1}$}}
\beq
\ket{\Psi_{44-1}} &=& \frac{1}{4}\left ( {\scriptstyle \ket{000d00u0}} - {\scriptstyle \ket{000du000}} + {\scriptstyle \ket{000u00d0}} - {\scriptstyle \ket{000ud000}} + {\scriptstyle \ket{00d0000u}} - {\scriptstyle \ket{00d00u00}}\right . \nonumber \\                        \nonumber 
                                  && \hspace{0.5cm} \left .  + {\scriptstyle \ket{00u0000d}} - {\scriptstyle \ket{00u00d00}} - {\scriptstyle \ket{0d0000u0}} + {\scriptstyle \ket{0d00u000}} - {\scriptstyle \ket{0u0000d0}} + {\scriptstyle \ket{0u00d000}}\right . \nonumber \\ 
                                  && \hspace{0.5cm} \left . -{\scriptstyle \ket{d000000u}} + {\scriptstyle \ket{d0000u00}} - {\scriptstyle \ket{u000000d}} + {\scriptstyle \ket{u0000d00}} \nonumber \right )
\nonumber \\ 
 {\bf H}_{44}&=&
 \left ( 
\begin{array}{c}
0
\end{array} 
 \right ) 
\nonumber 
\nonumber \eeq 
\subsection*{Subspace No. 45 with {\boldmath $N=2$, $s=2$, $m_s=0$, $\Gamma_{10-2}$}}
\beq
\ket{\Psi_{45-1}} &=& \frac{1}{4}\left ( {\scriptstyle \ket{00000d0u}} + {\scriptstyle \ket{00000u0d}} + {\scriptstyle \ket{0000d0u0}} + {\scriptstyle \ket{0000u0d0}} + {\scriptstyle \ket{000du000}} + {\scriptstyle \ket{000ud000}}\right . \nonumber \\                        \nonumber 
                                  && \hspace{0.5cm} \left .  + {\scriptstyle \ket{00d00u00}} + {\scriptstyle \ket{00u00d00}} - {\scriptstyle \ket{0d0000u0}} + {\scriptstyle \ket{0d0u0000}} - {\scriptstyle \ket{0u0000d0}} + {\scriptstyle \ket{0u0d0000}}\right . \nonumber \\ 
                                  && \hspace{0.5cm} \left . -{\scriptstyle \ket{d000000u}} + {\scriptstyle \ket{d0u00000}} - {\scriptstyle \ket{u000000d}} + {\scriptstyle \ket{u0d00000}} \nonumber \right )
\nonumber \\ 
 {\bf H}_{45}&=&
 \left ( 
\begin{array}{c}
0
\end{array} 
 \right ) 
\nonumber 
\nonumber \eeq 
\subsection*{Subspace No. 46 with {\boldmath $N=2$, $s=2$, $m_s=0$, $\Gamma_{10-3}$}}
\beq
\ket{\Psi_{46-1}} &=& -\frac{1}{4}\left ( {\scriptstyle \ket{00000d0u}} + {\scriptstyle \ket{00000u0d}} - {\scriptstyle \ket{0000d0u0}} - {\scriptstyle \ket{0000u0d0}} + {\scriptstyle \ket{000d00u0}} + {\scriptstyle \ket{000u00d0}}\right . \nonumber \\                      \nonumber 
                                   && \hspace{0.5cm} \left . -{\scriptstyle \ket{00d0000u}} - {\scriptstyle \ket{00u0000d}} - {\scriptstyle \ket{0d00u000}} + {\scriptstyle \ket{0d0u0000}} - {\scriptstyle \ket{0u00d000}} + {\scriptstyle \ket{0u0d0000}}\right . \nonumber \\ 
                                   && \hspace{0.5cm} \left .  + {\scriptstyle \ket{d0000u00}} - {\scriptstyle \ket{d0u00000}} + {\scriptstyle \ket{u0000d00}} - {\scriptstyle \ket{u0d00000}} \nonumber \right )
\nonumber \\ 
 {\bf H}_{46}&=&
 \left ( 
\begin{array}{c}
0
\end{array} 
 \right ) 
\nonumber 
\nonumber \eeq 
\subsection*{Subspace No. 47 with {\boldmath $N=2$, $s=2$, $m_s=1$, $\Gamma_{4-1}$}}
\beq
\ket{\Psi_{47-1}} &=& \frac{1}{2 \sqrt{2}}\left ( {\scriptstyle \ket{000000uu}} - {\scriptstyle \ket{0000uu00}} - {\scriptstyle \ket{000u000u}} + {\scriptstyle \ket{00u000u0}}\right . \nonumber \\                   \nonumber 
                                           && \hspace{0.5cm} \left . -{\scriptstyle \ket{00uu0000}} + {\scriptstyle \ket{0u000u00}} - {\scriptstyle \ket{u000u000}} + {\scriptstyle \ket{uu000000}} \nonumber \right )
\nonumber \\ 
\ket{\Psi_{47-2}} &=& \frac{1}{2 \sqrt{2}}\left ( {\scriptstyle \ket{00000u0u}} - {\scriptstyle \ket{0000u0u0}} - {\scriptstyle \ket{000uu000}} + {\scriptstyle \ket{00u00u00}}\right . \nonumber \\                     \nonumber 
                                           && \hspace{0.5cm} \left .  + {\scriptstyle \ket{0u0000u0}} - {\scriptstyle \ket{0u0u0000}} - {\scriptstyle \ket{u000000u}} + {\scriptstyle \ket{u0u00000}} \nonumber \right )
\nonumber \\ 
 {\bf H}_{47}&=&
 \left ( 
\begin{array}{cc}
\frac{1}{2} (J+8 W)&
2 t \\ 
2 t&
0
\end{array} 
 \right ) 
\nonumber 
\nonumber \eeq 
\subsection*{Subspace No. 48 with {\boldmath $N=2$, $s=2$, $m_s=1$, $\Gamma_{4-2}$}}
\beq
\ket{\Psi_{48-1}} &=& -\frac{1}{2 \sqrt{2}}\left ( {\scriptstyle \ket{00000u0u}} + {\scriptstyle \ket{0000u0u0}} - {\scriptstyle \ket{000u00u0}} - {\scriptstyle \ket{00u0000u}}\right . \nonumber \\                     \nonumber 
                                            && \hspace{0.5cm} \left .  + {\scriptstyle \ket{0u00u000}} - {\scriptstyle \ket{0u0u0000}} + {\scriptstyle \ket{u0000u00}} - {\scriptstyle \ket{u0u00000}} \nonumber \right )
\nonumber \\ 
\ket{\Psi_{48-2}} &=& -\frac{1}{2 \sqrt{2}}\left ( {\scriptstyle \ket{00000uu0}} + {\scriptstyle \ket{0000u00u}} - {\scriptstyle \ket{000u000u}} - {\scriptstyle \ket{00u000u0}}\right . \nonumber \\                     \nonumber 
                                            && \hspace{0.5cm} \left .  + {\scriptstyle \ket{0u000u00}} - {\scriptstyle \ket{0uu00000}} + {\scriptstyle \ket{u000u000}} - {\scriptstyle \ket{u00u0000}} \nonumber \right )
\nonumber \\ 
 {\bf H}_{48}&=&
 \left ( 
\begin{array}{cc}
0&
2 t \\ 
2 t&
\frac{1}{2} (J+8 W)
\end{array} 
 \right ) 
\nonumber 
\nonumber \eeq 
\subsection*{Subspace No. 49 with {\boldmath $N=2$, $s=2$, $m_s=1$, $\Gamma_{4-3}$}}
\beq
\ket{\Psi_{49-1}} &=& \frac{1}{2 \sqrt{2}}\left ( {\scriptstyle \ket{000000uu}} + {\scriptstyle \ket{00000uu0}} - {\scriptstyle \ket{0000u00u}} + {\scriptstyle \ket{0000uu00}}\right . \nonumber \\                     \nonumber 
                                           && \hspace{0.5cm} \left .  + {\scriptstyle \ket{00uu0000}} + {\scriptstyle \ket{0uu00000}} - {\scriptstyle \ket{u00u0000}} + {\scriptstyle \ket{uu000000}} \nonumber \right )
\nonumber \\ 
\ket{\Psi_{49-2}} &=& -\frac{1}{2 \sqrt{2}}\left ( {\scriptstyle \ket{000u00u0}} - {\scriptstyle \ket{000uu000}} - {\scriptstyle \ket{00u0000u}} + {\scriptstyle \ket{00u00u00}}\right . \nonumber \\                   \nonumber 
                                            && \hspace{0.5cm} \left . -{\scriptstyle \ket{0u0000u0}} + {\scriptstyle \ket{0u00u000}} + {\scriptstyle \ket{u000000u}} - {\scriptstyle \ket{u0000u00}} \nonumber \right )
\nonumber \\ 
 {\bf H}_{49}&=&
 \left ( 
\begin{array}{cc}
\frac{1}{2} (J+8 W)&
2 t \\ 
2 t&
0
\end{array} 
 \right ) 
\nonumber 
\nonumber \eeq 
\subsection*{Subspace No. 50 with {\boldmath $N=2$, $s=2$, $m_s=1$, $\Gamma_{5-1}$}}
\beq
\ket{\Psi_{50-1}} &=& \frac{1}{2 \sqrt{2}}\left ( {\scriptstyle \ket{000000uu}} - {\scriptstyle \ket{00000uu0}} + {\scriptstyle \ket{0000u00u}} + {\scriptstyle \ket{0000uu00}}\right . \nonumber \\                     \nonumber 
                                           && \hspace{0.5cm} \left .  + {\scriptstyle \ket{00uu0000}} - {\scriptstyle \ket{0uu00000}} + {\scriptstyle \ket{u00u0000}} + {\scriptstyle \ket{uu000000}} \nonumber \right )
\nonumber \\ 
\ket{\Psi_{50-2}} &=& -\frac{1}{2 \sqrt{2}}\left ( {\scriptstyle \ket{000u00u0}} + {\scriptstyle \ket{000uu000}} - {\scriptstyle \ket{00u0000u}} - {\scriptstyle \ket{00u00u00}}\right . \nonumber \\                     \nonumber 
                                            && \hspace{0.5cm} \left .  + {\scriptstyle \ket{0u0000u0}} + {\scriptstyle \ket{0u00u000}} - {\scriptstyle \ket{u000000u}} - {\scriptstyle \ket{u0000u00}} \nonumber \right )
\nonumber \\ 
 {\bf H}_{50}&=&
 \left ( 
\begin{array}{cc}
\frac{1}{2} (J+8 W)&
2 t \\ 
2 t&
0
\end{array} 
 \right ) 
\nonumber 
\nonumber \eeq 
\subsection*{Subspace No. 51 with {\boldmath $N=2$, $s=2$, $m_s=1$, $\Gamma_{5-2}$}}
\beq
\ket{\Psi_{51-1}} &=& \frac{1}{2 \sqrt{2}}\left ( {\scriptstyle \ket{000000uu}} - {\scriptstyle \ket{0000uu00}} + {\scriptstyle \ket{000u000u}} - {\scriptstyle \ket{00u000u0}}\right . \nonumber \\                   \nonumber 
                                           && \hspace{0.5cm} \left . -{\scriptstyle \ket{00uu0000}} - {\scriptstyle \ket{0u000u00}} + {\scriptstyle \ket{u000u000}} + {\scriptstyle \ket{uu000000}} \nonumber \right )
\nonumber \\ 
\ket{\Psi_{51-2}} &=& \frac{1}{2 \sqrt{2}}\left ( {\scriptstyle \ket{00000u0u}} - {\scriptstyle \ket{0000u0u0}} + {\scriptstyle \ket{000uu000}} - {\scriptstyle \ket{00u00u00}}\right . \nonumber \\                   \nonumber 
                                           && \hspace{0.5cm} \left . -{\scriptstyle \ket{0u0000u0}} - {\scriptstyle \ket{0u0u0000}} + {\scriptstyle \ket{u000000u}} + {\scriptstyle \ket{u0u00000}} \nonumber \right )
\nonumber \\ 
 {\bf H}_{51}&=&
 \left ( 
\begin{array}{cc}
\frac{1}{2} (J+8 W)&
2 t \\ 
2 t&
0
\end{array} 
 \right ) 
\nonumber 
\nonumber \eeq 
\subsection*{Subspace No. 52 with {\boldmath $N=2$, $s=2$, $m_s=1$, $\Gamma_{5-3}$}}
\beq
\ket{\Psi_{52-1}} &=& -\frac{1}{2 \sqrt{2}}\left ( {\scriptstyle \ket{00000u0u}} + {\scriptstyle \ket{0000u0u0}} + {\scriptstyle \ket{000u00u0}} + {\scriptstyle \ket{00u0000u}}\right . \nonumber \\                   \nonumber 
                                            && \hspace{0.5cm} \left . -{\scriptstyle \ket{0u00u000}} - {\scriptstyle \ket{0u0u0000}} - {\scriptstyle \ket{u0000u00}} - {\scriptstyle \ket{u0u00000}} \nonumber \right )
\nonumber \\ 
\ket{\Psi_{52-2}} &=& -\frac{1}{2 \sqrt{2}}\left ( {\scriptstyle \ket{00000uu0}} + {\scriptstyle \ket{0000u00u}} + {\scriptstyle \ket{000u000u}} + {\scriptstyle \ket{00u000u0}}\right . \nonumber \\                   \nonumber 
                                            && \hspace{0.5cm} \left . -{\scriptstyle \ket{0u000u00}} - {\scriptstyle \ket{0uu00000}} - {\scriptstyle \ket{u000u000}} - {\scriptstyle \ket{u00u0000}} \nonumber \right )
\nonumber \\ 
 {\bf H}_{52}&=&
 \left ( 
\begin{array}{cc}
0&
2 t \\ 
2 t&
\frac{1}{2} (J+8 W)
\end{array} 
 \right ) 
\nonumber 
\nonumber \eeq 
\subsection*{Subspace No. 53 with {\boldmath $N=2$, $s=2$, $m_s=1$, $\Gamma_{7}$}}
\beq
\ket{\Psi_{53-1}} &=& -\frac{1}{2 \sqrt{3}}\left ( {\scriptstyle \ket{000000uu}} - {\scriptstyle \ket{00000uu0}} + {\scriptstyle \ket{0000u00u}} + {\scriptstyle \ket{0000uu00}} + {\scriptstyle \ket{000u000u}} - {\scriptstyle \ket{00u000u0}}\right . \nonumber \\                   \nonumber 
                                            && \hspace{0.5cm} \left . -{\scriptstyle \ket{00uu0000}} + {\scriptstyle \ket{0u000u00}} + {\scriptstyle \ket{0uu00000}} - {\scriptstyle \ket{u000u000}} - {\scriptstyle \ket{u00u0000}} - {\scriptstyle \ket{uu000000}} \nonumber \right )
\nonumber \\ 
\ket{\Psi_{53-2}} &=& -\frac{1}{2}\left ( {\scriptstyle \ket{000u0u00}} - {\scriptstyle \ket{00u0u000}} + {\scriptstyle \ket{0u00000u}} - {\scriptstyle \ket{u00000u0}} \nonumber \right ) \nonumber 
\nonumber \\ 
 {\bf H}_{53}&=&
 \left ( 
\begin{array}{cc}
\frac{1}{2} (J+8 W)&
0 \\ 
0&
0
\end{array} 
 \right ) 
\nonumber 
\nonumber \eeq 
\subsection*{Subspace No. 54 with {\boldmath $N=2$, $s=2$, $m_s=1$, $\Gamma_{8-1}$}}
\beq
\ket{\Psi_{54-1}} &=& -\frac{1}{2 \sqrt{6}}\left ( {\scriptstyle \ket{000000uu}} - {\scriptstyle \ket{00000uu0}} + {\scriptstyle \ket{0000u00u}} + {\scriptstyle \ket{0000uu00}}\right . \nonumber \\                   \nonumber 
                                            && \hspace{0.5cm} \left . -{\scriptstyle \ket{00uu0000}} + {\scriptstyle \ket{0uu00000}} - {\scriptstyle \ket{u00u0000}} - {\scriptstyle \ket{uu000000}} \nonumber \right )
\\ 
 && +\frac{1}{\sqrt{6}}\left ( {\scriptstyle \ket{000u000u}} - {\scriptstyle \ket{00u000u0}} + {\scriptstyle \ket{0u000u00}} - {\scriptstyle \ket{u000u000}} \nonumber \right )
\nonumber \\ 
 {\bf H}_{54}&=&
 \left ( 
\begin{array}{c}
\frac{1}{2} (J+8 W)
\end{array} 
 \right ) 
\nonumber 
\nonumber \eeq 
\subsection*{Subspace No. 55 with {\boldmath $N=2$, $s=2$, $m_s=1$, $\Gamma_{8-2}$}}
\beq
\ket{\Psi_{55-1}} &=& -\frac{1}{2 \sqrt{2}}\left ( {\scriptstyle \ket{000000uu}} + {\scriptstyle \ket{00000uu0}} - {\scriptstyle \ket{0000u00u}} + {\scriptstyle \ket{0000uu00}}\right . \nonumber \\                   \nonumber 
                                            && \hspace{0.5cm} \left . -{\scriptstyle \ket{00uu0000}} - {\scriptstyle \ket{0uu00000}} + {\scriptstyle \ket{u00u0000}} - {\scriptstyle \ket{uu000000}} \nonumber \right )
\nonumber \\ 
 {\bf H}_{55}&=&
 \left ( 
\begin{array}{c}
\frac{1}{2} (J+8 W)
\end{array} 
 \right ) 
\nonumber 
\nonumber \eeq 
\subsection*{Subspace No. 56 with {\boldmath $N=2$, $s=2$, $m_s=1$, $\Gamma_{9-1}$}}
\beq
\ket{\Psi_{56-1}} &=& \frac{1}{2 \sqrt{2}}\left ( {\scriptstyle \ket{00000u0u}} + {\scriptstyle \ket{0000u0u0}} - {\scriptstyle \ket{000uu000}} - {\scriptstyle \ket{00u00u00}}\right . \nonumber \\                     \nonumber 
                                           && \hspace{0.5cm} \left .  + {\scriptstyle \ket{0u0000u0}} + {\scriptstyle \ket{0u0u0000}} + {\scriptstyle \ket{u000000u}} + {\scriptstyle \ket{u0u00000}} \nonumber \right )
\nonumber \\ 
\ket{\Psi_{56-2}} &=& \frac{1}{2}\left ( {\scriptstyle \ket{00000uu0}} + {\scriptstyle \ket{0000u00u}} + {\scriptstyle \ket{0uu00000}} + {\scriptstyle \ket{u00u0000}} \nonumber \right ) \nonumber 
\nonumber \\ 
\ket{\Psi_{56-3}} &=& -\frac{1}{2}\left ( {\scriptstyle \ket{000u0u00}} + {\scriptstyle \ket{00u0u000}} - {\scriptstyle \ket{0u00000u}} - {\scriptstyle \ket{u00000u0}} \nonumber \right ) \nonumber 
\nonumber \\ 
 {\bf H}_{56}&=&
 \left ( 
\begin{array}{ccc}
0&
2 \sqrt{2} t&
2 \sqrt{2} t \\ 
2 \sqrt{2} t&
\frac{1}{2} (J+8 W)&
0 \\ 
2 \sqrt{2} t&
0&
0
\end{array} 
 \right ) 
\nonumber 
\nonumber \eeq 
\subsection*{Subspace No. 57 with {\boldmath $N=2$, $s=2$, $m_s=1$, $\Gamma_{9-1}$}}
\beq
\ket{\Psi_{57-1}} &=& -\frac{1}{2}\left ( {\scriptstyle \ket{000000uu}} - {\scriptstyle \ket{0000uu00}} + {\scriptstyle \ket{00uu0000}} - {\scriptstyle \ket{uu000000}} \nonumber \right ) \nonumber 
\nonumber \\ 
\ket{\Psi_{57-2}} &=& -\frac{1}{2 \sqrt{2}}\left ( {\scriptstyle \ket{00000u0u}} - {\scriptstyle \ket{0000u0u0}} - {\scriptstyle \ket{000u00u0}} + {\scriptstyle \ket{00u0000u}}\right . \nonumber \\                     \nonumber 
                                            && \hspace{0.5cm} \left .  + {\scriptstyle \ket{0u00u000}} + {\scriptstyle \ket{0u0u0000}} - {\scriptstyle \ket{u0000u00}} - {\scriptstyle \ket{u0u00000}} \nonumber \right )
\nonumber \\ 
\ket{\Psi_{57-3}} &=& \frac{1}{2}\left ( {\scriptstyle \ket{000u0u00}} - {\scriptstyle \ket{00u0u000}} - {\scriptstyle \ket{0u00000u}} + {\scriptstyle \ket{u00000u0}} \nonumber \right ) \nonumber 
\nonumber \\ 
 {\bf H}_{57}&=&
 \left ( 
\begin{array}{ccc}
\frac{1}{2} (J+8 W)&
2 \sqrt{2} t&
0 \\ 
2 \sqrt{2} t&
0&
2 \sqrt{2} t \\ 
0&
2 \sqrt{2} t&
0
\end{array} 
 \right ) 
\nonumber 
\nonumber \eeq 
\subsection*{Subspace No. 58 with {\boldmath $N=2$, $s=2$, $m_s=1$, $\Gamma_{9-3}$}}
\beq
\ket{\Psi_{58-1}} &=& \frac{1}{2}\left ( {\scriptstyle \ket{000u000u}} + {\scriptstyle \ket{00u000u0}} + {\scriptstyle \ket{0u000u00}} + {\scriptstyle \ket{u000u000}} \nonumber \right ) \nonumber 
\nonumber \\ 
\ket{\Psi_{58-2}} &=& \frac{1}{2 \sqrt{2}}\left ( {\scriptstyle \ket{000u00u0}} + {\scriptstyle \ket{000uu000}} + {\scriptstyle \ket{00u0000u}} + {\scriptstyle \ket{00u00u00}}\right . \nonumber \\                     \nonumber 
                                           && \hspace{0.5cm} \left .  + {\scriptstyle \ket{0u0000u0}} + {\scriptstyle \ket{0u00u000}} + {\scriptstyle \ket{u000000u}} + {\scriptstyle \ket{u0000u00}} \nonumber \right )
\nonumber \\ 
\ket{\Psi_{58-3}} &=& \frac{1}{2}\left ( {\scriptstyle \ket{000u0u00}} + {\scriptstyle \ket{00u0u000}} + {\scriptstyle \ket{0u00000u}} + {\scriptstyle \ket{u00000u0}} \nonumber \right ) \nonumber 
\nonumber \\ 
 {\bf H}_{58}&=&
 \left ( 
\begin{array}{ccc}
\frac{1}{2} (J+8 W)&
2 \sqrt{2} t&
0 \\ 
2 \sqrt{2} t&
0&
2 \sqrt{2} t \\ 
0&
2 \sqrt{2} t&
0
\end{array} 
 \right ) 
\nonumber 
\nonumber \eeq 
\subsection*{Subspace No. 59 with {\boldmath $N=2$, $s=2$, $m_s=1$, $\Gamma_{10-1}$}}
\beq
\ket{\Psi_{59-1}} &=& \frac{1}{2 \sqrt{2}}\left ( {\scriptstyle \ket{000u00u0}} - {\scriptstyle \ket{000uu000}} + {\scriptstyle \ket{00u0000u}} - {\scriptstyle \ket{00u00u00}}\right . \nonumber \\                   \nonumber 
                                           && \hspace{0.5cm} \left . -{\scriptstyle \ket{0u0000u0}} + {\scriptstyle \ket{0u00u000}} - {\scriptstyle \ket{u000000u}} + {\scriptstyle \ket{u0000u00}} \nonumber \right )
\nonumber \\ 
 {\bf H}_{59}&=&
 \left ( 
\begin{array}{c}
0
\end{array} 
 \right ) 
\nonumber 
\nonumber \eeq 
\subsection*{Subspace No. 60 with {\boldmath $N=2$, $s=2$, $m_s=1$, $\Gamma_{10-2}$}}
\beq
\ket{\Psi_{60-1}} &=& \frac{1}{2 \sqrt{2}}\left ( {\scriptstyle \ket{00000u0u}} + {\scriptstyle \ket{0000u0u0}} + {\scriptstyle \ket{000uu000}} + {\scriptstyle \ket{00u00u00}}\right . \nonumber \\                   \nonumber 
                                           && \hspace{0.5cm} \left . -{\scriptstyle \ket{0u0000u0}} + {\scriptstyle \ket{0u0u0000}} - {\scriptstyle \ket{u000000u}} + {\scriptstyle \ket{u0u00000}} \nonumber \right )
\nonumber \\ 
 {\bf H}_{60}&=&
 \left ( 
\begin{array}{c}
0
\end{array} 
 \right ) 
\nonumber 
\nonumber \eeq 
\subsection*{Subspace No. 61 with {\boldmath $N=2$, $s=2$, $m_s=1$, $\Gamma_{10-3}$}}
\beq
\ket{\Psi_{61-1}} &=& -\frac{1}{2 \sqrt{2}}\left ( {\scriptstyle \ket{00000u0u}} - {\scriptstyle \ket{0000u0u0}} + {\scriptstyle \ket{000u00u0}} - {\scriptstyle \ket{00u0000u}}\right . \nonumber \\                   \nonumber 
                                            && \hspace{0.5cm} \left . -{\scriptstyle \ket{0u00u000}} + {\scriptstyle \ket{0u0u0000}} + {\scriptstyle \ket{u0000u00}} - {\scriptstyle \ket{u0u00000}} \nonumber \right )
\nonumber \\ 
 {\bf H}_{61}&=&
 \left ( 
\begin{array}{c}
0
\end{array} 
 \right ) 
\nonumber 
\nonumber \eeq 

\section{Groundstate of the Hubbard model for $U=\infty$ 
and $N_e=6$}
\label{appendix2a}
In the following we give the groundstate of the pure Hubbard model
with infinite on-site interaction $U$ and  six electrons, i.e. two holes
in the half filled cluster. 
\beq
\ket{\Psi_{GS}} &=& \ket{s=0,m_s=0,\Gamma_1,E_0=-2\sqrt{4+\sqrt{3}}\,t}\\
&=&\,C_1\ket{\Phi_{1}}+ \,C_2\,\ket{\Phi_{2}}+ \,C_3\,\ket{\Phi_{3}}+ \,C_4\,\ket{\Phi_{4}}
\nonumber 
\\
&&+ \,C_5\,\ket{\Phi_{5}}
+ \,C_6\,\ket{\Phi_{6}}
+ \,C_7\,\ket{\Phi_{7}}
+ \,C_8\,\ket{\Phi_{8}}
+ \,C_9\,\ket{\Phi_{9}}
\nonumber 
\eeq
\beq
\ket{\Phi_{1}}&=&\left ( {\scriptstyle \ket{00ddduuu}} - {\scriptstyle \ket{00dduduu}} - {\scriptstyle \ket{00dudduu}} - {\scriptstyle \ket{00duuudd}} + {\scriptstyle \ket{00uddduu}} + {\scriptstyle \ket{00uduudd}}\right . \nonumber \\
&& \hspace{0.5cm} \left .  + {\scriptstyle \ket{00uududd}} - {\scriptstyle \ket{00uuuddd}} + {\scriptstyle \ket{0dd0duuu}} - {\scriptstyle \ket{0dd0uuud}} - {\scriptstyle \ket{0ddu0udu}} + {\scriptstyle \ket{0du0duud}}\right . \nonumber \\ 
&& \hspace{0.5cm} \left .  + {\scriptstyle \ket{0du0uddu}} - {\scriptstyle \ket{0dud0duu}} - {\scriptstyle \ket{0dud0uud}} + {\scriptstyle \ket{0duu0dud}} - {\scriptstyle \ket{0ud0duud}} - {\scriptstyle \ket{0ud0uddu}}\right . \nonumber \\ 
&& \hspace{0.5cm} \left . -{\scriptstyle \ket{0udd0udu}} + {\scriptstyle \ket{0udu0ddu}} + {\scriptstyle \ket{0udu0udd}} + {\scriptstyle \ket{0uu0dddu}} - {\scriptstyle \ket{0uu0uddd}} + {\scriptstyle \ket{0uud0dud}}\right . \nonumber \\ 
&& \hspace{0.5cm} \left .  + {\scriptstyle \ket{d00duduu}} - {\scriptstyle \ket{d00duudu}} + {\scriptstyle \ket{d00uduud}} + {\scriptstyle \ket{d00uuddu}} - {\scriptstyle \ket{d0dud0uu}} - {\scriptstyle \ket{d0duu0du}}\right . \nonumber \\ 
&& \hspace{0.5cm} \left . -{\scriptstyle \ket{d0udu0ud}} + {\scriptstyle \ket{d0uud0du}} + {\scriptstyle \ket{dd00uudu}} - {\scriptstyle \ket{dd00uuud}} - {\scriptstyle \ket{dd0udu0u}} + {\scriptstyle \ket{dddu0uu0}}\right . \nonumber \\ 
&& \hspace{0.5cm} \left .  + {\scriptstyle \ket{ddduuu00}} - {\scriptstyle \ket{ddu0udu0}} + {\scriptstyle \ket{ddudu00u}} - {\scriptstyle \ket{dduduu00}} - {\scriptstyle \ket{dduu00du}} + {\scriptstyle \ket{dduu00ud}}\right . \nonumber \\ 
&& \hspace{0.5cm} \left . -{\scriptstyle \ket{dduudu00}} + {\scriptstyle \ket{dduuud00}} - {\scriptstyle \ket{du00dduu}} - {\scriptstyle \ket{du00uudd}} - {\scriptstyle \ket{du0ddu0u}} + {\scriptstyle \ket{du0udd0u}}\right . \nonumber \\ 
&& \hspace{0.5cm} \left .  + {\scriptstyle \ket{du0udu0d}} - {\scriptstyle \ket{dud0duu0}} - {\scriptstyle \ket{dud0uud0}} + {\scriptstyle \ket{dudd00uu}} - {\scriptstyle \ket{duddu00u}} + {\scriptstyle \ket{duu0dud0}}\right . \nonumber \\ 
&& \hspace{0.5cm} \left .  + {\scriptstyle \ket{duud0du0}} - {\scriptstyle \ket{duud0ud0}} + {\scriptstyle \ket{duudd00u}} - {\scriptstyle \ket{duudu00d}} + {\scriptstyle \ket{duuu00dd}} + {\scriptstyle \ket{duuu0dd0}}\right . \nonumber \\ 
&& \hspace{0.5cm} \left . -{\scriptstyle \ket{u00dduud}} - {\scriptstyle \ket{u00duddu}} + {\scriptstyle \ket{u00uddud}} - {\scriptstyle \ket{u00ududd}} - {\scriptstyle \ket{u0ddu0ud}} + {\scriptstyle \ket{u0dud0du}}\right . \nonumber \\ 
&& \hspace{0.5cm} \left .  + {\scriptstyle \ket{u0udd0ud}} + {\scriptstyle \ket{u0udu0dd}} + {\scriptstyle \ket{ud00dduu}} + {\scriptstyle \ket{ud00uudd}} - {\scriptstyle \ket{ud0dud0u}} - {\scriptstyle \ket{ud0duu0d}}\right . \nonumber \\ 
&& \hspace{0.5cm} \left .  + {\scriptstyle \ket{ud0uud0d}} - {\scriptstyle \ket{udd0udu0}} - {\scriptstyle \ket{uddd00uu}} - {\scriptstyle \ket{uddd0uu0}} + {\scriptstyle \ket{uddu0du0}} - {\scriptstyle \ket{uddu0ud0}}\right . \nonumber \\ 
&& \hspace{0.5cm} \left .  + {\scriptstyle \ket{uddud00u}} - {\scriptstyle \ket{udduu00d}} + {\scriptstyle \ket{udu0ddu0}} + {\scriptstyle \ket{udu0udd0}} - {\scriptstyle \ket{uduu00dd}} + {\scriptstyle \ket{uduud00d}}\right . \nonumber \\ 
&& \hspace{0.5cm} \left .  + {\scriptstyle \ket{uu00dddu}} - {\scriptstyle \ket{uu00ddud}} + {\scriptstyle \ket{uu0dud0d}} + {\scriptstyle \ket{uud0dud0}} - {\scriptstyle \ket{uudd00du}} + {\scriptstyle \ket{uudd00ud}}\right . \nonumber \\ 
&& \hspace{0.5cm} \left . -{\scriptstyle \ket{uudddu00}} + {\scriptstyle \ket{uuddud00}} - {\scriptstyle \ket{uudud00d}} + {\scriptstyle \ket{uududd00}} - {\scriptstyle \ket{uuud0dd0}} - {\scriptstyle \ket{uuuddd00}} \nonumber \right )
\eeq
\beq
\ket{\Phi_{2}}&=&
\left ( {\scriptstyle \ket{00dduudu}} - {\scriptstyle \ket{00dduuud}} + {\scriptstyle \ket{00uudddu}} - {\scriptstyle \ket{00uuddud}} - {\scriptstyle \ket{0dd0uduu}} + {\scriptstyle \ket{0dd0uudu}}\right . \nonumber \\                        \nonumber 
             && \hspace{0.5cm} \left .  + {\scriptstyle \ket{0ddu0duu}} - {\scriptstyle \ket{0duu0ddu}} + {\scriptstyle \ket{0udd0uud}} - {\scriptstyle \ket{0uu0ddud}} + {\scriptstyle \ket{0uu0dudd}} - {\scriptstyle \ket{0uud0udd}}\right . \nonumber \\ 
             && \hspace{0.5cm} \left . -{\scriptstyle \ket{d00dduuu}} + {\scriptstyle \ket{d00duuud}} + {\scriptstyle \ket{d0udd0uu}} - {\scriptstyle \ket{d0uud0ud}} + {\scriptstyle \ket{dd00duuu}} - {\scriptstyle \ket{dd00uduu}}\right . \nonumber \\ 
             && \hspace{0.5cm} \left .  + {\scriptstyle \ket{dd0uud0u}} + {\scriptstyle \ket{dddu00uu}} - {\scriptstyle \ket{ddduu00u}} + {\scriptstyle \ket{ddu0duu0}} - {\scriptstyle \ket{ddud00uu}} - {\scriptstyle \ket{ddud0uu0}}\right . \nonumber \\ 
             && \hspace{0.5cm} \left .  + {\scriptstyle \ket{du0duu0d}} + {\scriptstyle \ket{dudd0uu0}} + {\scriptstyle \ket{dudduu00}} - {\scriptstyle \ket{duu0ddu0}} - {\scriptstyle \ket{duuud00d}} + {\scriptstyle \ket{duuudd00}}\right . \nonumber \\ 
             && \hspace{0.5cm} \left . -{\scriptstyle \ket{u00udddu}} + {\scriptstyle \ket{u00uuddd}} + {\scriptstyle \ket{u0ddu0du}} - {\scriptstyle \ket{u0duu0dd}} - {\scriptstyle \ket{ud0udd0u}} + {\scriptstyle \ket{udd0uud0}}\right . \nonumber \\ 
             && \hspace{0.5cm} \left .  + {\scriptstyle \ket{udddu00u}} - {\scriptstyle \ket{uddduu00}} - {\scriptstyle \ket{uduu0dd0}} - {\scriptstyle \ket{uduudd00}} + {\scriptstyle \ket{uu00dudd}} - {\scriptstyle \ket{uu00uddd}}\right . \nonumber \\ 
             && \hspace{0.5cm} \left . -{\scriptstyle \ket{uu0ddu0d}} - {\scriptstyle \ket{uud0udd0}} + {\scriptstyle \ket{uudu00dd}} + {\scriptstyle \ket{uudu0dd0}} - {\scriptstyle \ket{uuud00dd}} + {\scriptstyle \ket{uuudd00d}} \nonumber \right )
\eeq
\beq
\ket{\Phi_{3}} &=&  
\left ( {\scriptstyle \ket{00dududu}} - {\scriptstyle \ket{00duduud}} + {\scriptstyle \ket{00ududdu}} - {\scriptstyle \ket{00ududud}} + {\scriptstyle \ket{0ddu0uud}} + {\scriptstyle \ket{0ddudu0u}}\right . \nonumber \\                        \nonumber 
             && \hspace{0.5cm} \left . -{\scriptstyle \ket{0dduuu0d}} - {\scriptstyle \ket{0du0udud}} + {\scriptstyle \ket{0du0uudd}} + {\scriptstyle \ket{0dudud0u}} + {\scriptstyle \ket{0duduu0d}} - {\scriptstyle \ket{0duu0udd}}\right . \nonumber \\ 
             && \hspace{0.5cm} \left .  + {\scriptstyle \ket{0duudu0d}} - {\scriptstyle \ket{0duuud0d}} - {\scriptstyle \ket{0ud0dduu}} + {\scriptstyle \ket{0ud0dudu}} + {\scriptstyle \ket{0udd0duu}} + {\scriptstyle \ket{0udddu0u}}\right . \nonumber \\ 
             && \hspace{0.5cm} \left . -{\scriptstyle \ket{0uddud0u}} - {\scriptstyle \ket{0ududd0u}} - {\scriptstyle \ket{0ududu0d}} - {\scriptstyle \ket{0uud0ddu}} + {\scriptstyle \ket{0uuddd0u}} - {\scriptstyle \ket{0uudud0d}}\right . \nonumber \\ 
             && \hspace{0.5cm} \left . -{\scriptstyle \ket{d00ududu}} + {\scriptstyle \ket{d00uuudd}} + {\scriptstyle \ket{d0duduu0}} + {\scriptstyle \ket{d0duuud0}} + {\scriptstyle \ket{d0udu0du}} + {\scriptstyle \ket{d0ududu0}}\right . \nonumber \\ 
             && \hspace{0.5cm} \left . -{\scriptstyle \ket{d0uduud0}} - {\scriptstyle \ket{d0uudud0}} - {\scriptstyle \ket{d0uuu0dd}} + {\scriptstyle \ket{d0uuudd0}} + {\scriptstyle \ket{dd0u0udu}} - {\scriptstyle \ket{dd0u0uud}}\right . \nonumber \\ 
             && \hspace{0.5cm} \left .  + {\scriptstyle \ket{dd0uuu0d}} - {\scriptstyle \ket{ddu0u0du}} + {\scriptstyle \ket{ddu0u0ud}} + {\scriptstyle \ket{ddu0uud0}} - {\scriptstyle \ket{dduu0ud0}} - {\scriptstyle \ket{dduuu00d}}\right . \nonumber \\ 
             && \hspace{0.5cm} \left .  + {\scriptstyle \ket{du00dudu}} - {\scriptstyle \ket{du00uddu}} - {\scriptstyle \ket{du0d0duu}} + {\scriptstyle \ket{du0d0udu}} + {\scriptstyle \ket{du0dud0u}} - {\scriptstyle \ket{du0u0ddu}}\right . \nonumber \\ 
             && \hspace{0.5cm} \left . -{\scriptstyle \ket{du0u0udd}} + {\scriptstyle \ket{dud0d0uu}} + {\scriptstyle \ket{dud0u0du}} + {\scriptstyle \ket{dudu00du}} + {\scriptstyle \ket{dudu0ud0}} - {\scriptstyle \ket{dudud00u}}\right . \nonumber \\ 
             && \hspace{0.5cm} \left .  + {\scriptstyle \ket{dududu00}} - {\scriptstyle \ket{duu0d0du}} + {\scriptstyle \ket{duu0u0dd}} - {\scriptstyle \ket{duu0udd0}} - {\scriptstyle \ket{duud00du}} + {\scriptstyle \ket{duudud00}}\right . \nonumber \\ 
             && \hspace{0.5cm} \left . -{\scriptstyle \ket{u00ddduu}} + {\scriptstyle \ket{u00dudud}} + {\scriptstyle \ket{u0ddd0uu}} - {\scriptstyle \ket{u0ddduu0}} + {\scriptstyle \ket{u0ddudu0}} - {\scriptstyle \ket{u0dud0ud}}\right . \nonumber \\ 
             && \hspace{0.5cm} \left .  + {\scriptstyle \ket{u0duddu0}} - {\scriptstyle \ket{u0dudud0}} - {\scriptstyle \ket{u0udddu0}} - {\scriptstyle \ket{u0ududd0}} + {\scriptstyle \ket{ud00duud}} - {\scriptstyle \ket{ud00udud}}\right . \nonumber \\ 
             && \hspace{0.5cm} \left .  + {\scriptstyle \ket{ud0d0duu}} + {\scriptstyle \ket{ud0d0uud}} - {\scriptstyle \ket{ud0u0dud}} + {\scriptstyle \ket{ud0u0udd}} - {\scriptstyle \ket{ud0udu0d}} - {\scriptstyle \ket{udd0d0uu}}\right . \nonumber \\ 
             && \hspace{0.5cm} \left .  + {\scriptstyle \ket{udd0duu0}} + {\scriptstyle \ket{udd0u0ud}} + {\scriptstyle \ket{uddu00ud}} - {\scriptstyle \ket{uddudu00}} - {\scriptstyle \ket{udu0d0ud}} - {\scriptstyle \ket{udu0u0dd}}\right . \nonumber \\ 
             && \hspace{0.5cm} \left . -{\scriptstyle \ket{udud00ud}} - {\scriptstyle \ket{udud0du0}} + {\scriptstyle \ket{ududu00d}} - {\scriptstyle \ket{ududud00}} + {\scriptstyle \ket{uu0d0ddu}} - {\scriptstyle \ket{uu0d0dud}}\right . \nonumber \\ 
             && \hspace{0.5cm} \left . -{\scriptstyle \ket{uu0ddd0u}} - {\scriptstyle \ket{uud0d0du}} + {\scriptstyle \ket{uud0d0ud}} - {\scriptstyle \ket{uud0ddu0}} + {\scriptstyle \ket{uudd0du0}} + {\scriptstyle \ket{uuddd00u}} \nonumber \right )
\eeq
\beq
 \ket{\Phi_{4}} &=&
\left ( {\scriptstyle \ket{00duuddu}} - {\scriptstyle \ket{00udduud}} + {\scriptstyle \ket{0ddd0uuu}} - {\scriptstyle \ket{0du0dduu}} + {\scriptstyle \ket{0ud0uudd}} - {\scriptstyle \ket{0uuu0ddd}}\right . \nonumber \\                        \nonumber 
             && \hspace{0.5cm} \left . -{\scriptstyle \ket{d00udduu}} + {\scriptstyle \ket{d0ddu0uu}} + {\scriptstyle \ket{dd0duu0u}} + {\scriptstyle \ket{ddd0uuu0}} - {\scriptstyle \ket{dduu0du0}} - {\scriptstyle \ket{dduud00u}}\right . \nonumber \\ 
             && \hspace{0.5cm} \left .  + {\scriptstyle \ket{du00duud}} - {\scriptstyle \ket{duud00ud}} + {\scriptstyle \ket{duuddu00}} + {\scriptstyle \ket{u00duudd}} - {\scriptstyle \ket{u0uud0dd}} - {\scriptstyle \ket{ud00uddu}}\right . \nonumber \\ 
             && \hspace{0.5cm} \left .  + {\scriptstyle \ket{uddu00du}} - {\scriptstyle \ket{udduud00}} - {\scriptstyle \ket{uu0udd0d}} + {\scriptstyle \ket{uudd0ud0}} + {\scriptstyle \ket{uuddu00d}} - {\scriptstyle \ket{uuu0ddd0}} \nonumber \right )
\eeq
\beq
\ket{\Phi_{5}} &=&\left ( {\scriptstyle \ket{0d0dduuu}} - {\scriptstyle \ket{0d0duudu}} - {\scriptstyle \ket{0dduudu0}} - {\scriptstyle \ket{0dudd0uu}} - {\scriptstyle \ket{0dudduu0}} + {\scriptstyle \ket{0duud0du}}\right . \nonumber \\                        \nonumber 
             && \hspace{0.5cm} \left .  + {\scriptstyle \ket{0u0uddud}} - {\scriptstyle \ket{0u0uuddd}} - {\scriptstyle \ket{0uddu0ud}} + {\scriptstyle \ket{0uduu0dd}} + {\scriptstyle \ket{0uduudd0}} + {\scriptstyle \ket{0uuddud0}}\right . \nonumber \\ 
             && \hspace{0.5cm} \left .  + {\scriptstyle \ket{d0d0uduu}} - {\scriptstyle \ket{d0d0uuud}} - {\scriptstyle \ket{d0du0duu}} - {\scriptstyle \ket{d0duud0u}} - {\scriptstyle \ket{d0uddu0u}} + {\scriptstyle \ket{d0uu0dud}}\right . \nonumber \\ 
             && \hspace{0.5cm} \left . -{\scriptstyle \ket{dd0uudu0}} + {\scriptstyle \ket{ddduu0u0}} - {\scriptstyle \ket{ddu0du0u}} + {\scriptstyle \ket{ddud0u0u}} - {\scriptstyle \ket{du0du0ud}} + {\scriptstyle \ket{du0ud0ud}}\right . \nonumber \\ 
             && \hspace{0.5cm} \left .  + {\scriptstyle \ket{du0uddu0}} - {\scriptstyle \ket{dud00uud}} - {\scriptstyle \ket{dud0uu0d}} - {\scriptstyle \ket{duddu0u0}} + {\scriptstyle \ket{duu00dud}} + {\scriptstyle \ket{duuu0d0d}}\right . \nonumber \\ 
             && \hspace{0.5cm} \left . -{\scriptstyle \ket{u0dd0udu}} + {\scriptstyle \ket{u0duud0d}} + {\scriptstyle \ket{u0u0dddu}} - {\scriptstyle \ket{u0u0dudd}} + {\scriptstyle \ket{u0ud0udd}} + {\scriptstyle \ket{u0uddu0d}}\right . \nonumber \\ 
             && \hspace{0.5cm} \left . -{\scriptstyle \ket{ud0du0du}} - {\scriptstyle \ket{ud0duud0}} + {\scriptstyle \ket{ud0ud0du}} - {\scriptstyle \ket{udd00udu}} - {\scriptstyle \ket{uddd0u0u}} + {\scriptstyle \ket{udu00ddu}}\right . \nonumber \\ 
             && \hspace{0.5cm} \left .  + {\scriptstyle \ket{udu0dd0u}} + {\scriptstyle \ket{uduud0d0}} + {\scriptstyle \ket{uu0ddud0}} + {\scriptstyle \ket{uud0ud0d}} - {\scriptstyle \ket{uudu0d0d}} - {\scriptstyle \ket{uuudd0d0}} \nonumber \right )
\eeq
\beq
 \ket{\Phi_{6}} &=&\left ( {\scriptstyle \ket{0d0udduu}} - {\scriptstyle \ket{0d0uuddu}} - {\scriptstyle \ket{0dddu0uu}} - {\scriptstyle \ket{0ddduuu0}} + {\scriptstyle \ket{0dduu0du}} - {\scriptstyle \ket{0duuddu0}}\right . \nonumber \\                        \nonumber 
             && \hspace{0.5cm} \left .  + {\scriptstyle \ket{0u0dduud}} - {\scriptstyle \ket{0u0duudd}} + {\scriptstyle \ket{0udduud0}} - {\scriptstyle \ket{0uudd0ud}} + {\scriptstyle \ket{0uuud0dd}} + {\scriptstyle \ket{0uuuddd0}}\right . \nonumber \\ 
             && \hspace{0.5cm} \left . -{\scriptstyle \ket{d0dd0uuu}} - {\scriptstyle \ket{d0dduu0u}} + {\scriptstyle \ket{d0u0dduu}} - {\scriptstyle \ket{d0u0duud}} + {\scriptstyle \ket{d0ud0uud}} - {\scriptstyle \ket{d0uudd0u}}\right . \nonumber \\ 
             && \hspace{0.5cm} \left . -{\scriptstyle \ket{dd0du0uu}} - {\scriptstyle \ket{dd0duuu0}} + {\scriptstyle \ket{dd0ud0uu}} - {\scriptstyle \ket{ddd00uuu}} - {\scriptstyle \ket{ddd0uu0u}} + {\scriptstyle \ket{ddu00duu}}\right . \nonumber \\ 
             && \hspace{0.5cm} \left .  + {\scriptstyle \ket{dduu0d0u}} + {\scriptstyle \ket{dduud0u0}} + {\scriptstyle \ket{du0dduu0}} - {\scriptstyle \ket{duu0du0d}} + {\scriptstyle \ket{duud0u0d}} - {\scriptstyle \ket{duudd0u0}}\right . \nonumber \\ 
             && \hspace{0.5cm} \left .  + {\scriptstyle \ket{u0d0uddu}} - {\scriptstyle \ket{u0d0uudd}} + {\scriptstyle \ket{u0dduu0d}} - {\scriptstyle \ket{u0du0ddu}} + {\scriptstyle \ket{u0uu0ddd}} + {\scriptstyle \ket{u0uudd0d}}\right . \nonumber \\ 
             && \hspace{0.5cm} \left . -{\scriptstyle \ket{ud0uudd0}} + {\scriptstyle \ket{udd0ud0u}} - {\scriptstyle \ket{uddu0d0u}} + {\scriptstyle \ket{udduu0d0}} - {\scriptstyle \ket{uu0du0dd}} + {\scriptstyle \ket{uu0ud0dd}}\right . \nonumber \\ 
             && \hspace{0.5cm} \left .  + {\scriptstyle \ket{uu0uddd0}} - {\scriptstyle \ket{uud00udd}} - {\scriptstyle \ket{uudd0u0d}} - {\scriptstyle \ket{uuddu0d0}} + {\scriptstyle \ket{uuu00ddd}} + {\scriptstyle \ket{uuu0dd0d}} \nonumber \right )
\eeq
\beq
 \ket{\Phi_{7}} &=&
\left ( {\scriptstyle \ket{0d0uduud}} - {\scriptstyle \ket{0d0uuudd}} - {\scriptstyle \ket{0dduduu0}} - {\scriptstyle \ket{0dudu0ud}} - {\scriptstyle \ket{0dududu0}} + {\scriptstyle \ket{0duuu0dd}}\right . \nonumber \\                        \nonumber 
             && \hspace{0.5cm} \left .  + {\scriptstyle \ket{0u0ddduu}} - {\scriptstyle \ket{0u0duddu}} - {\scriptstyle \ket{0uddd0uu}} + {\scriptstyle \ket{0udud0du}} + {\scriptstyle \ket{0ududud0}} + {\scriptstyle \ket{0uududd0}}\right . \nonumber \\ 
             && \hspace{0.5cm} \left . -{\scriptstyle \ket{d0du0udu}} - {\scriptstyle \ket{d0dudu0u}} + {\scriptstyle \ket{d0u0uddu}} - {\scriptstyle \ket{d0u0uudd}} - {\scriptstyle \ket{d0udud0u}} + {\scriptstyle \ket{d0uu0udd}}\right . \nonumber \\ 
             && \hspace{0.5cm} \left . -{\scriptstyle \ket{dd0uuud0}} - {\scriptstyle \ket{ddu0uu0d}} + {\scriptstyle \ket{dduu0u0d}} + {\scriptstyle \ket{dduuu0d0}} - {\scriptstyle \ket{du0du0du}} + {\scriptstyle \ket{du0ud0du}}\right . \nonumber \\ 
             && \hspace{0.5cm} \left .  + {\scriptstyle \ket{du0udud0}} - {\scriptstyle \ket{dud00udu}} - {\scriptstyle \ket{dud0du0u}} + {\scriptstyle \ket{duu00ddu}} + {\scriptstyle \ket{duud0d0u}} - {\scriptstyle \ket{duudu0d0}}\right . \nonumber \\ 
             && \hspace{0.5cm} \left .  + {\scriptstyle \ket{u0d0dduu}} - {\scriptstyle \ket{u0d0duud}} - {\scriptstyle \ket{u0dd0duu}} + {\scriptstyle \ket{u0dudu0d}} + {\scriptstyle \ket{u0ud0dud}} + {\scriptstyle \ket{u0udud0d}}\right . \nonumber \\ 
             && \hspace{0.5cm} \left . -{\scriptstyle \ket{ud0du0ud}} - {\scriptstyle \ket{ud0dudu0}} + {\scriptstyle \ket{ud0ud0ud}} - {\scriptstyle \ket{udd00uud}} - {\scriptstyle \ket{uddu0u0d}} + {\scriptstyle \ket{uddud0u0}}\right . \nonumber \\ 
             && \hspace{0.5cm} \left .  + {\scriptstyle \ket{udu00dud}} + {\scriptstyle \ket{udu0ud0d}} + {\scriptstyle \ket{uu0dddu0}} + {\scriptstyle \ket{uud0dd0u}} - {\scriptstyle \ket{uudd0d0u}} - {\scriptstyle \ket{uuddd0u0}} \nonumber \right )
\eeq
\beq
 \ket{\Phi_{8}} &=&
\left ( {\scriptstyle \ket{0ddduu0u}} - {\scriptstyle \ket{0dduud0u}} + {\scriptstyle \ket{0duudd0u}} - {\scriptstyle \ket{0udduu0d}} + {\scriptstyle \ket{0uuddu0d}} - {\scriptstyle \ket{0uuudd0d}}\right . \nonumber \\                        \nonumber 
             && \hspace{0.5cm} \left .  + {\scriptstyle \ket{d0dduuu0}} - {\scriptstyle \ket{d0udduu0}} + {\scriptstyle \ket{d0uuddu0}} + {\scriptstyle \ket{dd0d0uuu}} - {\scriptstyle \ket{dd0u0duu}} + {\scriptstyle \ket{ddd0u0uu}}\right . \nonumber \\ 
             && \hspace{0.5cm} \left . -{\scriptstyle \ket{ddu0d0uu}} - {\scriptstyle \ket{du0d0uud}} + {\scriptstyle \ket{duu0d0ud}} - {\scriptstyle \ket{u0dduud0}} + {\scriptstyle \ket{u0duudd0}} - {\scriptstyle \ket{u0uuddd0}}\right . \nonumber \\ 
             && \hspace{0.5cm} \left .  + {\scriptstyle \ket{ud0u0ddu}} - {\scriptstyle \ket{udd0u0du}} + {\scriptstyle \ket{uu0d0udd}} - {\scriptstyle \ket{uu0u0ddd}} + {\scriptstyle \ket{uud0u0dd}} - {\scriptstyle \ket{uuu0d0dd}} \nonumber \right )
\eeq
\beq
\ket{\Phi_{9}} &=& 
\left ( {\scriptstyle \ket{0duddu0u}} - {\scriptstyle \ket{0uduud0d}} + {\scriptstyle \ket{d0duudu0}} - {\scriptstyle \ket{du0u0dud}} + {\scriptstyle \ket{dud0u0ud}} - {\scriptstyle \ket{u0uddud0}}\right . \nonumber \\  \nonumber 
            && \hspace{0.5cm} \left .  + {\scriptstyle \ket{ud0d0udu}} - {\scriptstyle \ket{udu0d0du}} \nonumber \right )
\eeq 

\beq
\,C_1 &=&\frac{\sqrt{\frac{1}{2} \left(4+\sqrt{3}\right)} \left(3+2 \sqrt{3}\right)}{180+84 \sqrt{3}}\nonumber \\
\,C_2 &=&-\frac{1}{12 \sqrt{2 \left(4+\sqrt{3}\right)}}\nonumber \\
\,C_3 &=&-\frac{\sqrt{\frac{1}{6} \left(4+\sqrt{3}\right)}}{60+28 \sqrt{3}}\nonumber \\
\,C_4 &=&\frac{\sqrt{\frac{1}{2} \left(4+\sqrt{3}\right)} \left(3+2 \sqrt{3}\right)}{90+42 \sqrt{3}}\nonumber \\
\,C_5 &=&-\frac{1}{12 \sqrt{2}}\nonumber \\
\,C_6 &=&\frac{1}{48} \left(\sqrt{2}+\sqrt{6}\right)\nonumber \\
\,C_7 &=&\frac{1}{48} \left(\sqrt{2}-\sqrt{6}\right)\nonumber \\
\,C_8 &=&\frac{51+29 \sqrt{3}}{12 \sqrt{2 \left(4+\sqrt{3}\right)} \left(15+7 \sqrt{3}\right)}\nonumber \\
\,C_9 &=&-\frac{1}{4 \sqrt{2 \left(4+\sqrt{3}\right)}}\nonumber \eeq


\begin{thebibliography}{00}
\bibitem{Senechal00}
 S. Senechal, D. Perez and M. Pioro-Ladriere, Phys. Rev. Lett.\textbf{ 84},
 522 (2000).
\bibitem{Maier05}  
 T. Maier, M. Jarrell, T. Pruschke and M. H. Hettler, Rev. Mod. Phys. \textbf{77}, 1027 (2005).
\bibitem{Wang05}  W. Z. Wang, Phys. Rev. B \textbf{72}, 125116 (2005).
\bibitem{Potthoff03} M. Potthoff, M. Aichhorn, and C. Dahnken 
Phys. Rev. Lett. 91, 206402 (2003)
\bibitem{Balzer08} M. Balzer, W. Hanke, and M. Potthoff,
Phys. Rev. B\textbf{ 77}, 045133 (2008)
\bibitem{Potthoff07}  
M. Potthoff and M. Balzer, Phys. Rev. B \textbf{75}, 125112 (2007)
\bibitem{Schumann02} R. Schumann, Ann. Phys. (Leipzig) \textbf{11}, 49 (2002)
\bibitem{Kocharian06a} A.M. Kocharian, G.W. Fernando, K. Palandage, and J.W. Davenport, Phys. Rev.
B \textbf{74}, 024511 (2006).
\bibitem{Tsai06} W. F. Tsai and S. A. Kivelson, Phys. Rev. \textbf{B 73}, 214510 (2006).
\bibitem{Kocharian06b} A. N. Kocharian, G. W. Fernando, K. Palandage, and J. W. Davenport, J.
Magn. Magn. Mater. 300, e585 (2006).
\bibitem{Palandage07} K. Palandage, G. W. Fernando, A. N. Kocharian, and J. W. Davenport, J.
Comput.-Aided Mater. Des. 14, 103 (2007)
\bibitem{Kocharian08} A. N. Kocharian, G. W. Fernando, K. Palandage, and J. W. Davenport 
Phys. Rev. B 78, 075431 (2008)
\bibitem{Tsai08} W.-F. Tsai, H. Yao, A. Läuchli, and S. A. Kivelson, Physical Review B 77 214502
(2008)
\bibitem{Kocharian09}A. N. Kocharian, G. W. Fernando, K. Palandage, and J. W. Davenport , Physics Letters A 373 (2009) 1074-1082
\bibitem{Georges99}A. Georges and Y. Meir, Phys. Rev. Lett. 82, 3508 (1999)
\bibitem{Craig04}N. J. Craig, J. M. Taylor, E. A. Lester, C. M. Marcus, M. P. Hanson, and A. C.
Gossard,  Science \textbf{304}, 565 (2004)
\bibitem{Chen04}J. C. Chen, A. M. Chang, and M. R. Melloch, 
Phys. Rev. Lett. \textbf{92}, 176801 (2004)
\bibitem{Vidan04} A. Vidan, R. M. Westervelt, M. Stopa, M. Hanson, and A. C. Gossard, 
Appl. Phys. Lett. \textbf{85}, 3602 (2004)
\bibitem{Vidan05} A. Vidan, R. M. Westervelt, M. Stopa, M. Hanson, and A. C. Gossard, 
J. Supercond. \textbf{18},  (2005)
\bibitem{Kikoin06} T. Kuzmenko, K. Kikoin, and Y. Avishai, Phys. Rev. Lett. \textbf{96}, 046601
(2006)
\bibitem{Zhao08}Zhao-tan Jiang and Qing-Zhen Han, Phys. Rev. B \textbf{78}, 035307 (2008) 
\bibitem{Cuniberti05} G. Cuniberti, G. Fagas, and K. Richter (eds.), 
Introducing Molecular Electronics, (Springer, Berlin, 2005)
\bibitem{Heersche06} H. B. Heersche, Z. de Groot, J. A. Folk, H. S. J. van der Zant, 
C. Romeike, M. R. Wegewijs, L. Zobbi, D. Barreca, E. Tondello, and A. Cornia, 
Phys. Rev. Lett. \textbf{96}, 206801 (2006)
\bibitem{Donarini06} A. Donarini, M. Grifoni, and K. Richter, Phys. Rev. Lett. \textbf{97}, 166801
(2006)
\bibitem{Falicov69} L. M. Falicov and R. A. Harris,  J. Chem. Phys. \textbf{51}, 3153 (1969) 
\bibitem{Falicov84} L. M. Falicov and R.M. Victora, Phys. Rev. B \textbf{30}, 1696 (1984).
\bibitem{Falicov88} 
L. Milans del Bosch and L. M. Falicov, Phys. Rev. B 37, 6073 (1988)
\bibitem{Callaway87}J. Callaway, D. P. Chen, and R. Tang, Phys. Rev. B 35, 3705 (1987).
\bibitem{Schumann08}R. Schumann, Annalen der Physik 17 221 (2008)\\ 
for eigenvectors see: arXiv:cond-mat/0701060
\bibitem{Schumann09}R. Schumann, arXiv0902.0900v1(cond-mat.str-el)
\bibitem{Lieb68} E.H. Lieb and F.Y. Wu, Phys. Rev. Lett. \textbf{20}, 1445 (1968).
\bibitem{Korepin00}Deguchi, F.H.L. Essler, F. G\"ohmann, A. Kl\"umper, 
 V.E. Korepin, and K. Kusakabe, Phys. Rep. \textbf{331}, 197 (2000).
\bibitem{FuldeBuch} P. Fulde, Electron Correlations in Molecules and
Solids (Springer, Berlin Heidelberg New York Tokyo), 1997, p. 243.
\bibitem{Davoudi06}  B. Davoudi and A.-M. S. Tremblay, 
Phys. Rev. B \textbf{74}, 035113 (2006).
\bibitem{Richter08}J. Richter, private communication
\bibitem{FreFal90} K. Freericks and L. M. Falicov, Phys. Rev. B \textbf{42}, 4960 (1990)
\bibitem{Caffarel98}M. Caffarel and R. Mosseri, Phys. Rev. \textbf{B 57 }, 12651 (1998).
\bibitem{Nagaoka65} Y. Nagaoka, Solid State Communications 3 (1965) 409-412
\bibitem{Tasaki89} H. Tasaki, Phys. Rev. B 40 (1989) 13, 9192-9193
\bibitem{Takahashi79}M. Takahashi, J. Phys. Soc. Japan \textbf{47}, 47, 1978
\bibitem{Schumann06} R. Schumann, Physica C \textbf{ 460-462} 1015  (2007)
\bibitem{Callaway91} L. Tan, Q. Li und J. Callaway, Phys. Rev. B 44, 341-350 (1991)
\bibitem{CornwellBook}J.F. Cornwell, Group Theory in Physics (Academic Press, London, 1984).
\end{thebibliography}
\end{document}